\documentclass[11pt]{article}
\usepackage{graphicx}
\usepackage{amsmath}
\usepackage{amssymb}
\usepackage{latexsym}
\usepackage{bm}
\usepackage{overpic}
\usepackage[dvips]{color}
\usepackage{cases}
\usepackage{lscape}
\usepackage{here}
\usepackage{ulem}
\usepackage{arydshln}
\usepackage{float}
\newcommand{\ds}{\displaystyle}
\usepackage[right]{lineno}
\newtheorem{theorem}{Theorem}
\newtheorem{lemma}{Lemma}
\newtheorem{remark}{Remark}
\newtheorem{example}{Example}

\date{\today}
\begin{document}
%
%
\begin{center}
{\Large\bf Analysis of the Convergence Speed of the Arimoto-Blahut\\[2mm]
Algorithm by the Second Order Recurrence Formula}\\[7mm]
\end{center}

{\normalsize

\hspace{50mm}Kenji~Nakagawa, Member, IEEE, 

\hspace{50mm}Yoshinori~Takei, Member, IEEE, 

\hspace{50mm}Shin-ichiro~Hara, non-member, 

\hspace{50mm}Kohei~Watabe, Member, IEEE}\\
\begin{center}
\today
\end{center}
\footnote[0]{The material in this paper was presented in part at 2018 International Symposium on Information Theory and its Applications (ISITA2018).\\
K. Nakagawa and K. Watabe are with the Department of Electrical and Electronics and Information Engineering, Nagaoka University of Technology, Nagaoka, Niigata 940-2188, Japan (e-mail: nakagawa@nagaokaut.ac.jp; k\_watabe@vos.nagaokaut.ac.jp). Y. Takei is with the National Institute of Technology, Akita College, Akita 011-8511, Japan (e-mail: ytakei@akita-nct.ac.jp). S. Hara is with the Department of General Education, Nagaoka University of Technology (e-mail: sinara@blade.nagaokaut.ac.jp).}
\begin{abstract}
In this paper, we investigate the convergence speed of the Arimoto-Blahut algorithm. For many channel matrices the convergence is exponential, but for some channel matrices it is slower than exponential. By analyzing the Taylor expansion of the defining function of the Arimoto-Blahut algorithm, we will make the conditions clear for the exponential or slower convergence. The analysis of the slow convergence is new in this paper. Based on the analysis, we will compare the convergence speed of the Arimoto-Blahut algorithm numerically with the values obtained in our theorems for several channel matrices. The purpose of this paper is a complete understanding of the convergence speed of the Arimoto-Blahut algorithm.
\end{abstract}



\baselineskip 5mm

\section{Introduction}
The Arimoto-Blahut algorithm \cite{ari}, \cite{bla} is an algorithm for calculating a sequence of input distributions $\{\bm\lambda^N\}_{N=0,1,\ldots}$ converging to $\bm\lambda^\ast$ that achieves the capacity $C$ of a discrete memoryless channel. In the algorithm, $\bm\lambda^{N+1}$ is obtained by a function of $\bm\lambda^N$, i.e., $\bm\lambda^{N+1}=F(\bm\lambda^N)$, where $F$ is a differentiable function from the set $\Delta(\cal X)$ of the input distributions to itself. $\bm\lambda^\ast$ is the fixed point of $F$. In this paper and \cite{nak4}, the convergence speed of the Arimoto-Blahut algorithm is analyzed by the Taylor expansion of $F(\bm\lambda)$ about $\bm\lambda=\bm\lambda^\ast$.

Previous researches \cite{ari}, \cite{mat}, \cite{yu} deal with the case where $\bm\lambda^\ast=(\lambda^\ast_1,\ldots,\lambda^\ast_m)$ is an interior point of $\Delta(\cal X)$, i.e., $\lambda^\ast_i>0,\,i=1,\ldots,m$. Let $J(\bm\lambda^\ast)$ be the Jacobian matrix of the function $F(\bm\lambda)$ at $\bm\lambda^\ast$. If $\bm\lambda^\ast$ is an interior point of $\Delta(\cal X)$, all the eigenvalues $\theta_1,\ldots,\theta_m$ of $J(\bm\lambda^\ast)$ satisfy $0\leq\theta_i<1,\,i=1,\ldots,m$, then $\bm\lambda^N\to\bm\lambda^\ast$ is the exponential convergence, so this case is relatively easy.

If the convergence of $\bm\lambda^N\to\bm\lambda^\ast$ is slower than exponential, then $\bm\lambda^\ast$ is on the boundary of $\Delta(\cal X)$. Therefore, we need to examine the behavior of a differentiable function at the boundary, however, this is a difficult problem in general. We obtained \cite{nak4} a necessary and sufficient condition for the $O(1/N)$ convergence in the case that the input alphabet size is $m=3$. We classified the input symbols, or equivalently, the indices of the input alphabet into three types, and showed that the existence of type-II indices is a necessary and sufficient condition for the $O(1/N)$ convergence. In this paper, we will generalize the results of \cite{nak4} to arbitrary $m\geq3$.

We will analyze the recurrence formula obtained by truncating the Taylor expansion of the function $F(\bm\lambda)$ up to the second order term, which we call the {\it second order recurrence formula}.  The results of this paper can be stated as follows. If there are no type-II indices, the convergence speed of the Arimoto-Blahut algorithm is exponential. If there are type-II indices, the convergence speed of the second order recurrence formula is $O(1/N)$ for some initial vector. We will consider the condition on the $O(1/N)$ convergence for any initial vector. Furthermore, we will consider the speed that the mutual information converges to the channel capacity $C$. We will prove that if type-III indices do not exist, then the convergence speed of $I(\bm\lambda^N,\Phi)\to C$ is faster than that of $\bm\lambda^N\to\bm\lambda^\ast$. Especially, if $\bm\lambda^N\to\bm\lambda^\ast$ is $O(1/N)$, then $I(\bm\lambda^N,\Phi)\to C$ is $O(1/N^2)$.

The reason of the slow convergence of the Arimoto-Blahut algorithm will be made clear by this paper. Since $\lambda_i^\ast=0$ for the type-II index $i$, if those input symbols with type-II indices are removed from the input alphabet, the channel capacity does not change, and the convergence of the Arimoto-Blahut algorithm becomes exponential. Based on the results of this study, it is expected to be applied to the acceleration of the Arimoto-Blahut algorithm.
\section{Related works}
There have been many related works on the Arimoto-Blahut algorithm. For example, the extension to different kinds of channels \cite{nai}, \cite{rez}, \cite{von}, the acceleration of the Arimoto-Blahut algorithm \cite{mat}, \cite{yu}, and the characterization of the Arimoto-Blahut algorithm by the divergence geometry \cite{csi2}, \cite{mat}, \cite{naj}, \cite{nak1},\,etc. If we focus on the analysis of the convergence speed of the Arimoto-Blahut algorithm, we see in \cite{ari}, \cite{mat}, \cite{yu} where the eigenvalues of the Jacobian matrix are calculated and the convergence speed is investigated in the case that $\bm\lambda^\ast$ is in the interior of $\Delta(\cal X)$.

In this paper, we consider the Taylor expansion of the defining function $F(\bm\lambda)$ of the Arimoto-Blahut algorithm. We will calculate not only the Jacobian matrix of the first order term of the Taylor expansion, but also the Hessian matrix of the second order term, and examine the convergence speed of the exponential or $O(1/N)$ based on the Jacobian and Hessian matrices. Investigation of the Hessian matrix is new in this paper and \cite{nak4}.

\section{Channel matrix and channel capacity}
Consider a discrete memoryless channel $X\rightarrow Y$ with the input source $X$ and the output source $Y$. Let ${\cal X}=\{x_1,\ldots,x_m\}$ be the input alphabet and ${\cal Y}=\{y_1,\ldots,y_n\}$ be the output alphabet. The conditional probability that the output symbol $y_j$ is received when the input symbol $x_i$ was transmitted is denoted by $P^i_j=P(Y=y_j|X=x_i),\,i=1,\ldots,m, j=1,\ldots,n,$ and the row vector $P^i$ is defined by $P^i=(P^i_1,\ldots,P^i_n),\,i=1,\ldots,m$. The channel matrix $\Phi$ is defined by\\[-3mm]
\begin{align}
\label{eqn:thechannelmatrix}
\Phi=\begin{pmatrix}
\,P^1\,\\
\vdots\\
\,P^m\,
\end{pmatrix}
=\begin{pmatrix}
\,P^1_1 & \ldots & P^1_n\,\\
\vdots & & \vdots\\
\,P^m_1 & \ldots & P^m_n\,
\end{pmatrix}.
\end{align}
We assume that for any $j\,(j=1,\ldots,n)$ there exist at least one $i\,(i=1,\ldots,m)$ with $P^i_j>0$. This means that there are no useless output symbols.

The set of input probability distributions on the input alphabet ${\cal X}$ is denoted by $\Delta({\cal X})\equiv\{\bm\lambda=(\lambda_1,\ldots,\lambda_m)|\lambda_i\geq0,i=1,\ldots,m,\sum_{i=1}^m\lambda_i=1\}$. The interior of $\Delta({\cal X})$ is denoted by $\Delta({\cal X})^\circ\equiv\{\bm\lambda=(\lambda_1,\ldots,\lambda_m)\in\Delta({\cal X})\,|\,\lambda_i>0,\,i=1,\ldots,m\}$. Similarly, the set of output probability distributions on the output alphabet ${\cal Y}$ is denoted by $\Delta({\cal Y})\equiv\{Q=(Q_1,\ldots,Q_n)|Q_j\geq0,j=1,\ldots,n,\sum_{j=1}^nQ_j=1\}$, and its interior $\Delta({\cal Y})^\circ$ is similarly defined.

Let $Q=\bm\lambda\Phi$ be the output distribution for the input distribution $\bm\lambda\in\Delta(\cal X)$ and write its components as $Q_j=\sum_{i=1}^m\lambda_iP^i_j,\,j=1,\ldots,n$, then the mutual information is defined by $I(\bm\lambda,\Phi)=\sum_{i=1}^m\sum_{j=1}^n\lambda_iP^i_j\log\left({P^i_j}/{Q_j}\right)$. The channel capacity $C$ is defined by
\begin{align}
\label{eqn:Cdefinition}
C=\max_{\bm\lambda\in\Delta({\cal X})}I(\bm\lambda,\Phi).
\end{align}
The Kullback-Leibler divergence $D(Q\|Q')$ for two output distributions $Q=(Q_1,\ldots,Q_n)$, $Q'=(Q'_1,\ldots,Q'_n)\in\Delta(\cal Y)$ is defined \cite{csi1} by
\begin{align}
D(Q\|Q')=\sum_{j=1}^nQ_j\log\ds\frac{Q_j}{Q'_j}.
\end{align}

An important proposition for investigating the convergence speed of the Arimoto-Blahut algorithm is the Kuhn-Tucker condition on the input distribution $\bm\lambda=\bm\lambda^\ast$ that achieves the maximum of (\ref{eqn:Cdefinition}).

\medskip

\noindent{\bf Theorem}\ (Kuhn-Tucker condition \cite{cov}) In the maximization problem (\ref{eqn:Cdefinition}), a necessary and sufficient condition for the input distribution $\bm\lambda^\ast=(\lambda^\ast_1,\ldots,\lambda^\ast_m)\in\Delta({\cal X})$ to achieve the maximum is that there is a certain constant $\tilde{C}$ with
\begin{align}
\label{eqn:Kuhn-Tucker}
D(P^i\|\bm\lambda^\ast\Phi)\left\{\begin{array}{ll}=\tilde{C}, & {\mbox{\rm for}}\ i\ {\mbox{\rm with}}\ \lambda^\ast_i>0,\\
\leq \tilde{C}, & {\mbox{\rm for}}\ i\ {\mbox{\rm with}}\ \lambda^\ast_i=0.
\end{array}\right.
\end{align}
In (\ref{eqn:Kuhn-Tucker}), $\tilde{C}$ is equal to the channel capacity $C$.

\medskip

Since this Kuhn-Tucker condition is a necessary and sufficient condition, all the information about the capacity-achieving input distribution $\bm\lambda^\ast$ can be derived from this condition.
\section{Arimoto-Blahut algorithm}
\subsection{Definition of the algorithm}
A sequence of input distributions $\{\bm\lambda^N=(\lambda^N_1,\ldots,\lambda^N_m)\}_ {N=0,1,\ldots}\subset\Delta({\cal X})$ is defined by the Arimoto-Blahut algorithm as follows \cite{ari}, \cite{bla}. First, let $\bm\lambda^0=(\lambda^0_1,\ldots,\lambda^0_m)$ be an initial distribution taken in $\Delta(\cal X)^\circ$, i.e., $\lambda^0_i>0,\,i=1,\ldots,m$. Then, the Arimoto-Blahut algorithm is given by the recurrence formula
\begin{align}
\lambda^{N+1}_i=\ds\frac{\lambda^N_i\exp D(P^i\|\bm\lambda^N\Phi)}{\ds\sum_{k=1}^m\lambda^N_k\exp D(P^k\|\bm\lambda^N\Phi)},\,i=1,\ldots,m,\,N=0,1,\ldots.\label{eqn:arimotoalgorithm}
\end{align}
On the convergence of this Arimoto-Blahut algorithm, the following results were obtained in Arimoto \cite{ari}, \cite{ari2}.

By defining 
\begin{align}
C(N+1,N)\equiv-\ds\sum_{i=1}^m\lambda^{N+1}_i\log\lambda^{N+1}_i+\ds\sum_{i=1}^m\sum_{j=1}^n\lambda^{N+1}_iP^i_j\log\ds\frac{\lambda^N_iP^i_j}{\ds\sum_{k=1}^m\lambda^N_kP^k_j},
\end{align}
he obtained the following theorems.

\noindent{\bf Theorem A1}\,\cite{ari} If the initial input distribution $\bm\lambda^0$ is in $\Delta({\cal X})^\circ$, then
\begin{align}
\lim_{N\to\infty}C(N+1,N)=C.
\end{align}

\noindent{\bf Theorem A2}\,\cite{ari} If $\bm\lambda^0$ is the uniform distribution, then
\begin{align}
0\leq C-C(N+1,N)\leq\ds\frac{\log m-h(\bm\lambda^\ast)}{N},
\end{align}
where $\bm\lambda^\ast$ is the capacity-achieving input distribution and $h(\bm\lambda^\ast)$ is the entropy of $\bm\lambda^\ast$.

\noindent{\bf Theorem A3}\,\cite{ari2} Assume that $\bm\lambda^\ast$ is unique and belongs to $\Delta({\cal X})^\circ$. Then, for sufficiently small arbitrary $\epsilon>0$, there exists $N_0=N_0(\epsilon)$ such that
\begin{align}
0\leq C-C(N+1,N)\leq\epsilon(\theta)^{N-N_0},\,N\geq N_0,
\end{align}
where $\theta$ is a constant with $0\leq\theta<1$ and is unrelated to $\epsilon$ and $N_0$, further, $(\theta)^N$ denotes the $N$th power of $\theta$.

\medskip

In \cite{ari}, he considered the Taylor expansion of $D(\bm\lambda^\ast\|\bm\lambda)$ by $\bm\lambda$, and that of $D(Q^\ast\|Q)$ by $Q$, however he did not consider the Taylor expansion of the function $F(\bm\lambda)$.  Further, in the above Theorem A3, he considered only the case $\bm\lambda^\ast\in\Delta({\cal X})^\circ$, where the convergence is exponential.

In Yu\,\cite{yu}, he considered the function $F(\bm\lambda)$ and the Taylor expansion of $F(\bm\lambda)$ about $\bm\lambda=\bm\lambda^\ast$. He calculated the eigenvalues of the Jacobian matrix $J(\bm\lambda^\ast)$, however he did not consider the Hessian matrix. Further, he considered only the case $\bm\lambda^\ast\in\Delta({\cal X})^\circ$ as in \cite{ari}, \cite{ari2}.

\subsection{Function from $\Delta({\cal X})$ to $\Delta({\cal X})$}

Let $F_i(\bm\lambda)$ be the defining function of the Arimoto-Blahut algorithm (\ref{eqn:arimotoalgorithm}), i.e., 
\begin{align}
F_i(\bm\lambda)=\ds\frac{\lambda_i\exp D(P^i\|\bm\lambda\Phi)}{\ds\sum_{k=1}^m\lambda_k\exp D(P^k\|\bm\lambda\Phi)},\,i=1,\ldots,m.\label{eqn:Arimotofunction}
\end{align}
Define $F(\bm\lambda)\equiv(F_1(\bm\lambda),\ldots,F_m(\bm\lambda))$, then $F(\bm\lambda)$ is a differentiable function from $\Delta(\cal X)$ to $\Delta(\cal X)$, and (\ref{eqn:arimotoalgorithm}) is represented by $\bm\lambda^{N+1}=F(\bm\lambda^N)$.

In this paper, for the analysis of the convergence speed, we assume 
\begin{align}
{\rm rank}\,\Phi=m.\label{eqn:rankmdefinition}
\end{align}
Concerning this assumption, we see that in \cite{ari}, \cite{ari2}, for the analysis of the convergence speed, the uniqueness of the capacity-achieving $\bm\lambda^\ast$ is assumed, which is a necessary condition for (\ref{eqn:rankmdefinition}), in fact, we have
\begin{lemma}
\label{lem:capacityachievinglambdaisunique}
The capacity-achieving input distribution $\bm\lambda^\ast$ is unique.
\end{lemma}
{\bf Proof:} By Csisz\'{a}r\cite{csi1},\,p.137,\,eq.\,(37), for arbitrary $Q\in\Delta(\cal Y)$,%
\begin{align}
\ds\sum_{i=1}^m\lambda_iD(P^i\|Q)=I(\bm\lambda,\Phi)+D(\bm\lambda\Phi\|Q).\label{eqn:CKequality}
\end{align}
By the assumption (\ref{eqn:rankmdefinition}), we see that there exists $Q^0\in\Delta(\cal Y)$ \cite{nak2} with 
\begin{align}
D(P^1\|Q^0)=\ldots=D(P^m\|Q^0)\equiv C^0.
\end{align}
Substituting $Q=Q^0$ into (\ref{eqn:CKequality}), we have $C^0=I(\bm\lambda,\Phi)+D(\bm\lambda\Phi\|Q^0)$. Because $C^0$ is a constant,
\begin{align}
\max_{\bm\lambda\in\Delta(\cal X)}I(\bm\lambda,\Phi)\Longleftrightarrow\min_{\bm\lambda\in\Delta(\cal X)}D(\bm\lambda\Phi\|Q^0).\label{eqn:maxequalmin}
\end{align}
Define $W\equiv\{\bm\lambda\Phi\,|\,\bm\lambda\in\Delta(\cal X)\}$, then $W$ is a closed convex set, thus by Cover\,\cite{cov},\,p.297,\, Theorem 12.6.1, $Q=Q^\ast$ that achieves $\min_{Q\in W}D(Q\|Q^0)$ exists and is unique. By the assumption (\ref{eqn:rankmdefinition}), the mapping $\Delta\ni\bm\lambda\mapsto\bm\lambda\Phi\in W$ is one to one, therefore, $\bm\lambda^\ast$ with $Q^\ast=\bm\lambda^\ast\Phi$ is unique.\hfill$\blacksquare$
\begin{remark}
\rm Due to the equivalence (\ref{eqn:maxequalmin}), the Arimoto-Blahut algorithm can be obtained by Csisz\'{a}r \cite{csi2}, Chapter 4, ``Minimizing information distance from a single measure'', Theorem 5.
\end{remark}
\begin{lemma}
\label{lem:capacity_achieving_lambda_is_the_fixed_point}
The capacity-achieving input distribution $\bm\lambda^\ast$ is the fixed point of the function $F(\bm\lambda)$. That is, $\bm\lambda^\ast=F(\bm\lambda^\ast)$.
\end{lemma}
{\bf Proof:} In the Kuhn-Tucker condition (\ref{eqn:Kuhn-Tucker}), let us define $m_1$ as the number of indices $i$ with $\lambda^\ast_i>0$, i.e., 
\begin{align}
\lambda^\ast_i\left\{\begin{array}{ll}>0, & i=1,\ldots,m_1,\\=0, & i=m_1+1,\ldots,m,\end{array}\right.\label{eqn:m1definition}
\end{align}
by reordering the input symbols (if necessary), then 
\begin{align}
D(P^i\|\bm\lambda^\ast\Phi)\left\{\begin{array}{ll}=C, & i=1,\ldots,m_1,\\\leq C, & i=m_1+1,\ldots,m.\end{array}\right.
\end{align}
We have
\begin{align}
\ds\sum_{k=1}^m\lambda^\ast_k\exp D(P^k\|\bm\lambda^\ast\Phi)=\ds\sum_{k=1}^{m_1}\lambda^\ast_ke^C=e^C,\label{eqn:yobitekikeisan}
\end{align}
hence by (\ref{eqn:Arimotofunction}),\,(\ref{eqn:m1definition}),\,(\ref{eqn:yobitekikeisan}),
\begin{align}
F_i(\bm\lambda^\ast)&=\left\{\begin{array}{ll}e^{-C}\lambda^\ast_ie^C, & i=1,\ldots,m_1,\\0, & i=m_1+1,\ldots,m,\end{array}\right.\\
&=\lambda^\ast_i,\,i=1,\ldots,m,
\end{align}
which shows $F(\bm\lambda^\ast)=\bm\lambda^\ast$.\hfill$\blacksquare$

\medskip

The sequence $\left\{\bm\lambda^N\right\}_{N=0,1,\ldots}$ of the Arimoto-Blahut algorithm converges to the fixed point $\bm\lambda^\ast$, i.e., $\bm\lambda^N\to\bm\lambda^\ast,\,N\to\infty$.
We will investigate the convergence speed by using the Taylor expansion of $F(\bm\lambda)$ about $\bm\lambda=\bm\lambda^\ast$.

Now, we define two kinds of convergence speed for investigating $\bm\lambda^N\to\bm\lambda^\ast$. 
\begin{itemize}
\item[(i)] Exponential convergence\\
$\bm\lambda^N\to\bm\lambda^\ast$ is the {\it exponential convergence} if
\begin{align}
\|\bm\lambda^N-\bm\lambda^\ast\|<K(\theta)^N,\,K>0,\,0\leq\theta<1,\,N=0,1,\ldots,
\end{align}
where $\|\bm\lambda\|$ denotes the Euclidean norm $\|\bm\lambda\|=\left(\lambda_1^2+\ldots+\lambda_m^2\right)^{1/2}$.
\item[(ii)] $O(1/N)$ convergence\\
$\bm\lambda^N\to\bm\lambda^\ast$ is the {\it $O(1/N)$ convergence} if 
\begin{align}
\lim_{N\to\infty}N\left(\lambda^N_i-\lambda^\ast_i\right)=K_i\neq0,\,i=1,\ldots,m.
\end{align}
\end{itemize}
\subsection{Type of index}
Now, we classify the indices $i\,(i=1,\ldots,m)$ in the Kuhn-Tucker condition (\ref{eqn:Kuhn-Tucker}) in more detail into the following 3 types.
\begin{align}
\label{eqn:Kuhn-Tucker2}
D(P^i\|\bm\lambda^\ast\Phi)\left\{\begin{array}{ll}=C, & {\mbox{\rm for}}\ i\ {\mbox{\rm with}}\ \lambda^\ast_i>0\ \mbox{\rm [type-I]},\\
=C, & {\mbox{\rm for}}\ i\ {\mbox{\rm with}}\ \lambda^\ast_i=0\ \mbox{\rm [type-II]},\\
<C, & {\mbox{\rm for}}\ i\ {\mbox{\rm with}}\ \lambda^\ast_i=0\ \mbox{\rm [type-III]}.
\end{array}\right.
\end{align}
Let us define the sets of indices as follows.
\begin{align}
&{\rm all\ the\ indices:}\ {\cal I}\equiv\{1,\ldots,m\},\label{eqn:allset}\\
&\mbox{\rm type-I\ indices:}\ {\cal I}_{\rm I}\equiv\{1,\ldots,m_1\},\label{eqn:type1set}\\
&\mbox{\rm type-II\ indices:}\ {\cal I}_{\rm II}\equiv\{m_1+1,\ldots,m_1+m_2\},\label{eqn:type2set}\\
&\mbox{\rm type-III\ indices:}\ {\cal I}_{\rm III}\equiv\{m_1+m_2+1,\ldots,m\}.\label{eqn:type3set}
\end{align}
We have $|{\cal I}|=m$, $|{\cal I}_{\rm I}|=m_1$, $|{\cal I}_{\rm II}|=m_2$, $|{\cal I}_{\rm III}|=m-m_1-m_2\equiv m_3$, further, ${\cal I}={\cal I}_{\rm I}\cup{\cal I}_{\rm II}\cup{\cal I}_{\rm III}$ and $m=m_1+m_2+m_3$. ${\cal I}_{\rm I}$ is not empty and $|{\cal I}_{\rm I}|=m_1\geq2$ for any channel matrix, but ${\cal I}_{\rm II}$ and ${\cal I}_{\rm III}$ may be empty for some channel matrix.
\subsection{Examples of convergence speed}
Let us consider the difference of convergence speed  of the Arimoto-Blahut algorithm depending on the channel matrices.

For many channel matrices $\Phi$, the convergence is exponential, but for some special $\Phi$ the convergence is very slow. Let us consider the following examples with input alphabet size $m=3$ and output alphabet size $n=3$ taking types-I, II and III into account.

\begin{example}
\label{exa:CM_Phi(1)}
\rm (only type-I) If only type-I indices exist, then $\lambda^\ast_i>0,\,i=1,2,3$, hence $Q^\ast\equiv\bm\lambda^\ast\Phi$ is in the interior of $\triangle P^1P^2P^3$. See Fig. \ref{fig:1}. As a concrete channel matrix of this example, let us consider
\begin{align}
\label{eqn:Phi1}
\Phi^{(1)}=\begin{pmatrix}
\,0.800 & 0.100 & 0.100\,\\
\,0.100 & 0.800 & 0.100\,\\
\,0.250 & 0.250 & 0.500\,
\end{pmatrix}.
\end{align}
For this $\Phi^{(1)}$, we have $\bm\lambda^\ast=(0.431,0.431,0.138)$ and $Q^\ast=(0.422,0.422,0.156)$. The vertices of the large triangle in Fig. \ref{fig:1} are $\bm{e}_1=(1,0,0),\,\bm{e}_2=(0,1,0),\,\bm{e}_3=(0,0,1)$. We have $D(P^i\|Q^\ast)=C,\,i=1,2,3$, then considering the analogy to Euclidean geometry, $\triangle P^1P^2P^3$ can be regarded as an ``acute triangle''.
\begin{figure}[t]
\begin{center}
\begin{overpic}[width=8.8cm]{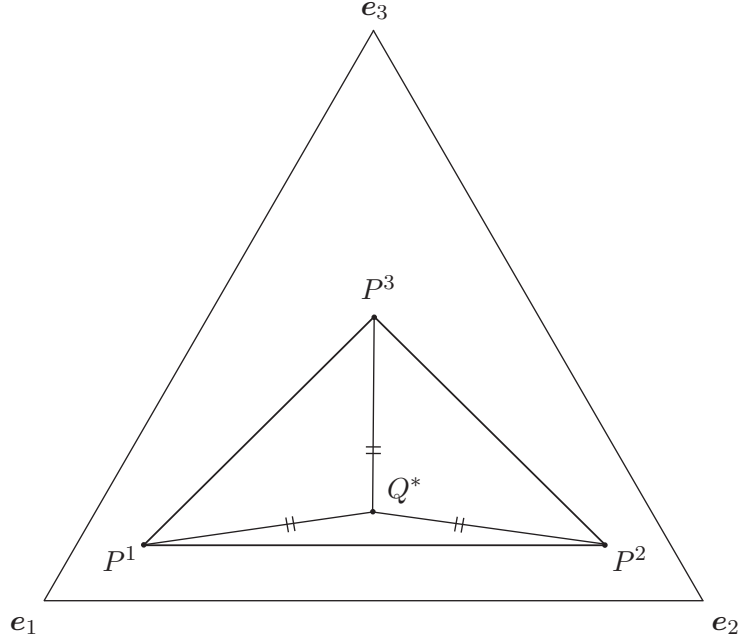}
\put(-5,-4){$\bm{e}_1$}
\put(101,-4){$\bm{e}_2$}
\put(48,89){$\bm{e}_3$}
\put(52,16){$Q^\ast$}
\put(9,5){$P^1$}
\put(86,5){$P^2$}
\put(48,46){$P^3$}
\end{overpic}
\medskip
\caption{Positional relation of row vectors $P^1,P^2,P^3$ of $\Phi^{(1)}$ and $Q^\ast$ in Example \ref{exa:CM_Phi(1)}.}
\label{fig:1}
\end{center}
\end{figure}
\end{example}

\begin{example}
\label{exa:CM_Phi(2)}
\rm (types-I and II) If there are type-I and type-II indices, we can assume $\lambda^\ast_1>0,\lambda^\ast_2>0,\lambda^\ast_3=0$ without loss of generality, hence $Q^\ast$ is on the side $P^1P^2$ and $D(P^i\|Q^\ast)=C,\,i=1,2,3$. See Fig. \ref{fig:2}. As a concrete channel matrix of this example, let us consider
\begin{align}
\label{eqn:Phi2}
\Phi^{(2)}=\begin{pmatrix}
\,0.800 & 0.100 & 0.100\,\\
\,0.100 & 0.800 & 0.100\,\\
\,0.300 & 0.300 & 0.400\,
\end{pmatrix}.
\end{align}
For this $\Phi^{(2)}$, we have $\bm\lambda^\ast=(0.500,0.500,0.000)$ and $Q^\ast=(0.450,0.450,0.100)$. Considering the analogy to Euclidean geometry, $\triangle P^1P^2P^3$ can be regarded as a ``right triangle''. 

\begin{figure}[t]
\begin{center}
 \medskip
\begin{overpic}[width=8.8cm]{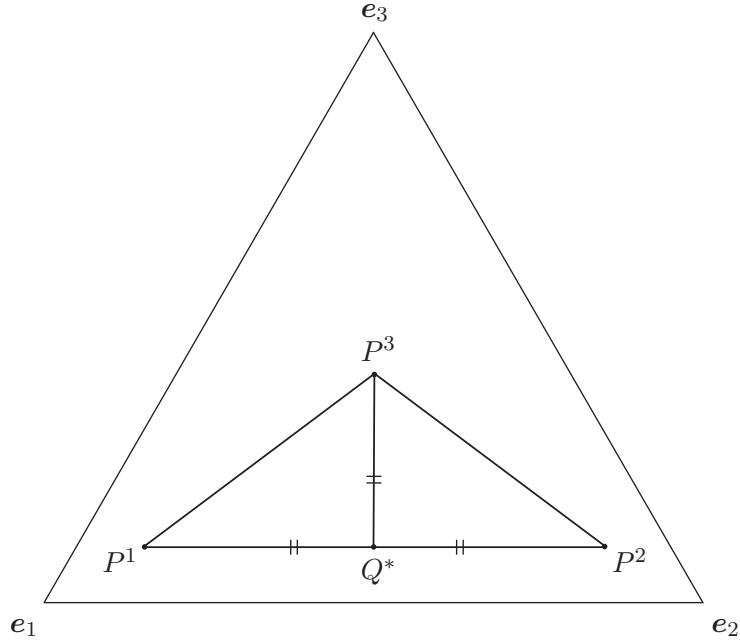}
\put(-5,-4){$\bm{e}_1$}
\put(101,-4){$\bm{e}_2$}
\put(48,89){$\bm{e}_3$}
\put(48,4){$Q^\ast$}
\put(9,5){$P^1$}
\put(86,5){$P^2$}
\put(48,37){$P^3$}
\end{overpic}
\medskip
\caption{Positional relation of row vectors $P^1,P^2,P^3$ of $\Phi^{(2)}$ and $Q^\ast$ in Example \ref{exa:CM_Phi(2)}.}
\label{fig:2}
\end{center}
\end{figure}
\end{example}

\begin{example}
\label{exa:CM_Phi(3)}
\rm (types-I and III) If there are type-I and type-III indices, we can assume $\lambda^\ast_1>0,\lambda^\ast_2>0,\lambda^\ast_3=0$ without loss of generality, hence $Q^\ast$ is on the side $P^1P^2$ and $C=D(P^1\|Q^\ast)=D(P^2\|Q^\ast)>D(P^3\|Q^\ast)$. See Fig. \ref{fig:3}. As a concrete channel matrix of this example, let us consider
\begin{align}
\label{eqn:Phi3}
\Phi^{(3)}=\begin{pmatrix}
\,0.800 & 0.100 & 0.100\,\\
\,0.100 & 0.800 & 0.100\,\\
\,0.350 & 0.350 & 0.300\,
\end{pmatrix}.
\end{align}
For this $\Phi^{(3)}$, we have $\bm\lambda^\ast=(0.500,0.500,0.000)$ and $Q^\ast=(0.450,0.450,0.100)$. Considering the analogy to Euclidean geometry, $\triangle P^1P^2P^3$ can be regarded as an ``obtuse triangle''. 
\begin{figure}[t]
\begin{center}
\begin{overpic}[width=8.8cm]{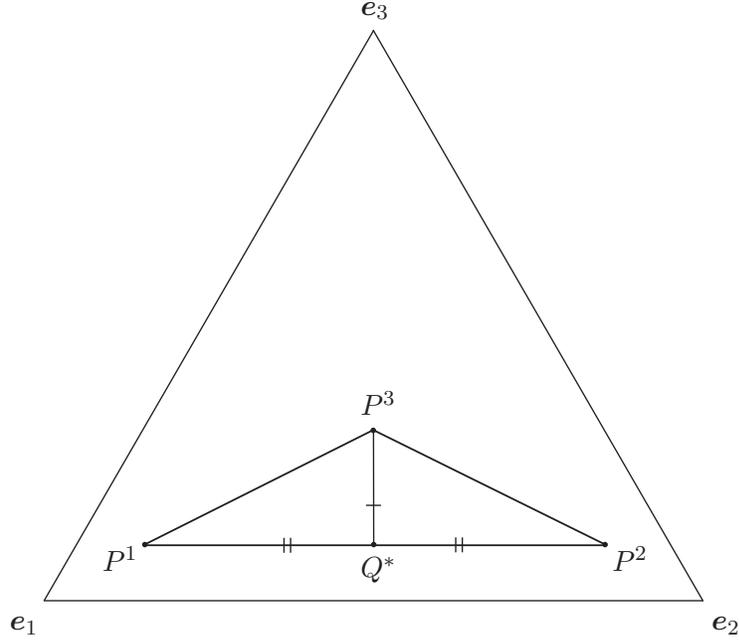}
\put(-5,-4){$\bm{e}_1$}
\put(101,-4){$\bm{e}_2$}
\put(48,89){$\bm{e}_3$}
\put(48,4){$Q^\ast$}
\put(9,5){$P^1$}
\put(86,5){$P^2$}
\put(48,28.5){$P^3$}
\end{overpic}
\medskip
\caption{Positional relation of row vectors $P^1,P^2,P^3$ of $\Phi^{(3)}$ and $Q^\ast$ in Example \ref{exa:CM_Phi(3)}.}
\label{fig:3}
\end{center}
\end{figure}
\end{example}

\begin{figure}[t]
\begin{center}
\begin{overpic}[width=9cm]{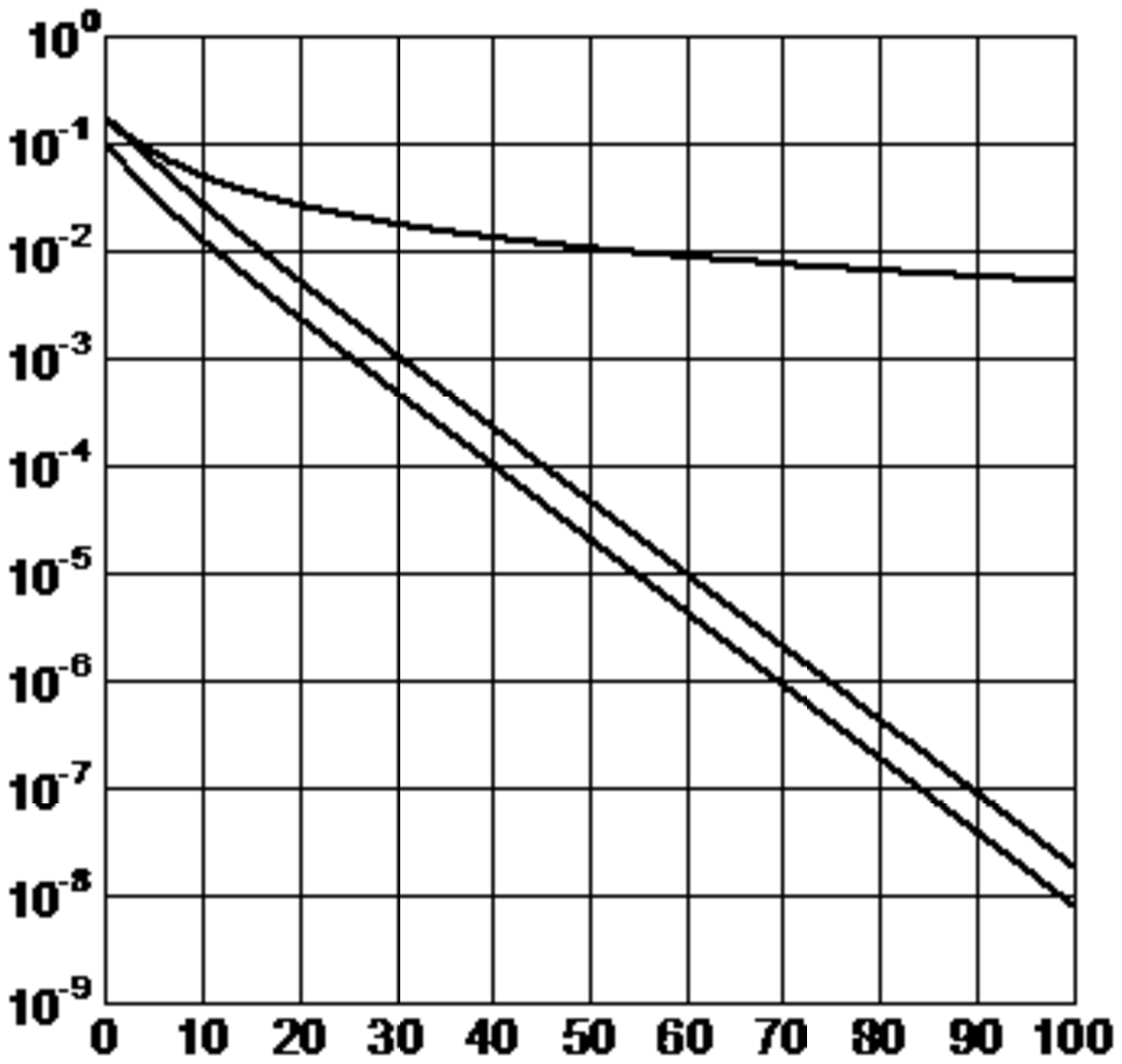}
%
\put(57,74){\rotatebox{60}{$\leftarrow$}}
\put(62,78){Example 2}
\put(62,74){(types-I and II)}
\put(58,47.5){\rotatebox{60}{$\leftarrow$}}
\put(62,50){Example 3}
\put(62,46){(types-I and III)}
\put(52,40){\rotatebox{60}{$\rightarrow$}}
\put(39,36){Example 1}
\put(39,31.5){(only type-I)}
\put(54,-3){$N$}
\put(-6,45){\rotatebox{90}{$|\lambda^N_1-\lambda^\ast_1|$}}
\end{overpic}
\caption{Comparison of the convergence speed in Examples \ref{exa:CM_Phi(1)},\ref{exa:CM_Phi(2)},\ref{exa:CM_Phi(3)}.}
\label{fig:lambda1}
\end{center}
\end{figure}

For the above $\Phi^{(1)},\Phi^{(2)},\Phi^{(3)}$, we show in Fig. \ref{fig:lambda1} the state of convergence of $|\lambda^N_1-\lambda^\ast_1|\to0$. By Fig. \ref{fig:lambda1}, we see that in Examples 1 and 3 the convergence is exponential, while in Example 2 the convergence is slower than exponential.

From the above three examples, it is inferred that the Arimoto-Blahut algorithm converges very slowly when type-II indices exist, and converges exponentially when type-II indices do not exist. We will analyze this phenomenon in the following.
\section{Taylor expansion of $F(\bm\lambda)$ about $\bm\lambda=\bm\lambda^\ast$}
We will examine the convergence speed of the Arimoto-Blahut algorithm by the Taylor expansion of $F(\bm\lambda)$ about the fixed point $\bm\lambda=\bm\lambda^\ast$. Taylor expansion of the function $F(\bm\lambda)=(F_1(\bm\lambda),\ldots,F_m(\bm\lambda))$ about $\bm\lambda=\bm\lambda^\ast$ is
\begin{align}
F(\bm\lambda)=F(\bm\lambda^\ast)+(\bm\lambda-\bm\lambda^\ast)J(\bm\lambda^\ast)+\ds\frac{1}{2!}(\bm\lambda-\bm\lambda^\ast)H(\bm\lambda^\ast)\,^t(\bm\lambda-\bm\lambda^\ast)+o\left(\|\bm\lambda-\bm\lambda^\ast\|^2\right),\label{eqn:Taylortenkai1}
\end{align}
where ${^t}\bm\lambda$ denotes the transpose of $\bm\lambda$, and $o\left(\|\bm\lambda\|^2\right)$ means $\lim_{\|\bm\lambda\|\to0}o\left(\|\bm\lambda\|^2\right)/\|\bm\lambda\|^2=0$.

In (\ref{eqn:Taylortenkai1}), $J(\bm\lambda^\ast)$ is the Jacobian matrix of $F(\bm\lambda)$ at $\bm\lambda=\bm\lambda^\ast$, i.e.,
\begin{align}
J(\bm\lambda^\ast)&=\left(\left.\ds\frac{\partial F_i}{\partial\lambda_{i'}}\right|_{\bm\lambda=\bm\lambda^\ast}\right)_{i',i=1,\ldots,m}.\label{eqn:Jacobiseibun}
\end{align}

In this paper, we assume that the input probability distribution $\bm\lambda$ is a row vector, thus the Jacobian matrix $J(\bm\lambda^\ast)$ is\\[-7mm]
\begin{align}
&\hspace{30mm}\leftarrow i\rightarrow\nonumber\\[0mm]
J(\bm\lambda^\ast)&=\begin{array}{c}\uparrow\\ i'\\\downarrow\end{array}
\hspace{-1mm}\begin{pmatrix}
\,\left.\ds\frac{\partial F_1}{\partial\lambda_1}\right|_{\bm\lambda=\bm\lambda^\ast} & \ldots & \left.\ds\frac{\partial F_m}{\partial\lambda_1}\right|_{\bm\lambda=\bm\lambda^\ast}\,\\
\vdots & & \vdots\\
\,\left.\ds\frac{\partial F_1}{\partial\lambda_m}\right|_{\bm\lambda=\bm\lambda^\ast} & \ldots & \left.\ds\frac{\partial F_m}{\partial\lambda_m}\right|_{\bm\lambda=\bm\lambda^\ast}\,
\end{pmatrix}\in\mathbb R^{m\times m},\label{eqn:rowvectorJacobimatrix}
\end{align}
i.e., $\partial F_i/\partial\lambda_{i'}|_{\bm\lambda=\bm\lambda^\ast}$ is the $(i',i)$ component. Note that our $J(\bm\lambda^\ast)$ is the transpose of the usual Jacobian matrix corresponding to column vector. 

Because $\sum_{i=1}^mF_i(\bm\lambda)=1$ by (\ref{eqn:Arimotofunction}), we have by (\ref{eqn:Jacobiseibun}),
\begin{lemma}
\label{lem:rowsumofJis0}
Every row sum of $J(\bm\lambda^\ast)$ is equal to $0$.
\end{lemma}

\medskip

In (\ref{eqn:Taylortenkai1}), $H(\bm\lambda^\ast)\equiv(H_1(\bm\lambda^\ast),\ldots,H_m(\bm\lambda^\ast))$, where $H_i(\bm\lambda^\ast)$ is the Hessian matrix of $F_i(\bm\lambda)$ at $\bm\lambda=\bm\lambda^\ast$, i.e.,
\begin{align}
H_i(\bm\lambda^\ast)=\left(\left.\ds\frac{\partial^2F_i}{\partial\lambda_{i'}\partial\lambda_{i''}}\right|_{\bm\lambda=\bm\lambda^\ast}\right)_{i',i''=1,\ldots,m},\label{eqn:Hesseseibun}
\end{align}
and $(\bm\lambda-\bm\lambda^\ast)H(\bm\lambda^\ast)\,^t(\bm\lambda-\bm\lambda^\ast)$ is an abbreviated expression of the $m$ dimensional row vector $\left((\bm\lambda-\bm\lambda^\ast)H_1(\bm\lambda^\ast)\,^t(\bm\lambda-\bm\lambda^\ast),\ldots,(\bm\lambda-\bm\lambda^\ast)H_m(\bm\lambda^\ast)\,^t(\bm\lambda-\bm\lambda^\ast)\right).$

\begin{remark}
\label{rem:justify}
\rm $\lambda_1,\ldots,\lambda_m$ satisfy the constraint $\sum_{i=1}^m\lambda_i=1$, but in (\ref{eqn:Taylortenkai1}),\,(\ref{eqn:Jacobiseibun}),\,(\ref{eqn:Hesseseibun}) we consider $\lambda_1,\ldots,\lambda_m$ as independent (or constraint free) variables to have the Taylor series approximation (\ref{eqn:Taylortenkai1}). This approximation is justified as follows. By the Kuhn-Tucker condition (\ref{eqn:Kuhn-Tucker}), $D(P^i\|Q^\ast)\leq C<\infty,\,i=1,\ldots,m$, hence by the assumption put below (\ref{eqn:thechannelmatrix}), we have $Q^\ast_j>0,\,j=1,\ldots,n$. See \cite{ari}. For $\epsilon>0$, define ${\cal Q}^\ast_\epsilon\equiv\{Q=(Q_1,\ldots,Q_n)\in{\mathbb R}^n\,|\,\|Q-Q^\ast\|<\epsilon\}$, i.e., ${\cal Q}^\ast_\epsilon$ is an open ball in $\mathbb{R}^n$ centered at $Q^\ast$ with radius $\epsilon$. Note that $Q\in{\cal Q}^\ast_\epsilon$ is free from the constraint $\sum_{j=1}^nQ_j=1$. Taking $\epsilon>0$ sufficiently small, we can have $Q_j>0,j=1,\ldots,n$, for any $Q\in{\cal Q}^\ast_\epsilon$. The function $F(\bm\lambda)$ is defined for $\bm\lambda$ with $Q_j=\left(\bm\lambda\Phi\right)_j>0,\,j=1,\ldots,n$, even if some $\lambda_i<0$. Therefore, the domain of definition of $F(\bm\lambda)$ can be extended to $\Phi^{-1}\left({\cal Q}^\ast_\epsilon\right)\subset\mathbb{R}^m$, where $\Phi^{-1}\left({\cal Q}^\ast_\epsilon\right)$ is the inverse image of ${\cal Q}^\ast_\epsilon$ by the mapping $\mathbb{R}^m\ni\bm\lambda\mapsto\bm\lambda\Phi\in\mathbb{R}^n$. $\Phi^{-1}\left({\cal Q}^\ast_\epsilon\right)$ is an open neighborhood of $\bm\lambda^\ast$ in $\mathbb{R}^m$. Then $F(\bm\lambda)$ is a function of $\bm\lambda=(\lambda_1,\ldots,\lambda_m)\in\Phi^{-1}\left({\cal Q}^\ast_\epsilon\right)$ as independent variables (free from the constraint $\sum_{i=1}^m\lambda_i=1$). We can consider (\ref{eqn:Taylortenkai1}) to be the Taylor expansion by independent variables $\lambda_1,\ldots,\lambda_m$, then substituting $\bm\lambda\in\Delta({\cal X})\cap\Phi^{-1}\left({\cal Q}^\ast_\epsilon\right)$ into (\ref{eqn:Taylortenkai1}) to obtain the approximation for $F(\bm\lambda)$ about $\bm\lambda=\bm\lambda^\ast$.
\end{remark}

\medskip

Now, substituting $\bm\lambda=\bm\lambda^N$ into (\ref{eqn:Taylortenkai1}), then by $F(\bm\lambda^\ast)=\bm\lambda^\ast$ and $F(\bm\lambda^N)=\bm\lambda^{N+1}$, we have
\begin{align}
\bm\lambda^{N+1}=\bm\lambda^\ast+(\bm\lambda^N-\bm\lambda^\ast)J(\bm\lambda^\ast)+\ds\frac{1}{2!}(\bm\lambda^N-\bm\lambda^\ast)H(\bm\lambda^\ast)\,^t(\bm\lambda^N-\bm\lambda^\ast)+o\left(\|\bm\lambda^N-\bm\lambda^\ast\|^2\right).\label{eqn:Taylortenkai2}
\end{align}
Then, by putting $\bm\mu^N\equiv\bm\lambda^N-\bm\lambda^\ast$, (\ref{eqn:Taylortenkai2}) becomes
\begin{align}
\bm\mu^{N+1}=\bm\mu^NJ(\bm\lambda^\ast)+\ds\frac{1}{2!}\bm\mu^NH(\bm\lambda^\ast)\,{^t}\bm\mu^N+o\left(\|\bm\mu^N\|^2\right).\label{eqn:Taylortenkai3}
\end{align}
Then, we will investigate the convergence $\bm\mu^N\to\bm0,\,N\to\infty$, based on the Taylor expansion (\ref{eqn:Taylortenkai3}). Let $\mu^N_i\equiv\lambda^N_i-\lambda^\ast_i,\,i=1,\ldots,m$, denote the components of $\bm\mu^N=\bm\lambda^N-\bm\lambda^\ast$, and write $\bm\mu^N$ by components as $\bm\mu^N=(\mu^N_1,\ldots,\mu^N_m)$, then we have $\sum_{i=1}^m\mu^N_i=0,\,N=0,1,\ldots$, because $\sum_{i=1}^m\lambda^N_i=\sum_{i=1}^m\lambda^\ast_i=1$.
\subsection{The Jacobian matrix $J(\bm\lambda^\ast)$}
Let us consider the Jacobian matrix $J(\bm\lambda^\ast)$. We are assuming ${\rm rank}\,\Phi=m$ in (\ref{eqn:rankmdefinition}), hence $m\leq n$. We will calculate the components (\ref{eqn:Jacobiseibun}) of $J(\bm\lambda^\ast)$.

Defining $D_i\equiv D(P^i\|\bm\lambda\Phi)$ and $F_i\equiv F_i(\bm\lambda),\,i=1,\ldots,m$, we can write (\ref{eqn:Arimotofunction}) as
\begin{align}
F_i=\ds\frac{\lambda_ie^{D_i}}{\ds\sum_{k=1}^m\lambda_ke^{D_k}},\,i=1,\ldots,m.\label{eqn:teigikansuFi}
\end{align}
From (\ref{eqn:teigikansuFi}) it follows that
\begin{align}
F_i\ds\sum_{k=1}^m\lambda_ke^{D_k}=\lambda_ie^{D_i},\label{eqn:Fibunboharau}
\end{align}
then differentiating both sides of (\ref{eqn:Fibunboharau}) with respect to $\lambda_{i'}$, we have
\begin{align}
\ds\frac{\partial F_i}{\partial\lambda_{i'}}\ds\sum_{k=1}^m\lambda_ke^{D_k}+F_i\ds\frac{\partial}{\partial\lambda_{i'}}\ds\sum_{k=1}^m\lambda_ke^{D_k}=\delta_{i'i}e^{D_i}+\lambda_ie^{D_i}\ds\frac{\partial D_i}{\partial\lambda_{i'}},\label{eqn:dFi}
\end{align}
where $\delta_{i'i}$ is the Kronecker delta.

Before substituting $\bm\lambda=\bm\lambda^\ast=(\lambda^\ast_1,\ldots,\lambda^\ast_m)$ into the both sides of (\ref{eqn:dFi}), we define the following symbols. Remember that the integer $m_1$ was defined in (\ref{eqn:m1definition}). See also (\ref{eqn:type1set}).

Let us define
\begin{align}
Q^\ast&\equiv Q(\bm\lambda^\ast)=\bm\lambda^\ast\Phi,\\
Q_j^\ast&\equiv Q(\bm\lambda^\ast)_j=\ds\sum_{i=1}^m\lambda_i^\ast P_j^i=\ds\sum_{i=1}^{m_1}\lambda_i^\ast P_j^i,\,j=1,\ldots,n,\\
D_i^\ast&\equiv D(P^i\|Q^\ast),\,i=1,\ldots,m,\\
D_{i',i}^\ast&\left.\equiv\ds\frac{\partial D_i}{\partial\lambda_{i'}}\right|_{\bm\lambda=\bm\lambda^\ast},\,i',i=1,\ldots,m,\label{eqn:Diidefinition}\\
F_i^\ast&\equiv F_i(\bm\lambda^\ast),\,i=1,\ldots,m.
\end{align}
\begin{lemma}
\label{lem:shoryou}We have
\begin{align}
&\left.\ds\sum_{k=1}^m\lambda_ke^{D_k}\right|_{\bm\lambda=\bm\lambda^\ast}=e^C,\label{eqn:lem3-1}\\[2mm]
&\ds\frac{\partial D_i}{\partial\lambda_{i'}}=-\ds\sum_{j=1}^n\ds\frac{P_j^{i'}P_j^i}{Q_j},\,i',i=1,\ldots,m,\label{eqn:lem3-2}\\[2mm]
&\left.\ds\frac{\partial}{\partial\lambda_{i'}}\ds\sum_{k=1}^m\lambda_ke^{D_k}\right|_{\bm\lambda=\bm\lambda^\ast}=e^{D_{i'}^\ast}-e^C,\,i'=1,\ldots,m,\label{eqn:lem3-3}\\[2mm]
&F^\ast_i=\lambda^\ast_i,\,i=1,\ldots,m.\label{eqn:lem3-4}
\end{align}
\end{lemma}
{\bf Proof:} Eq. (\ref{eqn:lem3-1}) was proved in (\ref{eqn:yobitekikeisan}). Eq. (\ref{eqn:lem3-2}) is proved by simple calculation. Eq. (\ref{eqn:lem3-3}) is proved as follows.
\begin{align}
\left.\ds\frac{\partial}{\partial\lambda_{i'}}\ds\sum_{k=1}^m\lambda_ke^{D_k}\right|_{\bm\lambda=\bm\lambda^\ast}
&=\left.\ds\sum_{k=1}^m\left(\delta_{i'k}e^{D_k}+\lambda_ke^{D_k}\ds\frac{\partial D_k}{\partial\lambda_{i'}}\right)\right|_{\bm\lambda=\bm\lambda^\ast}\\
&=e^{D_{i'}^\ast}+\ds\sum_{k=1}^{m_1}\lambda_k^\ast e^C\left(-\ds\sum_{j=1}^n\ds\frac{P_j^kP_j^{i'}}{Q_j^\ast}\right)\\
&=e^{D_{i'}^\ast}-e^C\ds\sum_{j=1}^nP_j^{i'}\ds\frac{1}{Q_j^\ast}\ds\sum_{k=1}^{m_1}\lambda_k^\ast P_j^k\\
&=e^{D_{i'}^\ast}-e^C.
\end{align}
Note that $Q^\ast_j>0,\,j=1,\ldots,n$, from Remark \ref{rem:justify}. Eq. (\ref{eqn:lem3-4}) is the result of Lemma \ref{lem:capacity_achieving_lambda_is_the_fixed_point}.\hfill$\blacksquare$

\medskip

Substituting the results of Lemma \ref{lem:shoryou} into (\ref{eqn:dFi}), we have
\begin{align}
\left.\ds\frac{\partial F_i}{\partial\lambda_{i'}}\right|_{\bm\lambda=\bm\lambda^\ast}e^C+\lambda_i^\ast\left(e^{D_{i'}^\ast}-e^C\right)=\delta_{i'i}e^{D^\ast_i}+\lambda_i^\ast e^{D^\ast_i}D^\ast_{i',i}.
\end{align}

Consequently, we have
\begin{theorem}
\label{the:1}The components of the Jacobian matrix $J(\bm\lambda^\ast)$ are given as follows.
\begin{align}
\left.\ds\frac{\partial F_i}{\partial\lambda_{i'}}\right|_{\bm\lambda=\bm\lambda^\ast}&=e^{D_i^\ast-C}\left(\delta_{i'i}+\lambda_i^\ast D^\ast_{i',i}\right)+\lambda^\ast_i\left(1-e^{D^\ast_{i'}-C}\right),\,i',i\in{\cal I},\label{eqn:theorem1-0}\\[1mm]
&=\left\{\begin{array}{l}
\delta_{i'i}+\lambda_i^\ast\left(D^\ast_{i',i}+1-e^{D_{i'}^\ast-C}\right),\,i'\in{\cal I},\,i\in{\cal I}_{\rm I},\\[2mm]
\delta_{i'i},\,i'\in{\cal I},\,i\in{\cal I}_{\rm II},\\[2mm]
e^{D^\ast_i-C}\delta_{i'i},\,i'\in{\cal I},\,i\in{\cal I}_{\rm III},
\end{array}\right.\label{eqn:theorem1-1}
\end{align}
where the sets of indices ${\cal I}$, ${\cal I}_{\rm I}$, ${\cal I}_{\rm II}$, ${\cal I}_{\rm III}$ were defined in $(\ref{eqn:allset})$-$(\ref{eqn:type3set})$. Note that $D^\ast_i=C$ for $i\in{\cal I}_{\rm I}\cup{\cal I}_{\rm II}$ and $\lambda^\ast_i=0$ for $i\in{\cal I}_{\rm II}\cup{\cal I}_{\rm III}$.
\end{theorem}
%
%
%
%
\subsection{Eigenvalues of the Jacobian matrix $J(\bm\lambda^\ast)$}
From (\ref{eqn:theorem1-1}), we see that the Jacobian matrix $J(\bm\lambda^\ast)$ is of the form

\begin{align}
&J(\bm\lambda^\ast)\equiv\begin{pmatrix}
\,J^{\rm I} & O & O\,\\[1mm]
\,\ast & J^{\rm II} & O\,\\[1mm]
\,\ast & O & J^{\rm III}
\end{pmatrix},\label{eqn:J1AJ2}\\
&J^{\rm I}\equiv\left(\partial F_i/\partial\lambda_{i'}|_{\bm\lambda=\bm\lambda^\ast}\right)_{i,i'\in{\cal I}_{\rm I}}\in{\mathbb R}^{m_1\times m_1},\label{eqn:Jstructure1}\\
&J^{\rm II}\equiv\left(\partial F_i/\partial\lambda_{i'}|_{\bm\lambda=\bm\lambda^\ast}\right)_{i,i'\in{\cal I}_{\rm II}}=I\ ({\rm the\ identity\ matrix})\in{\mathbb R}^{m_2\times m_2},\label{eqn:Jstructure2}\\
&J^{\rm III}\equiv\left(\partial F_i/\partial\lambda_{i'}|_{\bm\lambda=\bm\lambda^\ast}\right)_{i,i'\in{\cal I}_{\rm III}}={\rm diag}\left(e^{D_i^\ast-C},\,i\in{\cal I}_{\rm III}\right)\in{\mathbb R}^{m_3\times m_3},\label{eqn:Jstructure3}\\
&O\ \mbox{\rm denotes\ the\ all-zero\ matrix\ of\ appropriate\ size.}\nonumber
\end{align}

\noindent In (\ref{eqn:Jstructure3}), ${\rm diag}\left(e^{D_i^\ast-C},\,i\in{\cal I}_{\rm III}\right)$ denotes the diagonal matrix with diagonal components $e^{D^\ast_i-C},\,i\in{\cal I}_{\rm III}$. Here, $e^{D^\ast_i-C}<1,\,i\in{\cal I}_{\rm III}$ holds from type-III in (\ref{eqn:Kuhn-Tucker2}).

\bigskip

Let $\{\theta_1,\ldots,\theta_m\}\equiv\{\theta_i\,|\,i\in{\cal I}\}$ be the set of eigenvalues of $J(\bm\lambda^\ast)$. By (\ref{eqn:J1AJ2}), the eigenvalues of $J(\bm\lambda^\ast)$ are the eigenvalues of $J^{\rm I}$, $J^{\rm II}$, $J^{\rm III}$, hence we can put

$\{\theta_i\,|\,i\in{\cal I}_{\rm I}\}$: the set of eigenvalues of $J^{\rm I}$,

$\{\theta_i\,|\,i\in{\cal I}_{\rm II}\}$: the set of eigenvalues of $J^{\rm II}$,

$\{\theta_i\,|\,i\in{\cal I}_{\rm III}\}$: the set of eigenvalues of $J^{\rm III}$.

\bigskip

We will evaluate the eigenvalues of $J^{\rm I}$, $J^{\rm II}$ and $J^{\rm III}$ as follows.
\subsection{Eigenvalues of $J^{\rm I}$}
First, we consider the eigenvalues of $J^{\rm I}$. Let $J^{\rm I}_{i'i}$ be the $(i',i)$ component of $J^{\rm I}$, then by (\ref{eqn:theorem1-1}),
\begin{align}
J^{\rm I}_{i'i}=\delta_{i'i}+\lambda_i^\ast D_{i',i}^\ast,\ i',i\in{\cal I}_{\rm I}.\label{eqn:J1seibun}
\end{align}
Let $I\in{\mathbb R}^{m_1\times m_1}$ denote the identity matrix and define $B\equiv I-J^{\rm I}$. Let $B_{i'i}$ be the $(i',i)$ component of $B$, then from (\ref{eqn:J1seibun}),
\begin{align}
B_{i'i}&=-\lambda^\ast_iD^\ast_{i',i}\\[1mm]
&=\lambda^\ast_i\ds\sum_{j=1}^n\ds\frac{P_j^{i'}P_j^i}{Q_j^\ast},\,i',i\in{\cal I}_{\rm I}.\label{eqn:componentofB}
\end{align}
Let $\{\beta_i\,|\,i\in{\cal I}_{\rm I}\}$ be the set of eigenvalues of $B$, then we have $\theta_i=1-\beta_i,\,i\in{\cal I}_{\rm I}$. In order to calculate the eigenvalues of $B$, we will define the following matrices. Similar calculations are performed in \cite{yu}.

Let us define
\begin{align}
\Phi_1&\equiv\begin{pmatrix}P^1\\\vdots\\P^{m_1}\end{pmatrix}\in\mathbb {R}^{m_1\times n},\\[3mm]
\Gamma&\equiv\left(-D_{i',i}^\ast\right)=\left(\ds\sum_{j=1}^n\ds\frac{P_j^{i'}P_j^i}{Q_j^\ast}\right)_{i',i\in{\cal I}_{\rm I}}\in\mathbb {R}^{m_1\times m_1},\\[3mm]
\Lambda&\equiv{\rm diag}\left(\lambda_1^\ast,\ldots,\lambda_{m_1}^\ast\right)\in\mathbb {R}^{m_1\times m_1}.\label{eqn:diagonalLambda}
\end{align}
Furthermore, define
\begin{align}
\sqrt{\Lambda}&\equiv{\rm diag}\left(\sqrt{\lambda_1^\ast},\ldots,\sqrt{\lambda_{m_1}^\ast}\right)\in\mathbb {R}^{m_1\times m_1},\\
\Omega&\equiv{\rm diag}\left((Q_1^\ast)^{-1},\ldots,(Q_n^\ast)^{-1}\right)\in\mathbb {R}^{n\times n},\\
\sqrt\Omega&\equiv{\rm diag}\left((Q_1^\ast)^{-1/2},\ldots,(Q_n^\ast)^{-1/2}\right)\in\mathbb {R}^{n\times n}.
\end{align}
Then, we have, by calculation,
\begin{align}
\sqrt\Lambda B\sqrt\Lambda^{-1}&=\sqrt\Lambda\Gamma\sqrt\Lambda\\
&=\sqrt\Lambda\Phi_1\Omega\ {^t}\Phi_1\ {^t}\sqrt\Lambda\\
&=\sqrt\Lambda\Phi_1\sqrt\Omega\ {^t}\sqrt\Omega\ {^t}\Phi_1\ {^t}\sqrt\Lambda\\
&=\sqrt\Lambda\Phi_1\sqrt\Omega\ {^t}\!\left(\sqrt\Lambda\Phi_1\sqrt\Omega\right).\label{eqn:L-1BL}
\end{align}
From (\ref{eqn:m1definition}), $\sqrt\Lambda$ is a regular matrix and from the assumption (\ref{eqn:rankmdefinition}), ${\rm rank}\,\Phi_1=m_1$. Therefore, by $m_1\leq m\leq n$, we have ${\rm rank}\,\sqrt\Lambda\Phi_1\sqrt\Omega=m_1$, and thus from (\ref{eqn:L-1BL}), $\sqrt\Lambda B\sqrt\Lambda^{-1}$ is symmetric and positive\ definite. In particular, all the eigenvalues $\beta_1,\ldots,\beta_{m_1}$ of $B$ are positive. Without loss of generality, let $\beta_1\geq\ldots\geq\beta_{m_1}>0$.
By (\ref{eqn:componentofB}), every component of $B$ is non-negative and by Lemma \ref{lem:rowsumofJis0}, every row sum of $B$ is equal to 1, hence by the Perron-Frobenius theorem \cite{hor}

\begin{align}
1=\beta_1\geq\beta_2\geq\ldots\geq\beta_{m_1}>0.
\end{align}
Because $\theta_i=1-\beta_i,\,i\in{\cal I}_{\rm I}$, we have
\begin{align}
0=\theta_1\leq\theta_2\leq\ldots\leq\theta_{m_1}<1,
\end{align}
therefore,
\begin{theorem}
\label{the:eigenvaluesofJ1}
The eigenvalues of $J^{\rm I}$ satisfy
\begin{align}
0\leq\theta_i<1,\,i\in{\cal I}_{\rm I}.\label{eqn:J1nokoyuchi}
\end{align}
\end{theorem}
\subsection{Eigenvalues of $J^{\rm II}$}
Second, we consider the eigenvalues of $J^{\rm II}$. From (\ref{eqn:J1AJ2}),\,(\ref{eqn:Jstructure2}), we have
\begin{theorem}
\label{the:eigenvaluesofJ2}
The eigenvalues of $J^{\rm II}$ satisfy
\begin{align}
\theta_i=1,\,i\in{\cal I}_{\rm II}.\label{eqn:J2nokoyuchi}
\end{align}
\end{theorem}
\subsection{Eigenvalues of $J^{\rm III}$}
Third, we consider the eigenvalues of $J^{\rm III}$. From (\ref{eqn:J1AJ2}),\,(\ref{eqn:Jstructure3}), we have
\begin{theorem}
\label{the:eigenvaluesofJ3}
The eigenvalues of $J^{\rm III}$ are $\theta_i=e^{D^\ast_i-C},\,D^\ast_i<C,\,i\in{\cal I}_{\rm III}$, hence
\begin{align}
0<\theta_i<1,\,i\in{\cal I}_{\rm III}.\label{eqn:J3nokoyuchi}
\end{align}
\end{theorem}

\begin{remark}
\rm From the above consideration, we know that all the eigenvalues of the Jacobian matrix $J(\bm\lambda^\ast)$ are real.
\end{remark}

\begin{lemma}
\label{lem:Jlambdadiagonalizable}
Assume that the eigenvalues of $J^{\rm I}$ and $J^{\rm III}$ are distinct, i.e., 
\begin{align}
\theta_i\neq\theta_{i'},\,i\in{\cal I}_{\rm I},\,i'\in{\cal I}_{\rm III},\label{eqn:eigenvaluesaredistinct}
\end{align}
then $J(\bm\lambda^\ast)$ is diagonalizable. Especially, if ${\cal I}_{\rm III}=\emptyset$ then $(\ref{eqn:eigenvaluesaredistinct})$ holds, and thus $J(\bm\lambda^\ast)$ is diagonalizable.
\end{lemma}
\noindent{\bf Proof:} See Appendix \ref{sec:proooflemmaJlambdadiagonalizable}.\hfill$\blacksquare$
\section{On the exponential convergence}
We obtained in Theorems \ref{the:eigenvaluesofJ1}, \ref{the:eigenvaluesofJ2} and \ref{the:eigenvaluesofJ3} the evaluation for the eigenvalues of $J(\bm\lambda^\ast)$. Let $\theta_{\rm max}\equiv\max_{i\in{\cal I}}\theta_i$ be the maximum eigenvalue of $J(\bm\lambda^\ast)$, then by Theorems \ref{the:eigenvaluesofJ1}, \ref{the:eigenvaluesofJ2} and \ref{the:eigenvaluesofJ3}, we have $0\leq\theta_{\rm max}<1$ if ${\cal I}_{\rm II}$ is empty and $\theta_{\rm max}=1$ if ${\cal I}_{\rm II}$ is not empty. First, we show that the convergence is exponential if ${\cal I}_{\rm II}$ is empty.
\begin{theorem}
\label{the:exponentialconvergence}
Assume ${\cal I}_{\rm II}=\emptyset$, then for any $\theta$ with $\theta_{\rm max}<\theta<1$, there exist $\delta>0$ and $K>0$, such that for arbitrary initial distribution $\bm\lambda^0$ with $\|\bm\lambda^0-\bm\lambda^\ast\|<\delta$, we have
\begin{align}
\|\bm\mu^N\|=\|\bm\lambda^N-\bm\lambda^\ast\|<K(\theta)^N,\,N=0,1,\ldots,
\end{align}
i.e., the convergence is exponential.
\end{theorem}
{\bf Proof:} See Appendix \ref{sec:exponentialconvergence}.\hfill$\blacksquare$



%
\section{On the $O(1/N)$ convergence}
We will consider the second order recurrence formula obtained by truncating the Taylor expansion of $F(\bm\lambda)$ up to the second order term and analyze the $O(1/N)$ convergence of the sequence defined by the second order recurrence formula.
\subsection{The Hessian matrix $H_i(\bm\lambda^\ast)$}
If $0\leq\theta_{\rm max}<1$, then the convergence speed of $\bm\lambda^N\to\bm\lambda^\ast$ is determined by the Jacobian matrix $J(\bm\lambda^\ast)$ due to Theorem \ref{the:exponentialconvergence}. But, if $\theta_{\rm max}=1$, the convergence speed is not determined only by $J(\bm\lambda^\ast)$, hence we must investigate the Hessian matrix. In \cite{ari}, \cite{yu}, the Jacobian matrix is considered, but the Hessian matrix is not considered in the past literature.

Now, we will calculate the Hessian matrix
\begin{align}
H_i(\bm\lambda^\ast)=\left(\left.\ds\frac{\partial^2F_i}{\partial\lambda_{i'}\partial\lambda_{i''}}\right|_{\bm\lambda=\bm\lambda^\ast}\right)_{i',i''\in{\cal I}},\,i\in{\cal I}
\end{align}
of $F_i(\bm\lambda)$ at $\bm\lambda=\bm\lambda^\ast$. We have
\begin{theorem}
\label{the:Hessecomponents}The components of the Hessian matrix $H_i(\bm\lambda^\ast),\,i=1,\ldots,m$, are given as follows.
\begin{align}
\left.\dfrac{\partial^2F_i}{\partial\lambda_{i'}\partial\lambda_{i''}}\right|_{\bm\lambda=\bm\lambda^\ast}&=e^{D^\ast_i-C}\Big\{\delta_{ii'}D^\ast_{i,i''}+\delta_{ii''}D^\ast_{i,i'}+\lambda^\ast_i\left(D^\ast_{i,i'}D^\ast_{i,i''}+D^\ast_{i,i',i''}\right)\nonumber\\
&\ \ \ +\left(\delta_{ii'}+\lambda^\ast_iD^\ast_{i,i'}\right)\left(1-e^{D^\ast_{i''}-C}\right)+\left(\delta_{ii''}+\lambda^\ast_iD^\ast_{i,i''}\right)\left(1-e^{D^\ast_{i'}-C}\right)\Big\}\nonumber\\
&\ \ \ +2\lambda^\ast_i\left(1-e^{D^\ast_{i'}-C}\right)\left(1-e^{D^\ast_{i''}-C}\right)-\lambda^\ast_i\Big(e^{D^\ast_{i'}-C}D^\ast_{i',i''}+e^{D^\ast_{i''}-C}D^\ast_{i',i''}\nonumber\\
&\ \ \ +E_{i',i''}-D^\ast_{i',i''}\Big),\,i,i',i''\in{\cal I}.
\end{align}
where $D^\ast_{i,i',i''}\equiv\partial^2D_i/\partial\lambda_{i'}\partial\lambda_{i{''}}|_{\bm\lambda=\bm\lambda^\ast}$ and $E_{i',i''}\equiv\sum_{k=1}^{m_1}\lambda_k^\ast D_{k,i'}^\ast D_{k,i''}^\ast$.

Especially, if $i\in{\cal I}_{\rm II}$, then $\lambda^\ast_i=0$ by $(\ref{eqn:Kuhn-Tucker2})$, thus
\begin{align}
\left.\ds\frac{\partial^2F_i}{\partial\lambda_{i'}\partial\lambda_{i''}}\right|_{\bm\lambda=\bm\lambda^\ast}
=\delta_{ii''}\left(1-e^{D_{i'}^\ast-C}+D_{i,i'}^\ast\right)+\delta_{ii'}\left(1-e^{D_{i''}^\ast-C}+D_{i,i''}^\ast\right).\label{eqn:Hesse1}
\end{align}
Further, if $i\in{\cal I}_{\rm II}$ and $D^\ast_{i'}=C$ holds for arbitrary $i'\in{\cal I}$, then by $(\ref{eqn:Hesse1})$, we have
\begin{align}
\left.\ds\frac{\partial^2F_i}{\partial\lambda_{i'}\partial\lambda_{i''}}\right|_{\bm\lambda=\bm\lambda^\ast}=\delta_{ii'}D_{i,i''}^\ast+\delta_{ii''}D_{i,i'}^\ast,\ i',i''\in{\cal I}.
\end{align}
\end{theorem}
\noindent{\bf Proof:} See Appendix \ref{sec:Hessiancomponents}.\hfill$\blacksquare$
\subsection{Analysis of the $O(1/N)$ convergence}
We consider a recurrence formula obtained by truncating the Taylor expansion (\ref{eqn:Taylortenkai3}) up to the second order term and write the variables as $\bar{\bm\mu}^N=(\bar{\mu} _1^N,\ldots,\bar{\mu}_m^N)$. That is, we have
\begin{align}
\bar{\bm\mu}^{N+1}=\bar{\bm\mu}^NJ(\bm\lambda^\ast)+\ds\frac{1}{2!}\bar{\bm\mu}^NH(\bm\lambda^\ast)\,{^t}\bar{\bm\mu}^N.\label{eqn:second_order_recurrence_formula}
\end{align}
The recurrence formula (\ref{eqn:second_order_recurrence_formula}) is called the {\it second order recurrence formula} of the Taylor expansion (\ref{eqn:Taylortenkai3}). We investigate the convergence speed of $\bar{\bm\mu}^N\to\bm0$. The convergence speed of $\bar{\bm\mu}^N\to\bm0$ seems to be the same as that of the original $\bm\mu^N\to\bm0$, but the proof is not obtained. Numerical comparison will be done in Chapter \ref{sec:numerical_evaluation}. In this chapter we will prove that, if ${\cal I}_{\rm II}\neq\emptyset$, there exists an initial vector $\bar{\bm\mu}^0$ such that $\bar{\bm\mu}^ N\to\bm0$ is the $O(1/N)$ convergence. Furthermore, we will consider the condition that $\bar{\bm\mu}^N\to\bm0$ is the $O(1/N)$ convergence for arbitrary initial vector $\bar{\bm\mu}^0$. The $O(1/N)$ convergence will be proved by the following three steps.

\medskip

\begin{description}
\item[Step 1:] Represent $\bar\mu^N_i$ with types-I and III indices by $\bar\mu^N_i$ with type-II indices.
\item[Step 2:] Obtain the recurrence formula satisfied by $\bar\mu^N_i$ with type-II indices.
\item[Step 3:] Prove that the convergence of $\bar\mu^N_i$ with type-II indices is $O(1/N)$ for some initial vector $\bar{\bm\mu}^0$.
\end{description}
\subsection{Step 1}
Here, we consider the types-I and III indices together. Then, put ${\cal I}_{\rm I}\cup{\cal I}_{\rm III}=\{1,\ldots,m'\}$ and ${\cal I}_{\rm II}=\{m'+1,\ldots,m\}$. We have $m_2=m-m'$ and $|{\cal I}_{\rm II}|=m_2$. The purpose of the step 1 is to represent $\bar{\bm\mu}^N_{\rm I,III}\equiv(\bar\mu^N_1,\ldots,\bar\mu^N_{m'})$ by $\bar{\bm\mu}^N_{\rm II}\equiv(\bar{\mu}^N_{m'+1},\ldots,\bar{\mu}^N_m)$.

In the Jacobian matrix, by changing the order of $J^{\rm II}$ and $J^{\rm III}$ we have
\begin{align}
J(\bm\lambda^\ast)=\begin{pmatrix}
\,J^{\rm I} & O & O\,\\[1mm]
\,\ast & J^{\rm III} & O\,\\[1mm]
\,\ast & O & J^{\rm II}
\end{pmatrix}.
\end{align}
Then by defining 
\begin{align}
\label{eqn:J'JIJIII}
J'\equiv\begin{pmatrix}
\,J^{\rm I} & O \,\\[1mm]
\,\ast & J^{\rm III}
\end{pmatrix},
\end{align}
we have
\begin{align}
\label{eqn:JJ'JII}
J(\bm\lambda^\ast)=\begin{pmatrix}
\,J' & O \,\\[1mm]
\,\ast & J^{\rm II}
\end{pmatrix}.
\end{align}
The eigenvalues of $J(\bm\lambda^\ast)$ are $\theta_i,\,i=1,\ldots,m$, and then the eigenvalues of $J'$ are $\theta_i,\,i=1,\ldots,m'$ with $0\leq\theta_i<1$, and those of $J^{\rm II}$ are $\theta_i,\,i=m'+1,\ldots,m$ with $\theta_i=1$.

\medskip

Now, let $\bm a_i$ be a right eigenvector of $J(\bm\lambda^\ast)$ for $\theta_i$ and define
\begin{align}
A\equiv(\bm a_1,\ldots,\bm a_m)\in{\mathbb R}^{m\times m}.
\end{align}
Under the assumption (\ref{eqn:eigenvaluesaredistinct}), by choosing the eigenvectors $\bm a_1,\ldots,\bm a_m$ appropriately we can make $A$ a regular matrix. In fact, because $J(\bm\lambda^\ast)$ is diagonalizable by Lemma \ref{lem:Jlambdadiagonalizable}, the direct sum of all the eigenspaces spans the whole ${\mathbb R}^m$ (See \cite{sat},\,p.161,\,Example 4).

For $i=m'+1,\ldots,m$, define
\begin{align}
\bm{e}_i=(0,\ldots,0,\stackrel{i\,\text{th}}{\stackrel{\vee}{1}}\hspace{-1mm},\ 0,\ldots,0)\in{\mathbb R}^m,\ i=m'+1,\ldots,m,
\end{align}
then because $\theta_i=1$, we can take
\begin{align}
{\bm a}_i={^t}\bm{e}_i,\,i=m'+1,\ldots,m.\label{eqn:eigenvectorfor1}
\end{align}
Therefore, we have
\begin{align}
A&=\left(\bm a_1,\ldots,\bm a_{m'},{^t}\bm e_{m'+1},\ldots,{^t}\bm e_m\right)\\
&=\begin{pmatrix}
\begin{array}{l:c}
\begin{matrix}\hspace{1mm}a_{11}&\hspace{1mm}\ldots&\hspace{3mm}a_{m'1}\\
\hspace{3.5mm}\vdots&&\vdots\\
\hspace{2mm}a_{1m'}&\hspace{2mm}\ldots&\hspace{2mm}a_{m'm'}
\end{matrix}&O\\[7mm]
\hdashline\\[-4mm]
\begin{matrix}a_{1,m'+1}&\ldots&a_{m',m'+1}\\\vdots&&\vdots\\a_{1m}&\ldots&a_{m'm}\end{matrix}&
\begin{matrix}1&\ldots&0\\\vdots&\ddots&\vdots\\0&\ldots&1\end{matrix}
\end{array}
\end{pmatrix}\\
&\equiv\begin{pmatrix}A_1&O\\A_2&I\end{pmatrix},\label{eqn:A1OA2I}
\end{align}
where
\begin{align}
A_1\equiv\begin{pmatrix}a_{11}&\ldots&a_{m'1}\\\vdots&&\vdots\\a_{1m'}&\ldots&a_{m'm'}\end{pmatrix},\ \ A_2\equiv\begin{pmatrix}a_{1,m'+1}&\ldots&a_{m',m'+1}\\\vdots&&\vdots\\a_{1m}&\ldots&a_{m'm}\end{pmatrix}.\label{eqn:A1A2definition}
\end{align}
Because $A$ is regular, $A_1$ is also regular by (\ref{eqn:A1OA2I}). $J(\bm\lambda^\ast)$ is diagonalized by $A$, i.e., $A^{-1}J(\bm\lambda^\ast)A=\Theta$, where
\begin{align}
\Theta=\begin{pmatrix}\Theta_1&O\\O&I\end{pmatrix},\ \Theta_1=\begin{pmatrix}\theta_1&&O\\&\ddots&\\O&&\theta_{m'}\end{pmatrix},\,0\leq\theta_i<1,\,i=1,\ldots,m'.\label{eqn:Theta_and_Theta1}
\end{align}
Calculating by using only the first order term of the Taylor expansion (\ref{eqn:Taylortenkai3}), we have
\begin{align}
\bm\mu^{N+1}A&=\bm\mu^NJ(\bm\lambda^\ast)A\\
&=\bm\mu^NA\Theta,\label{eqn:bmmu{N+1}A=bmmuNATheta}
\end{align}
thus by $\bm\mu^N=(\bm\mu^N_{\rm I,III},\bm\mu^N_{\rm II})$, (\ref{eqn:bmmu{N+1}A=bmmuNATheta}) and  (\ref{eqn:A1OA2I}),
\begin{align}
\bm\mu^{N+1}_{\rm I,III}A_1+\bm\mu^{N+1}_{\rm II}A_2&=\left(\bm\mu^N_{\rm I,III}A_1+\bm\mu^N_{\rm II}A_2\right)\Theta_1.\label{eqn:bmmu{N+1}{I,III}A1+bmmu{N+1}{II}A2}
\end{align}
Hence, if the second and higher order terms are negligible, we have
\begin{align}
\bm\mu^N_{\rm I,III}A_1+\bm\mu^N_{\rm II}A_2\to\bm0\ ({\mbox{\rm exponentially}}),\ N\to\infty.\label{eqn:bmmuN{I,III}A1+bmmuN{II}A2tobm0}
\end{align}
To show that the second and higher terms are negligible, let us consider $\bm\mu^{N+1}\bm{a}_i$ as a function of $\bm\mu^N\bm{a}_1,\ldots,\bm\mu^N\bm{a}_{m'}$. If the Taylor expansion (\ref{eqn:Taylortenkai3}) satisfies that
\begin{align}
\bm\mu^{N+1}\bm{a}_i\ {\mbox{\rm is\ divisible\ by}}\ \bm\mu^N\bm{a}_i,\label{eqn:bmmu{N+1}bm{a}isdivisiblebybmmuNbm{a}i}
\end{align}
i.e., if $\bm\mu^N\bm{a}_i=0$ implies that $\bm\mu^{N+1}\bm{a}_i=0$, then, we have
\begin{align}
\bm\mu^{N+1}\bm{a}_i=\theta_i\bm\mu^N\bm{a}_i\left(1+o(|\bm\mu^N\bm{a}_i|)\right),\ N\to\infty,\ i=1,\ldots,m',
\end{align}
and hence (\ref{eqn:bmmuN{I,III}A1+bmmuN{II}A2tobm0}) holds. However, in general, it is difficult to prove (\ref{eqn:bmmu{N+1}bm{a}isdivisiblebybmmuNbm{a}i}). We will show later in Examples \ref{exa:CM_Phi(2)_again} and \ref{exa:CM_Phi(5)} that (\ref{eqn:bmmu{N+1}bm{a}isdivisiblebybmmuNbm{a}i}) holds.

In what follows, we assume (\ref{eqn:bmmuN{I,III}A1+bmmuN{II}A2tobm0}) and regard it as $\bm\mu^N_{\rm I,III}A_1+\bm\mu^N_{\rm II}A_2=\bm0.$ Then, we replace $\bm\mu^N$ by $\bar{\bm\mu}^N$ to have $\bar{\bm\mu}^N_{\rm I,III}A_1+\bar{\bm\mu}^N_{\rm II}A_2=\bm0$ and hence
\begin{align}
\bar{\bm\mu}^N_{\rm I,III}=-\bar{\bm\mu}^N_{\rm II}A_2A_1^{-1}.\label{eqn:bmmuN{I,III}A1+bmmuN{II}A2=bm0}
\end{align}
The validity of (\ref{eqn:bmmuN{I,III}A1+bmmuN{II}A2=bm0}) will be checked by numerical examples.
\subsection{Step 2}
The purpose of the step 2 is to obtain a recurrence formula satisfied by $\bar{\mu}_i^N,\,i=m'+1,\ldots,m$.

The $i$-th component of (\ref{eqn:second_order_recurrence_formula}) for $i=m'+1,\ldots,m$ is
\begin{align}
\bar{\mu}^{N+1}_i=\bar{\mu}^N_i+\dfrac{1}{2!}\bar{\bm\mu}^NH_i(\bm\lambda^\ast)\,{^t}\bar{\bm\mu}^N,\,i=m'+1,\ldots,m.\label{eqn:Taylor2nd}
\end{align}
We will represent the second term of the right hand side of (\ref{eqn:Taylor2nd}) by $\bar{\bm\mu}^N_{\rm II}$.

Let $H_{i,i'i''}$ be the $(i',i'')$ component of the Hessian matrix $H_i(\bm\lambda^\ast)$, then by Theorem \ref{the:Hessecomponents}, we have
\begin{align}
H_{i,i'i''}&=\left.\dfrac{\partial^2F_i}{\partial\lambda_{i'}\partial\lambda_{i''}}\right|_{\bm\lambda=\bm\lambda^\ast}\\
&=\delta_{ii''}\left(1-e^{D^\ast_{i'}-C}+D^\ast_{i,i'}\right)+\delta_{ii'}\left(1-e^{D^\ast_{i''}-C}+D^\ast_{i,i''}\right),\label{eqn:Hessiancomponent}\\
i&=m'+1,\ldots,m,\,i',i''=1,\ldots,m.\nonumber
\end{align}
Here, for the simplicity of symbols, define
\begin{align}
S_{ii'}\equiv1-e^{D^\ast_{i'}-C}+D^\ast_{i,i'},\,i=m'+1,\ldots,m,i'=1,\ldots,m,
\end{align}
then, we can write (\ref{eqn:Hessiancomponent}) as
\begin{align}
H_{i,i'i''}=\delta_{ii''}S_{ii'}+\delta_{ii'}S_{ii''}.\label{eqn:Hessiancomponent2}
\end{align}
Further, writing $A_1^{-1}\equiv\left(\zeta_{i'i''}\right),\,i',i''=1,\ldots,m'$, $T_{ii'}\equiv-\sum_{k,k'=1}^{m'}a_{i'k}\zeta_{kk'}S_{ik'},\,i,i'=m'+1,\ldots,m$, $r_{ii'}\equiv T_{ii'}+D^\ast_{i,i'},\,i,i'=m'+1,\ldots,m$, we have the following theorem.
\begin{theorem}
\label{the:1/Norderkihonzenkashiki}
$\{\bar{\mu}^N_i\},\,i=m'+1,\ldots,m$, satisfies the recurrence formula
\begin{align}
\bar{\mu}^{N+1}_i=\bar{\mu}^N_i+\bar{\mu}^N_i\ds\sum_{i'=m'+1}^mr_{ii'}\bar{\mu}^N_{i'},\,i=m'+1,\ldots,m.\label{eqn:1/Norderzenkashiki}
\end{align}
\end{theorem}
\noindent{\bf Proof:} See Appendix \ref{sec:proofofstep2}.\hfill$\blacksquare$

\medskip

The step 2 is achieved by (\ref{eqn:1/Norderzenkashiki}).

\subsection{Step 3}
\label{sec:step3}
The purpose of the step 3 is to prove that $\bar{\mu}_i^N\to0,\,i=1,\ldots,m$, is the $O(1/N)$ convergence.
%
We will define the canonical form of the recurrence formula (\ref{eqn:1/Norderzenkashiki}). Writing $R\equiv(r_{ii'}),\,i,i'=m'+1,\ldots,m$, and $\bm1=(1,\ldots,1)\in{\mathbb R}^{m_2}$, we consider the equation $R\,{^t}\bm\sigma=-{^t}\bm1$ for the variables $\bm\sigma=(\sigma_{m'+1},\ldots,\sigma_m)$. Assuming that $R$ is regular, we have
\begin{align}
{^t}\bm\sigma=-R^{-1}{^t}\bm1,\label{eqn:sigmadefinition}
\end{align}
and then we assume $\bm\sigma>0$, i.e., $\sigma_i>0, i=m'+1,\ldots,m$. Further, by putting
\begin{align}
\nu_i^N&\equiv\bar{\mu}_i^N/\sigma_i,\,i=m'+1,\ldots,m,\label{eqn:nuiNequivmuiN/sigmai}\\
p_{ii'}&\equiv-r_{ii'}\sigma_{i'},\,i,i'=m'+1,\ldots,m,
\end{align}
the recurrence formula (\ref{eqn:1/Norderzenkashiki}) becomes
\begin{align}
\nu_i^{N+1}&=\nu_i^N-\nu_i^N\sum_{i'=1}^mp_{ii'}\nu_{i'}^N,\,i=m'+1,\ldots,m,\label{eqn:canonical_recursion_formula}\\
{\rm where}\ {\bm p}_i&\equiv(p_{i,m'+1},\ldots,p_{i,m})\,{\rm\ is\ a\ probability\ vector}.
\end{align}
The recurrence formula (\ref{eqn:canonical_recursion_formula}) is called the {\it canonical form} of (\ref{eqn:1/Norderzenkashiki}).

\bigskip

For the analysis of (\ref{eqn:canonical_recursion_formula}), we prepare the following lemma.
\begin{lemma}
\label{lem:onevariablereccurence}
Let us define a positive sequence $\{\nu^N\}_{N=0,1\ldots}$ by the recurrence formula
\begin{align}
&\nu^{N+1}=\nu^N-\left(\nu^N\right)^2,\,N=0,1,\ldots,\label{eqn:1hensuzenkashiki}\\
&0<\nu^0\leq1/2.
\end{align}
Then we have
\begin{align}
\lim_{N\to\infty}N\nu^N=1.
\end{align}
\end{lemma}
\noindent{\bf Proof:} Since the function $g(\nu)\equiv\nu-\nu^2$ satisfies $0<g(\nu)<\nu$ for $0<\nu\leq1/2$, we see $0<\nu^{N+1}<\nu^N,\,N=0,1,\ldots$ by mathematical induction. Thus, $\nu^\infty\equiv\lim_{N\to\infty}\nu^N\geq0$ exists and by (\ref{eqn:1hensuzenkashiki}) $\nu^\infty=\nu^\infty-\left(\nu^\infty\right)^2$ holds, then we have $\nu^\infty=0$.

Next, by (\ref{eqn:1hensuzenkashiki}) we have
\begin{align}
\dfrac{1}{N}\left(\dfrac{1}{\nu^N}-\dfrac{1}{\nu^0}\right)&=\dfrac{1}{N}\sum_{l=0}^{N-1}\left(\dfrac{1}{\nu^{l+1}}-\dfrac{1}{\nu^l}\right)\\
&=\dfrac{1}{N}\sum_{l=0}^{N-1}\dfrac{1}{1-\nu^l}.\label{eqn:soukaheikin}
\end{align}
Applying the proposition that ``the arithmetic mean of a convergent sequence converges to the same limit as the original sequence'' (\cite{ahl},\,p.37) to the right hand side of (\ref{eqn:soukaheikin}), we have 
\begin{align}
\lim_{N\to\infty}\dfrac{1}{N\nu^N}&=\lim_{N\to\infty}\dfrac{1}{1-\nu^N}\\
&=1,
\end{align}
which proves the lemma.\hfill$\blacksquare$
\begin{lemma}
\label{lem:lim_{Ntoinfty}Nnu_iN=1}
In the canonical form $(\ref{eqn:canonical_recursion_formula})$, for the initial values $\nu_i^0=1/2,\,i=m'+1,\ldots,m$, we have
\begin{align}
\lim_{N\to\infty}N\nu_i^N=1,\,i=m'+1,\ldots,m.\label{eqn:lim_{Ntoinfty}Nnu_iN=1}
\end{align}
\end{lemma}
\noindent{\bf Proof:} By mathematical induction, we see that $\nu_{m'+1}^N=\ldots=\nu_m^N$ holds for $N=0,1,\ldots$, thus (\ref{eqn:canonical_recursion_formula}) becomes $\nu_i^{N+1}=\nu_i^N-\left(\nu_i^N\right)^2,\,i=m'+1,\ldots,m$. Therefore (\ref{eqn:lim_{Ntoinfty}Nnu_iN=1}) holds by Lemma \ref{lem:onevariablereccurence}.\hfill$\blacksquare$
\begin{theorem}
\label{the:1/Norderconvergence_specialinitial}
In $(\ref{eqn:1/Norderzenkashiki})$, for the initial values $\bar{\mu}_i^0=\sigma_i/2,\ i=m'+1,\ldots,m$, we have
\begin{align}
\lim_{N\to\infty}N\bar{\mu}_i^N=\sigma_i,\ i=m'+1,\ldots,m.\label{eqn:lim_{Ntoinfty}Nmu_iN=sigma_i}
\end{align}
Further, under the assumption $(\ref{eqn:bmmuN{I,III}A1+bmmuN{II}A2=bm0})$, we have
\begin{align}
\lim_{N\to\infty}N\bar{\mu}^N_i=-\left(\bm\sigma A_2A_1^{-1}\right)_i,\ i=1,\ldots,m',\label{eqn:NmuNi1...m'}
\end{align}
where $\bm\sigma$, $A_1$, $A_2$ were defined by $(\ref{eqn:sigmadefinition}),(\ref{eqn:A1A2definition})$.
\end{theorem}
\noindent{\bf Proof:} From $\nu_i^N=\bar{\mu}_i^N/\sigma_i,\ i=m'+1,\ldots,m$, and Lemma \ref{lem:lim_{Ntoinfty}Nnu_iN=1}, we obtain (\ref{eqn:lim_{Ntoinfty}Nmu_iN=sigma_i}). Further, by (\ref{eqn:bmmuN{I,III}A1+bmmuN{II}A2=bm0}), we obtain (\ref{eqn:NmuNi1...m'}).\hfill$\blacksquare$

\medskip

The step 3 is achieved by Theorem \ref{the:1/Norderconvergence_specialinitial}. Summarizing above, we have the following theorem.
\begin{theorem}
\label{the:maintheorem}
If type-II indices exist, then there exists an initial vector $\bar{\bm\mu}^0$ such that $\bar{\bm\mu}^N\to\bm0$ is the $O(1/N)$ convergence.
\end{theorem}
\begin{remark} In Theorem $\ref{the:maintheorem}$, we must choose a specific initial vector $\bar{\bm\mu}^0$, but we want to prove it for arbitrary initial distribution. If the existence of $\lim_{N\to\infty}N\mu^N_i,\,i=1,\ldots,m$, is proved for arbitrary $\bar{\bm\mu}^0$, then we have $(\ref{eqn:lim_{Ntoinfty}Nmu_iN=sigma_i})$ and $(\ref{eqn:NmuNi1...m'})$. However, the analysis for the convergence speed of the recurrence formula $(\ref{eqn:1/Norderzenkashiki})$ for arbitrary initial distribution is very difficult, so it has not been achieved. We will give a proof under the assumption that a conjecture holds.
\end{remark}

\subsection{On the initial distribution}
In this section, we will investigate the convergence of (\ref{eqn:canonical_recursion_formula}) for arbitrary initial distribution. 

Now, for the sake of simplicity, we will change the indices and symbols of the canonical form (\ref{eqn:canonical_recursion_formula}). Noting $m-m'=m_2$, we change the indices from $m'+1,\ldots,m$ to $1,\ldots,m_2$, and define
\begin{align}
\xi_i^N&\equiv\nu_{m'+i},\,i=1,\ldots,m_2,\\
q_{ii'}&\equiv p_{m'+i,m'+i'},\,i,i'=1,\ldots,m_2,
\end{align}
which are just shifting the indices. By the above change, the canonical form (\ref{eqn:canonical_recursion_formula}) becomes
\begin{align}
\xi_i^{N+1}&=\xi_i^N-\xi_i^N\sum_{i'=1}^{m_2}q_{ii'}\xi_{i'}^N,\,i=1,\ldots,m_2,\label{eqn:modified_standardform}\\
{\rm where}\ {\bm q}_i&\equiv\left(q_{i1},\ldots,q_{i,m_2}\right)\in{\mathbb R^{m_2}}\ \mbox{\rm is\ a\ probability\ vector}.
\end{align}
\subsection{Diagonally dominant condition}

For the probability vectors $\bm q_i,\,i=1,\cdots,m_2$, we assume
\begin{align}
q_{ii}>\sum_{i'=1,i'\neq i}^{m_2}q_{ii'},\,i=1,\ldots,m_2.\label{eqn:taikakuyuui2}
\end{align}
This assumption means that the matrix made by arranging the row vectors ${\bm q}_1,\ldots,{\bm q}_{m_2}$ vertically is diagonally dominant. We call (\ref{eqn:taikakuyuui2}) the {\it diagonally dominant condition}. By numerical calculation, we confirmed that (\ref{eqn:taikakuyuui2}) holds in all the cases of our examples.

\subsection{Conjecture}
Our goal is to prove that $\lim_{N\to\infty}N\xi^N_i=1,\,i=1,\ldots,m_2$ for any initial vector $\bm\xi^0$. We found that we can prove it if the following conjecture holds, however, the proof of this conjecture has not yet been obtained.

\bigskip

\noindent{\bf Conjecture}\ \ By reordering the indices if necessary, there exists an $N_0$ such that the following inequalities hold.
\begin{align}
\xi_1^N\geq\xi_2^N\geq\ldots\geq\xi_{m_2}^N,\,N\geq N_0.\label{eqn:daishoukankei}
\end{align}

Many numerical examples we observed seem to support this conjecture.

We have
\begin{theorem}
\label{the:1/Norderconvergenceinitialfree}
We assume that the sequence defined by the recurrence formula $(\ref{eqn:modified_standardform})$ satisfies $(\ref{eqn:taikakuyuui2})$ and $(\ref{eqn:daishoukankei})$. Then, for arbitrary initial values with $0<\xi_i^0\leq1/2,\,i=1,\ldots,m_2$, we have
\begin{align}
\lim_{N\to\infty}N\xi_i^N&=1,\,i=1,\ldots,m_2.
\end{align}
\end{theorem}
\noindent{\bf Proof:} See Appendix \ref{sec:proofof1/Norderconvergenceinitialfree}.\hfill$\blacksquare$

\subsection{Special cases where the conjecture holds}

Now, we consider some special cases where the conjecture (\ref{eqn:daishoukankei}) holds.

If $m_2=1$, then the variable is only $\xi_1^N$, hence we do not need to consider the inequality condition.

If $m_2=2$, the canonical form (\ref{eqn:modified_standardform}) becomes
\begin{align}
\xi_1^{N+1}&=\xi_1^N-q_{11}\left(\xi_1^N\right)^2-q_{12}\xi_1^N\xi_2^N,\\
\xi_2^{N+1}&=\xi_2^N-q_{21}\xi_2^N\xi_1^N-q_{22}\left(\xi_2^N\right)^2,\\
q_{11}&>0,\,q_{12}>0,\,q_{11}+q_{12}=1,\\
q_{21}&>0,\,q_{22}>0,\,q_{21}+q_{22}=1.
\end{align}
By calculation, we have
\begin{align}
\xi_1^{N+1}-\xi_2^{N+1}=\left(\xi_1^N-\xi_2^N\right)\left(1-q_{11}\xi_1^N-q_{22}\xi_2^N\right).
\end{align}
By Lemma \ref{lem:0<xiiN<=1/2} in Appendix \ref{sec:proofof1/Norderconvergenceinitialfree}, we have $1-q_{11}\xi_1^N-q_{22}\xi_2^N>0$ for any $N=0,1,\ldots$, hence if $\xi_1^0\geq\xi_2^0$ then $\xi_1^N\geq\xi_2^N,\,N=0,1,\ldots$. Thus, the conjecture (\ref{eqn:daishoukankei}) holds with $N_0=0$.

In the case of $m_2\geq3$, this problem is very difficult. We have a sufficient condition for (\ref{eqn:daishoukankei}) in the following lemma.
\begin{lemma}
\label{lem:daishou_kenkei_hozon_under_special_condition}
For $i'$, consider $q_{ii'},\,i=1,\ldots,m_2$. Assume that $q_{ii'}$ are equal for all $i$ except $i'$, i.e., $q_{1i'}=\ldots=q_{i'-1,i'}=q_{i'+1,i'}=\ldots=q_{m_2,i'}$. Then, the conjecture $(\ref{eqn:daishoukankei})$ holds.
\end{lemma}
\noindent{\bf Proof:} By calculation, we have
\begin{align}
\xi_1^{N+1}-\xi_2^{N+1}=\left(\xi_1^N-\xi_2^N\right)\left(1-q_{11}\xi_1^N-q_{22}\xi_2^N-\sum_{i'=3}^{m_2}q_{1i'}\xi_{i'}^N\right).\label{eqn:tahensuinsubunkai}
\end{align}
The second factor of the right hand side of (\ref{eqn:tahensuinsubunkai}) is positive for all $N=0,1,\ldots$ by Lemma \ref{lem:0<xiiN<=1/2}, hence if $\xi^0_1\geq\xi^0_2$ then $\xi_1^N\geq\xi_2^N$ holds for $N=0,1,\ldots$. Similarly, by considering any pair $\xi_i^N$ and $\xi_{i'}^N$, we see that (\ref{eqn:daishoukankei}) holds. \hfill$\blacksquare$

\bigskip

Summarizing above, we have
\begin{theorem}
\label{the:1/N_order_convergence_under_special_condition}
Under the assumptions of Lemma $\ref{lem:daishou_kenkei_hozon_under_special_condition}$ and the diagonally dominant condition $(\ref{eqn:taikakuyuui2})$, the convegence of the recurrence formula $(\ref{eqn:modified_standardform})$ is $O(1/N)$ for arbitrary initial vector $\bm\xi^0=(\xi_1^0,\ldots,\xi_{m_2}^0)$ with $0<\xi_i\leq1/2,\,i=1,\ldots,m_2$, and 
\begin{align}
\lim_{N\to\infty}N\xi_i^N=1,\,i=1,\ldots,m_2.
\end{align}
\end{theorem}
\section{Numerical Evaluation}
\label{sec:numerical_evaluation}
Based on the analysis in the previous sections, we will evaluate numerically the convergence speed of the Arimoto-Blahut algorithm for several channel matrices.

In Examples \ref{exa:CM_Phi(1)_again} and \ref{exa:CM_Phi(4)} below, we will investigate the exponential convergence, where the capacity-achieving $\bm\lambda^\ast$ is in $\Delta({\cal X})^\circ$ (the interior of $\Delta({\cal X})$). In Example \ref{exa:CM_Phi(4)}, we will discuss how the convergence speed depends on the choice of the initial distribution $\bm\lambda^0$. Next, in Examples \ref{exa:CM_Phi(2)_again} and \ref{exa:CM_Phi(5)}, we will consider the $O(1/N)$ convergence. It will be confirmed that the convergence speed is accurately approximated by the values obtained in Theorem \ref{the:1/Norderconvergence_specialinitial}. In Example \ref{exa:CM_Phi(3)_again}, we will investigate the exponential convergence, where $\bm\lambda^\ast$ is on $\partial\Delta({\cal X})$ (the boundary of $\Delta({\cal X})$).

Here, in the case of exponential convergence, we will evaluate the values of the function
\begin{align}
L(N)\equiv-\ds\frac{1}{N}\log\|\bm\mu^N\|.
\end{align}
Based on the results of Theorem \ref{the:exponentialconvergence}, i.e., $\|\bm\mu^N\|=\|\bm\lambda^N-\bm\lambda^\ast\|<K(\theta)^N$, $\theta\doteqdot\theta_{\rm max}$, we will compare $L(N)$ for large $N$ with $-\log\theta_{\rm max}$ (or other values). 

On the other hand, in the case of $O(1/N)$ convergence, we will evaluate 
\begin{align}
N\bm\mu^N=(N\mu^N_1,\ldots,N\mu_m^N).
\end{align}
We will compare $N\bm\mu^N$ for large $N$ with the values obtained in Theorem \ref{the:1/Norderconvergence_specialinitial}.

\subsection{Exponential convergence where $\bm\lambda^\ast\in\Delta({\cal X})^\circ$}
\begin{example}
\label{exa:CM_Phi(1)_again}
\rm Consider the channel matrix $\Phi^{(1)}$ of (\ref{eqn:Phi1}), i.e.,
\begin{align}
\Phi^{(1)}=
\begin{pmatrix}
0.800 & 0.100 & 0.100\\
0.100 & 0.800 & 0.100\\
0.250 & 0.250 & 0.500
\end{pmatrix},
\end{align}
and an initial distribution $\bm\lambda^0=(1/3,1/3,1/3)$.
We have
\begin{align}
\bm\lambda^\ast&=(0.431,0.431,0.138),\\
Q^\ast&=(0.422,0.422,0.156),\\
J(\bm\lambda^\ast)&=
\begin{pmatrix}
\,0.308 & -0.191 & -0.117\,\cr
\,-0.191 & 0.308 & -0.117\,\cr
\,-0.369 & -0.369 & 0.738\,\cr
\end{pmatrix}.
\end{align}
The eigenvalues of $J(\bm\lambda^\ast)$ are $(\theta_1,\theta_2,\theta_3)=(0.000,0.500,$ $0.855)$. Then, $\theta_{\rm max}=\theta_3=0.855$. We have, for $N=500$,
\begin{align}
L(500)=0.161\doteqdot-\log\theta_{\rm max}=0.157.\label{eqn:example4speedcomparison}
\end{align}
We can see from Fig. \ref{fig:Phi(1)graphexpconvergence} that $L(N)$ for large $N$ is accurately approximated by the value $-\log\theta_{\rm max}$.
\begin{figure}[t]
\begin{center}
\begin{overpic}[width=8cm]{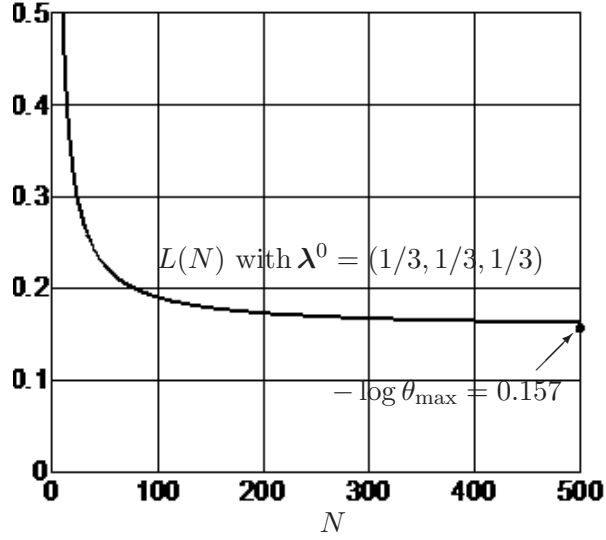}
\put(87,23){\vector(1,1){6}}
\put(53,18){$-\log\theta_{\rm max}=0.157$}
\put(51,-4){$N$}
\put(24,40){$L(N)\ \text{\rm with}\,\bm\lambda^0=(1/3,1/3,1/3)$}
\end{overpic}
\caption{Convergence of $L(N)$ in Example \ref{exa:CM_Phi(1)_again} with initial distribution $\bm\lambda^0=(1/3,1/3,1/3)$.}
\label{fig:Phi(1)graphexpconvergence}
\end{center}
\end{figure}
\end{example}

\begin{example}
\label{exa:CM_Phi(4)}
\rm Let us consider another channel matrix. Define
\begin{align}
\Phi^{(4)}\equiv
\begin{pmatrix}
\,0.793 & 0.196 & 0.011\,\\
0.196 & 0.793 & 0.011 \\
0.250 & 0.250 & 0.500
\end{pmatrix}.
\end{align}
We have
\begin{align}
\bm\lambda^\ast&=(0.352,0.352,0.296),\\
Q^\ast&=(0.422,0.422,0.156),\\
J(\bm\lambda^\ast)&=
\begin{pmatrix}
0.443 & -0.260 & -0.183\,\cr
-0.260 & 0.443 & -0.183\cr
\,-0.218 & -0.218 & 0.436
\end{pmatrix}.\label{eqn:example5J}
\end{align}
The eigenvalues of $J(\bm\lambda^\ast)$ are $(\theta_1,\theta_2,\theta_3)=(0.000,0.618,$ $0.702)$. Then, $\theta_{\rm max}=\theta_3=0.702$. Write the second largest eigenvalue as $\theta_{\rm sec}$, then $\theta_{\rm sec}=\theta_2=0.618$.

We show in Fig. \ref{fig:2initialsgraphexpconvergence} the graph of $L(N)$ with initial distribution ${\bm\lambda}^0_1\equiv(1/3,1/3,1/3)$ by the solid line, and the graph with initial distribution ${\bm\lambda}^0_2\equiv(1/2,1/3,1/6)$ by the dotted line.
\begin{figure}[t]
\begin{center}
\begin{overpic}[width=8cm]{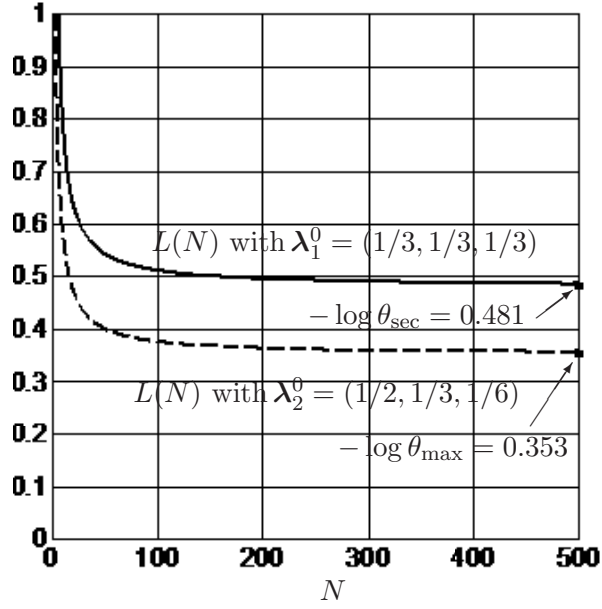}
\put(23,54){$L(N)\ \text{\rm with}\,\bm\lambda^0_1=(1/3,1/3,1/3)$}
\put(20,29){$L(N)\ \text{\rm with}\,\bm\lambda^0_2=(1/2,1/3,1/6)$}
\put(87,43){\vector(4,3){6}}
\put(49,42){$-\log\theta_{\rm sec}=0.481$}
\put(86,25){\vector(2,3){7}} 
\put(54,20){$-\log\theta_{\text{\rm max}}=0.353$}
\put(51,-4){$N$}
\end{overpic}
\caption{Convergence of $L(N)$ in Example \ref{exa:CM_Phi(4)} with initial distribution $\bm\lambda^0_1=(1/3,1/3,1/3)$ and $\bm\lambda^0_2=(1/2,1/3,1/6)$.}
\label{fig:2initialsgraphexpconvergence}
\end{center}
\end{figure}
The larger $L(N)$ the faster the convergence, hence the convergence with $\bm\lambda^0_1$ is faster than with $\bm\lambda^0_2$. The convergence speed varies depending on the choice of initial distribution. What kind of initial distribution yields faster convergence? We will investigate it below.

Let us define
\begin{align}
\bm\mu^0_1&\equiv\bm\lambda^0_1-\bm\lambda^\ast=(-0.019,-0.019,0.038),\label{eqn:initialmu0}\\
\bm\mu^0_2&\equiv\bm\lambda^0_2-\bm\lambda^\ast=(0.148,-0.019,-0.129).\label{eqn:initialbarmu0}
\end{align}

We will execute the following calculation by regarding $\bm\mu^{N+1}=\bm\mu^NJ(\bm\lambda^\ast),\,N=0,1,\ldots$ holds exactly.

Here, we will investigate for general $m,n$. We assume (\ref{eqn:eigenvaluesaredistinct}). Let $\bm b_{\rm max}$ be the left eigenvector of $J(\bm\lambda^\ast)$ for $\theta_{\rm max}$, and let $\bm b_{\rm max}^\perp$ be the orthogonal complement of $\bm b_{\rm max}$, i.e., $\bm b_{\rm max}^\perp\equiv\{\bm\mu\in\mathbb{R}^m\,|\,\bm\mu\,{^t}\bm b_{\rm max}=0\}.$
\begin{lemma}
\label{lem:thetasec}
If 
\begin{align}
\bm\mu^N\in\bm b_{\rm max}^\perp,\,N=0,1,\ldots,\label{eqn:innuperp}
\end{align}
then for any $\theta$ with $\theta_{\rm sec}<\theta<1$, we have $\|\bm\mu^N\|<K\left(\theta\right)^N,\,K>0,\,N=0,1,\ldots$.
\end{lemma}
{\bf Proof:}
See Appendix \ref{sec:proofofthetasec}.\hfill$\blacksquare$

\bigskip

Because $\theta_{\text{\rm sec}}<\theta_{\rm max}$, if (\ref{eqn:innuperp}) holds then the convergence speed is faster than $\theta_{\rm max}$ by Lemma \ref{lem:thetasec}. Next lemma gives a necessary and sufficient condition for guaranteeing (\ref{eqn:innuperp}). 
\begin{lemma}
\label{lem:migikoyuuvector}
$\bm\mu J(\bm\lambda^\ast)\in\bm b_{\rm max}^\perp$ holds for any $\bm\mu\in\bm b_{\rm max}^\perp$ if and only if ${^t}\bm b_{\rm max}$ is a right eigenvector for $\theta_{\rm max}$.
\end{lemma}
{\bf Proof:} See Appendix \ref{sec:proofofmigikoyuuvector}.\hfill$\blacksquare$

\bigskip

If ${^t}\bm b_{\rm max}$ is a right eigenvector, then by Lemma \ref{lem:migikoyuuvector}, any $\bm\mu^0\in\bm b_{\rm max}^\perp$ yields (\ref{eqn:innuperp}), hence the convergence becomes faster.

Now, we will evaluate the convergence speed for the initial distributions (\ref{eqn:initialmu0}) and (\ref{eqn:initialbarmu0}). For $J(\bm\lambda^\ast)$ of (\ref{eqn:example5J}), $\theta_{\rm max}=0.702$ and $\theta_{\rm sec}=0.618$. The left eigenvector for $\theta_{\rm max}$ is $\bm b_{\rm max}=(-0.500,0.500,0.000)$. We can confirm that ${^t}\bm b_{\rm max}$ is a right eigenvector for $\theta_{\rm max}$ and $\bm\mu^0_1\,{^t}\bm b_{\rm max}=0$, thus by Lemmas \ref{lem:thetasec} and \ref{lem:migikoyuuvector}, we have $\lim_{N\to\infty}L(N)\doteqdot-\log\theta_{\rm sec}$. Then by the solid line in Fig. \ref{fig:2initialsgraphexpconvergence}, we have, for $N=500$,
\begin{align}
L(500)=0.489\doteqdot-\log\theta_{\rm sec}=0.481.
\end{align}
On the other hand, we have $\bm\mu^0_2\,{^t}\bm b_{\rm max}\neq0$, thus by Lemma \ref{lem:migikoyuuvector}, we have $\lim_{N\to\infty}L(N)\doteqdot-\log\theta_{\rm max}$. Then by the dotted line, we have, for $N=500$,
\begin{align}
L(500)=0.360\doteqdot-\log\theta_{\rm max}=0.353.
\end{align}

Checking Example \ref{exa:CM_Phi(1)_again} in this way, we can see that $\bm b_{\rm max}=(-0.431,-0.431,0.862)$ is a left eigenvector for $\theta_{\rm max}=0.855$, but ${^t}\bm b_{\rm max}$ is not a right eigenvector. Thus, by Lemma \ref{lem:migikoyuuvector}, we have $\lim_{N\to\infty}L(N)\doteqdot-\log\theta_{\rm max}$ and (\ref{eqn:example4speedcomparison}).
\end{example}

\subsection{$O(1/N)$ convergence}
\begin{example}
\label{exa:CM_Phi(2)_again}
\rm Consider the channel matrix $\Phi^{(2)}$ of (\ref{eqn:Phi2}), i.e.,
\begin{align}
\Phi^{(2)}&=\begin{pmatrix}
\,0.800 & 0.100 & 0.100\,\\
\,0.100 & 0.800 & 0.100\,\\
\,0.300 & 0.300 & 0.400\,
\end{pmatrix},
\end{align}
and an initial distribution $\bm\lambda^0=(1/3,1/3,1/3)$. We have
\begin{align}
\bm\lambda^\ast&=(0.500,0.500,0.000),\\
Q^\ast&=(0.450,0.450,0.100),\\
D_{1,1}^\ast&=-1.544,\,D_{1,2}^\ast=-0.456,\,D_{1,3}^\ast=-1,\\
D_{2,2}^\ast&=-1.544,\,D_{2,3}^\ast=-1,\,D_{3,3}^\ast=-2,\\
J(\bm\lambda^\ast)
&=\begin{pmatrix}
\,1+\lambda_1^\ast D_{1,1}^\ast & \lambda_2^\ast D_{1,2}^\ast & 0\,\\
\,\lambda_1^\ast D_{2,1}^\ast & 1+\lambda_2^\ast D_{2,2}^\ast & 0 \\
\,\lambda_1^\ast D_{3,1}^\ast & \lambda_2^\ast D_{3,2}^\ast & 1\end{pmatrix}\\
&=\begin{pmatrix}
\,0.228 & -0.228 & 0.000\,\\
\,-0.228 & 0.228 & 0.000\,\\
\,-0.500 & -0.500 & 1.000\,
\end{pmatrix}.
\end{align}
The eigenvalues of $J(\bm\lambda^\ast)$ are $(\theta_1,\theta_2,\theta_3)=(0.000,0.456,1.000)$. 
\begin{align}
A=\begin{pmatrix}
\,1 & 1 & 0\,\\
\,1 & -1 & 0\,\\
\,1 & 0 & 1\,
\end{pmatrix},\ 
A_1=\begin{pmatrix}
\,1 & 1\,\\
\,1 & -1\,
\end{pmatrix},\ 
A_2=\begin{pmatrix}
\,1 & 0\,
\end{pmatrix}.\label{eqn:AA1A2}
\end{align}

By (\ref{eqn:AA1A2}), the eigenvectors $\bm{a}_1$ for $\theta_1=0$ and $\bm{a}_2$ for $\theta_2=0.456$ are
\begin{align}
\bm{a}_1=\begin{pmatrix}1\\1\\1\end{pmatrix},\ \bm{a}_2=\begin{pmatrix}1\\-1\\0\end{pmatrix}.
\end{align}
We will prove (\ref{eqn:bmmu{N+1}bm{a}isdivisiblebybmmuNbm{a}i}). First, for $\bm{a}_1$, $\bm\mu^{N+1}\bm{a}_1=\bm\mu^N\bm{a}_1=\bm0$, thus (\ref{eqn:bmmu{N+1}bm{a}isdivisiblebybmmuNbm{a}i}) is trivial. Next, for $\bm{a}_2$, we will prove that $\bm\mu^{N+1}\bm{a}_2=\mu^{N+1}_1-\mu^{N+1}_2$ is divisible by $\bm\mu^N\bm{a}_2=\mu^N_1-\mu^N_2$. In fact, by $\lambda^\ast_1=\lambda^\ast_2$ we have
\begin{align}
\mu^{N+1}_1-\mu^{N+1}_2&=\lambda^{N+1}_1-\lambda^{N+1}_2\\
&=F_1(\bm\lambda^N)-F_2(\bm\lambda^N)\\
&=\dfrac{\lambda^N_1\exp D(P^1\|\bm\lambda^N\Phi)-\lambda^N_2\exp D(P^2\|\bm\lambda^N\Phi)}{\ds\sum_{k=1}^3\lambda^N_k\exp D(P^k\|\bm\lambda^N\Phi)}.\label{eqn:F1-F2}
\end{align}
If we put $\lambda^N_1=\lambda^N_2$ in (\ref{eqn:F1-F2}), we have by calculation, $\mu^{N+1}_1-\mu^{N+1}_2=0$. Because $\lambda^N_1=\lambda^N_2$ is equivalent to $\mu^N_1=\mu^N_2$, we see that $\mu^N_1-\mu^N_2=0$ implies $\mu^{N+1}_1-\mu^{N+1}_2=0$, thus (\ref{eqn:bmmu{N+1}bm{a}isdivisiblebybmmuNbm{a}i}) holds and then we can consider (\ref{eqn:bmmuN{I,III}A1+bmmuN{II}A2=bm0}).

For $\bar{\bm\mu}^N_{\rm I,\,III}=(\bar{\mu}^N_1,\,\bar{\mu}^N_2),\ \bar{\bm\mu}^N_{\rm II}=(\bar{\mu}^N_3)$, we have $\bar{\bm\mu}^N_{\rm I,\,III}=-\bar{\bm\mu}^N_{\rm II}A_2A_1^{-1}$ by (\ref{eqn:bmmuN{I,III}A1+bmmuN{II}A2=bm0}), hence 
\begin{align}
\bar{\mu}^N_1=\bar{\mu}^N_2=-(1/2)\bar{\mu}^N_3.\label{eqn:mu_1=mu_2=-(1/2)mu_3}
\end{align}
Further, the Hessian matrix $H_3(\bm\lambda^\ast)$ is
\begin{align}
H_3(\bm\lambda^\ast)&=\begin{pmatrix}\,0 & 0 & D_{3,1}^\ast \,\\
0 & 0 & D_{3,2}^\ast \\
\,D_{3,1}^\ast & D_{3,2}^\ast & 2D_{3,3}^\ast\,\end{pmatrix}.\\
&=\begin{pmatrix}
\,0.000 & 0.000 & -1.000\,\\
\,0.000 & 0.000 & -1.000\,\\
\,-1.000 & -1.000 & -4.000\,\end{pmatrix},
\end{align}
then, we have by (\ref{eqn:mu_1=mu_2=-(1/2)mu_3}),
\begin{align}
\dfrac{1}{2}\bar{\bm\mu}^NH_3{^t}\bar{\bm\mu}^N=-\left(\bar{\mu}^N_3\right)^2,
\end{align}
and the second order recurrence formula
\begin{align}
\bar{\mu}_3^{N+1}=\bar{\mu}_3^N-\left(\bar{\mu}_3^N\right)^2.
\end{align}
By Lemma \ref{lem:onevariablereccurence} and (\ref{eqn:mu_1=mu_2=-(1/2)mu_3}), we have $\lim_{N\to\infty}N\bar{\mu}_3^N=1,\,\lim_{N\to\infty}N\bar{\mu}_1^N=\lim_{N\to\infty}N\bar{\mu}_2^N=-1/2$. 

By the numerical simulation, $N\bm\mu^N$ for $N=500$ is
\begin{align}
&N\bm\mu^N=(-0.510,-0.510,1.019)\\
&\doteqdot\ds\lim_{N\to\infty}N\bar{\bm\mu}^N=(-1/2,-1/2,1).\label{eqn:rironchiexample6}
\end{align}
See Fig. \ref{fig:Phi(2)graph1Nconvergence}. We can confirm that $N\bm\mu^N$ for large $N$ is close to the values obtained in Theorem \ref{the:1/Norderconvergence_specialinitial}.

\begin{figure}[t]
\begin{center}
\begin{overpic}[width=8cm]{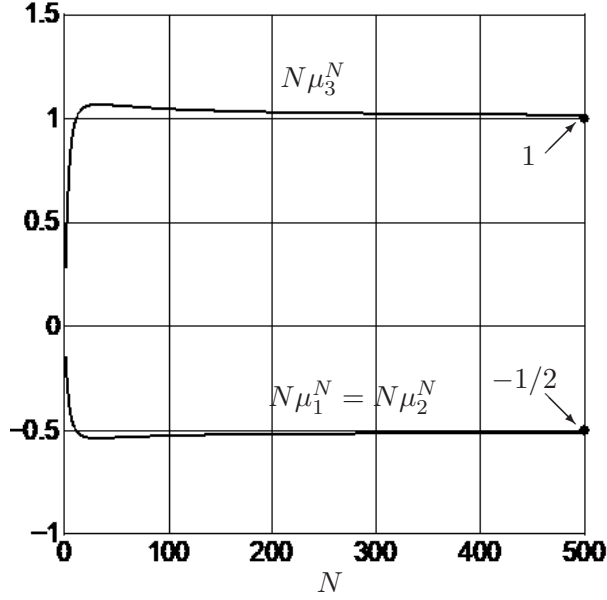}
\put(89,69){\vector(1,1){5}}
\put(85,67){$1$}
\put(89,29){\vector(1,-1){5}}
\put(80,30){$-1/2$}
\put(51,-4){$N$}
\put(45,80){$N\mu^N_3$}
\put(43,27){$N\mu^N_1=N\mu^N_2$}
\end{overpic}
\caption{Convergence of $N\mu^N_i$ in Example \ref{exa:CM_Phi(2)_again}.}
\label{fig:Phi(2)graph1Nconvergence}
\end{center}
\end{figure}
\end{example}
\begin{example}
\label{exa:CM_Phi(5)}
\rm Consider a channel matrix
\begin{align}
\Phi^{(5)}&=\begin{pmatrix}
0.6 & 0.1 & 0.1 & 0.1 & 0.1\\
0.1 & 0.6 & 0.1 & 0.1 & 0.1\\
s & s & t & 0.1 & 0.1\\
s & s & 0.1 & t & 0.1\\
s & s & 0.1 & 0.1 & t\\
\end{pmatrix},\\
s&\equiv0.238,\,t\equiv0.324,\,(2s+t+0.2=1),
\end{align}
and an initial distribution $\bm\lambda^0=(1/5,1/5,1/5,1/5,1/5)$. For this $\Phi^{(5)}$, we have
\begin{align}
\bm\lambda^\ast&=(0.5,\,0.5,\,0,\,0,\,0),\\
Q^\ast&=(0.35,\,0.35,\,0.1,\,0.1,\,0.1),\\
D^\ast_{1,1}&=-19/14,\,D^\ast_{1,2}=-9/14,\,D^\ast_{1,3}=-1,\,D^\ast_{1,4}=-1,\,D^\ast_{1,5}=-1,\\
D^\ast_{2,2}&=-19/14,\,D^\ast_{2,3}=-1,\,D^\ast_{2,4}=-1,\,D^\ast_{2,5}=-1,\\
D^\ast_{3,3}&=-1.576\equiv-\alpha,\,D^\ast_{3,4}=-1.072\equiv-\beta,\,D^\ast_{3,5}=-\beta,\\
D^\ast_{4,4}&=-\alpha,\,D^\ast_{4,5}=-\beta,\,D^\ast_{5,5}=-\alpha,\\[2mm]
J(\bm\lambda^\ast)
&=\begin{pmatrix}
\,9/28 & -9/28 & 0 & 0 & 0\,\\
\,-9/28 & 9/28 & 0 & 0 & 0\,\\
\,-1/2 & -1/2 & 1 & 0 & 0\,\\
\,-1/2 & -1/2 & 0 & 1 & 0\,\\
\,-1/2 & -1/2 & 0 & 0 & 1\,
\end{pmatrix}.
\end{align}
The eigenvalues of $J(\bm\lambda^\ast)$ are $(\theta_1,\theta_2,\theta_3,\theta_4,\theta_5)=(0,9/14,1,1,1)$ and
\begin{align}
A=\begin{pmatrix}
\,1 & 1 & 0 & 0 & 0\,\\
\,1 & -1 & 0 & 0 & 0\,\\
\,1 & 0 & 1 & 0 & 0\,\\
\,1 & 0 & 0 & 1 & 0\,\\
\,1 & 0 & 0 & 0 & 1\,
\end{pmatrix},\ 
A_1=\begin{pmatrix}
\,1 & 1\,\\
\,1 & -1\,
\end{pmatrix},\ 
A_2=\begin{pmatrix}
\,1 & 0\,\\
\,1 & 0\,\\
\,1 & 0\,
\end{pmatrix}.
\end{align}

We can prove (\ref{eqn:bmmu{N+1}bm{a}isdivisiblebybmmuNbm{a}i}) in a similar way as in the Example \ref{exa:CM_Phi(2)_again}.

For $\bar{\bm\mu}^N_{\rm I,\,III}=(\bar{\mu}^N_1,\,\bar{\mu}^N_2),\ \bar{\bm\mu}^N_{\rm II}=(\bar{\mu}^N_3,\,\bar{\mu}^N_4,\,\bar{\mu}^N_5)$, we have $\bar{\bm\mu}^N_{\rm I,\,III}=-\bar{\bm\mu}^N_{\rm II}A_2A_1^{-1}$ by (\ref{eqn:bmmuN{I,III}A1+bmmuN{II}A2=bm0}), hence
\begin{align}
\bar{\mu}^N_1=\bar{\mu}^N_2=-(\bar{\mu}^N_3+\bar{\mu}^N_4+\bar{\mu}^N_5)/2.\label{eqn:mu1=mu2=-(mu3+mu4+mu5)/2}
\end{align}
Further, the Hessian matrix $H_3(\bm\lambda^\ast)$ is
\begin{align}
H_3(\bm\lambda^\ast)
&=\begin{pmatrix}
\,0 & 0 & -1 & 0 & 0\,\\
\,0 & 0 & -1 & 0 & 0\,\\
\,-1 & -1 & -2\alpha & -\beta & -\beta\,\\
\,0 & 0 & -\beta & 0 & 0\,\\
\,0 & 0 & -\beta & 0 & 0\,
\end{pmatrix},
\end{align}
thus, we have by (\ref{eqn:mu1=mu2=-(mu3+mu4+mu5)/2})
\begin{align}
\dfrac{1}{2}\bar{\bm\mu}^NH_3(\bm\lambda^\ast)\,{^t}\bar{\bm\mu}^N=-(\alpha-1)\left(\bar{\mu}^N_3\right)^2-(\beta-1)\bar{\mu}^N_3\bar{\mu}^N_4-(1-\beta)\bar{\mu}^N_3\bar{\mu}^N_5.
\end{align}
Similarly,
\begin{align}
\dfrac{1}{2}\bar{\bm\mu}^NH_4(\bm\lambda^\ast)\,{^t}\bar{\bm\mu}^N&=-(\beta-1)\bar{\mu}^N_3\bar{\mu}^N_4-(\alpha-1)\left(\bar{\mu}^N_4\right)^2-(\beta-1)\bar{\mu}^N_4\bar{\mu}^N_5,\\
\dfrac{1}{2}\bar{\bm\mu}^NH_5(\bm\lambda^\ast)\,{^t}\bar{\bm\mu}^N&=-(\beta-1)\bar{\mu}^N_3\bar{\mu}^N_5-(\beta-1)\bar{\mu}^N_4\bar{\mu}^N_5-(\alpha-1)\left(\bar{\mu}^N_5\right)^2.
\end{align}
Therefore, by putting $\alpha'\equiv\alpha-1,\,\beta'\equiv\beta-1$, we have
\begin{align}
\bar{\mu}_3^{N+1}&=\bar{\mu}_3^N-\alpha'\left(\bar{\mu}_3^N\right)^2-\beta'\bar{\mu}_3^N\bar{\mu}_4^N-\beta'\bar{\mu}_3^N\bar{\mu}_5^N,\label{eqn:Phi6zenkashiki1}\\
\bar{\mu}_4^{N+1}&=\bar{\mu}_4^N-\beta'\bar{\mu}_3^N\bar{\mu}_4^N-\alpha'\left(\bar{\mu}_4^N\right)^2-\beta'\bar{\mu}_4^N\bar{\mu}_5^N,\label{eqn:Phi6zenkashiki2}\\
\bar{\mu}_5^{N+1}&=\bar{\mu}_5^N-\beta'\bar{\mu}_3^N\bar{\mu}_5^N-\beta'\bar{\mu}_4^N\bar{\mu}_5^N-\alpha'\left(\bar{\mu}_5^N\right)^2.\label{eqn:Phi6zenkashiki3}
\end{align}
From $(\ref{eqn:sigmadefinition})$, we have $\bm\sigma=(\sigma_3,\sigma_4,\sigma_5)=(1.389,1.389,1.389)$, so the canonical form for (\ref{eqn:Phi6zenkashiki1}), (\ref{eqn:Phi6zenkashiki2}), (\ref{eqn:Phi6zenkashiki3}) is 
\begin{align}
\nu_3^{N+1}&=\nu_3^N-0.8\left(\nu_3^N\right)^2-0.1\nu_3^N\mu_4^N-0.1\nu_3^N\nu_5^N,\label{eqn:Phi6canonical1}\\
\nu_4^{N+1}&=\nu_4^N-0.1\nu_3^N\nu_4^N-0.8\left(\nu_4^N\right)^2-0.1\nu_4^N\nu_5^N,\label{eqn:Phi6canonical2}\\
\nu_5^{N+1}&=\nu_5^N-0.1\nu_3^N\nu_5^N-0.1\nu_4^N\nu_5^N-0.8\left(\nu_5^N\right)^2.\label{eqn:Phi6canonical3}
\end{align}
Eqs. (\ref{eqn:Phi6canonical1}), (\ref{eqn:Phi6canonical2}), (\ref{eqn:Phi6canonical3}) satisfy the assumptions of Lemma $\ref{lem:daishou_kenkei_hozon_under_special_condition}$ and the diagonally dominant condition $(\ref{eqn:taikakuyuui2})$, then by Theorem $\ref{the:1/N_order_convergence_under_special_condition}$, they converge for arbitrary initial values and
\begin{align}
\lim_{N\to\infty}N\nu_i^N=1,\,i=3,4,5.
\end{align}
Therefore, by (\ref{eqn:nuiNequivmuiN/sigmai})
\begin{align}
\lim_{N\to\infty}N\bar{\mu}_i^N=\sigma_i=1.389,\,i=3,4,5,\label{eqn:lim{Ntoinfty}NmuiN=sigmai=1.389}
\end{align}
and by $(\ref{eqn:mu1=mu2=-(mu3+mu4+mu5)/2})$,
\begin{align}
\lim_{N\to\infty}N\bar{\mu}_i^N=-3\sigma_3/2=-2.083,\,i=1,2.\label{eqn:lim{Ntoinfty}NmuiN=-3sigma3/2=-2.083}
\end{align}

We will show in Fig. $\ref{fig:5-5}$ the comparison of the numerical results and the values of (\ref{eqn:lim{Ntoinfty}NmuiN=sigmai=1.389}), (\ref{eqn:lim{Ntoinfty}NmuiN=-3sigma3/2=-2.083}).
\begin{figure}[t]
\begin{center}
\begin{overpic}[width=8cm]{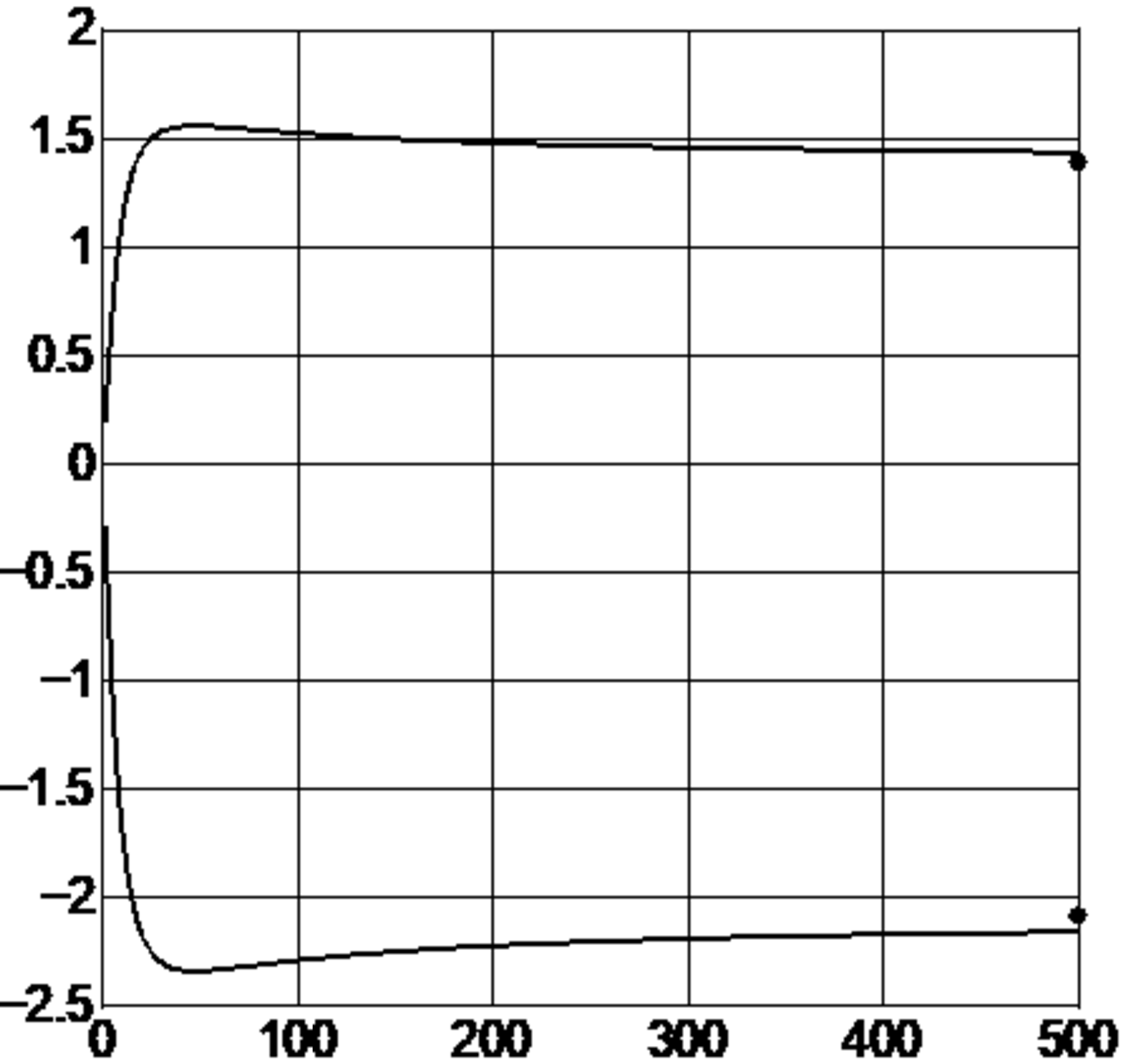}
\put(86,71.5){\vector(1,1){8}}
\put(70,67){$\sigma_3=1.389$}
\put(86.5,22.5){\vector(1,-1){8}}
\put(55.5,25){$-3\sigma_3/2=-2.083$}
\put(51,-4){$N$}
\put(25,74){$N\mu^N_3=N\mu^N_4=N\mu^N_5$}
\put(30,17){$N\mu^N_1=N\mu^N_2$}
\end{overpic}
\caption{Convergence of $N\mu^N_i$ in Example \ref{exa:CM_Phi(5)}.}
\label{fig:5-5}
\end{center}
\end{figure}
\end{example}
\subsection{Exponential convergence where $\bm\lambda^\ast\in\partial\Delta({\cal X})$}
\begin{example}
\label{exa:CM_Phi(3)_again}
\rm Consider the channel matrix $\Phi^{(3)}$ of (\ref{eqn:Phi3})
\begin{align}
\Phi^{(3)}&=\begin{pmatrix}
\,0.800 & 0.100 & 0.100\,\\
\,0.100 & 0.800 & 0.100\,\\
\,0.350 & 0.350 & 0.300\,
\end{pmatrix},
\end{align}
and an initial distribution $\bm\lambda^0=(1/3,1/3,1/3)$. We have
\begin{align}
\bm\lambda^\ast&=(0.500,0.500,0.000),\\
Q^\ast&=(0.450,0.450,0.100),\\
J(\bm\lambda^\ast)&=
\begin{pmatrix}
\,0.228 & -0.228 & 0.000\,\cr
\,-0.228 & 0.228 & 0.000\,\cr
\,-0.428 & -0.428 & 0.856\,\cr
\end{pmatrix}.\label{eqn:example8Jacobimatrix}
\end{align}
The eigenvalues of $J(\bm\lambda^\ast)$ are $(\theta_1,\theta_2,\theta_3)=(0.000,0.456,$ $0.856)$. Then, $\theta_{\rm max}=\theta_3=0.856$. We have for $N=500$
\begin{align}
L(500)=0.159\doteqdot-\log\theta_{\rm max}=0.155.
\end{align}
See Fig. \ref{fig:Phi(3)graphexpconvergence}.
\begin{figure}[t]
\begin{center}
\begin{overpic}[width=8cm]{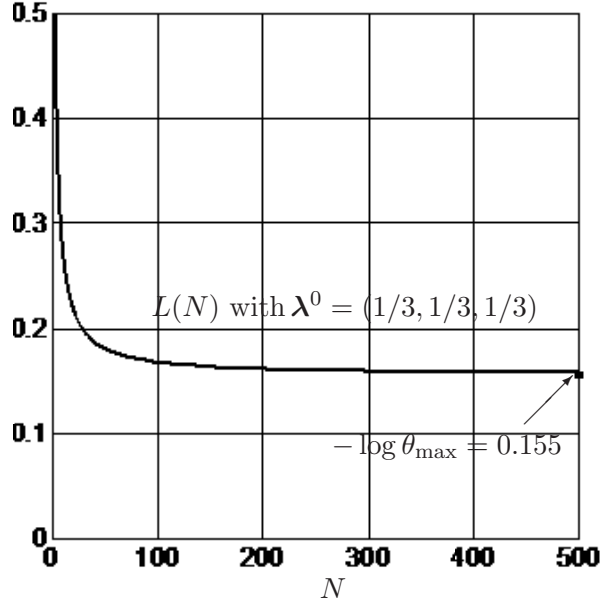}
\put(85,25){\vector(1,1){8}}
\put(53,20){$-\log\theta_{\rm max}=0.155$}
\put(51,-4){$N$}
\put(23,43){$L(N)\ \text{\rm with}\,\bm\lambda^0=(1/3,1/3,1/3)$}
\end{overpic}
\caption{Convergence of $L(N)$ in Example \ref{exa:CM_Phi(3)_again} with initial distribution $\bm\lambda^0=(1/3,1/3,1/3)$.}
\label{fig:Phi(3)graphexpconvergence}
\end{center}
\end{figure}

Extending this result, we have the following lemma.
\begin{lemma}
\label{lem:thetamaxisinI3}
Assume that type-II indices do not exist and $\theta_{\rm max}=\max_{i\in{\cal I}_{\rm III}}\theta_i$, i.e., the maximum eigenvalue of $J(\bm\lambda^\ast)$ is achieved in $J^{\rm III}$. Then, the convergence speed does not depend on the choice of initial distribution. In other words, $\lim_{N\to\infty}L(N)=-\log\theta_{\rm max}$ holds for arbitrary initial distribution, hence the convergence speed cannot be increased any more.
\end{lemma}
\noindent{\bf Proof:} Let $\theta_{\rm max}=\theta_{i^\ast},\,i^\ast\in{\cal I}_{\rm III}$. $J^{\rm III}$ is diagonal by (\ref{eqn:Jstructure3}), thus we can take ${^t}{\bm e}_{i^\ast}={^t}(0,\ldots,0,\stackrel{i^\ast\,\text{th}}{\stackrel{\vee}{1}}\hspace{-1mm},\ 0,\ldots,0)$ as a right eigenvector for $\theta_{i^\ast}$. However, ${\bm e}_{i^\ast}$ is not a left eigenvector for $\theta_{i^\ast}$. In fact, since every row sum of $J(\bm\lambda^\ast)$ is 0 by Lemma \ref{lem:rowsumofJis0}, putting $\bm1\equiv(1,\ldots,1)\in{\mathbb R}^m$, we have $J(\bm\lambda^\ast){^t}\bm1=\bm0$. If ${\bm e}_{i^\ast}$ were a left eigenvector for $\theta_{i^\ast}$, then $0={\bm e}_{i^\ast}J(\bm\lambda^\ast){^t}\bm1=\theta_{i^\ast}{\bm e}_{i^\ast}{^t}\bm1=\theta_{i^\ast}>0$, which is a contradiction. $\theta_{i^\ast}>0$ is due to Theorem \ref{the:eigenvaluesofJ3}. Therefore, by Theorems \ref{lem:thetasec} and \ref{lem:migikoyuuvector}, the convergence speed does not depend on the choice of initial distribution and $\lim_{N\to\infty}L(N)=-\log\theta_{\rm max}$.\hfill$\blacksquare$
\end{example}

\section{Convergence speed of $I(\bm\lambda^N,\Phi)\to C$}
Based on the results obtained so far, we will consider the convergence speed that the mutual information $I(\bm\lambda^N,\Phi)$ tends to $C$ as $N\to\infty$. We will show that if ${\cal I}_{\rm III}=\emptyset$ and $\bm\lambda^N\to\bm\lambda^\ast$ is the $O(1/N)$ convergence, then $I(\bm\lambda^N,\Phi)\to C$ is $O(1/N^2)$. Including this fact, we have the following theorem.
\begin{theorem}
\label{the:I(lambdaN,Phi)toC} Let ${\cal I}_{\rm III}=\emptyset$.

If $\|\bm\mu^N\|<K_1\left(\theta\right)^N,\,0\leq\theta_{\rm max}<\theta<1,\,K_1>0,\,N=0,1,\ldots$, then
\begin{align}
0<C-I(\bm\lambda^N,\Phi)<K_2\left(\theta\right)^{2N},\,K_2>0,\,N=0,1,\ldots.\label{eqn:IlambdaPhitoC_1}
\end{align}
While, if $\lim_{N\to\infty}N\mu^N_i=\sigma_i\neq0,\,i=1,\ldots,m$, then
\begin{align}
\lim_{N\to\infty}N^2\left(C-I(\bm\lambda^N,\Phi)\right)=\dfrac{1}{2}\sum_{j=1}^n\dfrac{1}{Q^\ast_j}\left(\sum_{i=1}^m\sigma_iP^i_j\right)^2.\label{eqn:IlambdaPhitoC_2}
\end{align}
Next, let ${\cal I}_{\rm III}\neq\emptyset$.

If $\|\bm\mu^N\|<K_1\left(\theta\right)^N,\,0\leq\theta_{\rm max}<\theta<1,\,K_1>0,\,N=0,1,\ldots$, then
\begin{align}
0<C-I(\bm\lambda^N,\Phi)<K_2\left(\theta\right)^N,\,K_2>0,\,N=0,1,\ldots.\label{eqn:IlambdaPhitoC_3}
\end{align}
While, if $\lim_{N\to\infty}N\mu^N_i=\sigma_i\neq0,\,i=1,\ldots,m$, then
\begin{align}
0<C-I(\bm\lambda^N,\Phi)<K/N,\,K>0,\,N=0,1,\ldots.\label{eqn:IlambdaPhitoC_4}
\end{align}
\end{theorem}

\noindent{\bf Proof:} See Appendix \ref{sec:proofofI(lambdaN,Phi)toC}.\hfill$\blacksquare$

\bigskip

\noindent We will show in the following tables the evaluation of the convergence speed of $I(\bm\lambda^N,\Phi^{(k)})\to C$ for $k=1,2,3$, where $\Phi^{(1)}$, $\Phi^{(2)}$ and $\Phi^{(3)}$ were defined in Examples \ref{exa:CM_Phi(1)}, \ref{exa:CM_Phi(2)} and \ref{exa:CM_Phi(3)}, and examined in Examples \ref{exa:CM_Phi(1)_again}, \ref{exa:CM_Phi(2)_again} and \ref{exa:CM_Phi(3)_again}, respectively.

\begin{table}[H]
\begin{center}
\caption{Convergence speed of $I(\bm\lambda^N,\Phi^{(1)})\to C$.}
\medskip
\begin{tabular}{|l|c|}
\hline
$-(1/N)\log\left(C-I(\bm\lambda^N,\Phi^{(1)})\right)\big|_{N=500}$ & 0.324\rule[-3mm]{0mm}{8.2mm}\\
\hline
$-2\log\theta_{\rm max}$ & 0.313\rule[-2mm]{0mm}{7mm}\\
\hline
\end{tabular}
\end{center}
\end{table}

\vspace{-12mm}

\begin{table}[H]
\begin{center}
\caption{Convergence speed of $I(\bm\lambda^N,\Phi^{(2)})\to C$.}
\medskip
\begin{tabular}{|l|c|}
\hline
$N^2\left(C-I(\bm\lambda^N,\Phi^{(2)})\right)\big|_{N=500}$ & 0.516\rule[-3mm]{0mm}{8.2mm}\\
\hline
eq. (\ref{eqn:IlambdaPhitoC_2}) & 0.500\rule[-2mm]{0mm}{7mm}\\
\hline
\end{tabular}
\end{center}
\end{table}

\vspace{-12mm}

\begin{table}[H]
\begin{center}
\caption{Convergence speed of $I(\bm\lambda^N,\Phi^{(3)})\to C$.}
\medskip
\begin{tabular}{|l|c|}
\hline
$-(1/N)\log\left(C-I(\bm\lambda^N,\Phi^{(3)})\right)\big|_{N=500}$ & 0.163 \rule[-3mm]{0mm}{8.2mm}\\
\hline
$-\log\theta_{\rm max}$ & 0.155 \rule[-2mm]{0mm}{7mm}\\
\hline
\end{tabular}
\end{center}
\end{table}
We can see from these tables that the convergence speed of $I\left(\bm\lambda^N,\Phi\right)\to C$ is accurately approximated by the results of Theorem \ref{the:I(lambdaN,Phi)toC}.
\section{Conclusion}
In this paper, we investigated the convergence speed of the Arimoto-Blahut algorithm. We showed that the capacity-achieving input distribution $\bm\lambda^\ast$ is the fixed point of $F(\bm\lambda)$, and analyzed the convergence speed by the Taylor expansion of $F(\bm\lambda)$ about $\bm\lambda=\bm\lambda^\ast$. We concretely calculated the Jacobian matrix $J$ of the first order term of the Taylor expansion and the Hessian matrix $H$ of the second order term. The analysis of the convergence speed by the Hessian matrix $H$ was done for the first time in this paper.

We showed that if type-II indices do not exist then the convergence of $\bm\lambda^N\to\bm\lambda^\ast$ is exponential and if type-II indices exist then the convergence of the second order recurrence formula obtained by truncating the Taylor expansion is $O(1/N)$ for some initial vector. Further, we considered the condition for the $O(1/N)$ convergence for arbitrary initial vector. Next, we considered the convergence speed of $I(\bm\lambda^N,\Phi)\to C$ and showed that the type-III indices concern the convergence speed. Especially, if there exist no type-III indices and $\bm\lambda^N\to\bm\lambda^\ast$ is $O(1/N)$, then $I(\bm\lambda^N,\Phi)\to C$ is $O(1/N^2)$.

Based on these analysis, the convergence speeds for several channel matrices were numerically evaluated. As a result, it was confirmed that the convergence speed of the Arimoto-Blahut algorithm is very accurately approximated by the values obtained by our theorems.


%


%

\newpage

\appendix

\section{Proof of Lemma \ref{lem:Jlambdadiagonalizable}}
\label{sec:proooflemmaJlambdadiagonalizable}
Since $\sqrt\Lambda B\sqrt\Lambda^{-1}$ is symmetric by (\ref{eqn:L-1BL}), it is diagonalizable, hence $J^{\rm I}$ is diagonalizable because $J^{\rm I}=I-B$. Therefore, there exists a regular matrix $\Pi\in{\mathbb R}^{m_1\times m_1}$ with 
\begin{align}
\Pi^{-1}J^{\rm I}\Pi=\Theta_1,\label{eqn:J1taikakuka}
\end{align}
where $\Theta_1$ is a diagonal matrix whose diagonal components are the eigenvalues of $J^{\rm I}$, i.e., $\Theta_1={\rm diag}(\theta_i,\,i\in{\cal I}_{\rm I})\in{\mathbb R}^{m_1\times m_1}$. We can write it by components as $\Theta_1=\left(\theta_i\delta_{i'i}\right),\,i',i\in{\cal I}_{\rm I}$.

Next, let $\Theta_2$ be a diagonal matrix whose diagonal components are the eigenvalues of $J^{\rm II}$ and $J^{\rm III}$, i.e., $\Theta_2={\rm diag}(\theta_i,\,i\in{\cal I}_{\rm II}\cup{\cal I}_{\rm III})\in{\mathbb R}^{(m_2+m_3)\times(m_2+m_3)}$. We have $ \Theta_2=\left(\theta_i\delta_{i'i}\right),\,i',i\in{\cal I}_{\rm II}\cup{\cal I}_{\rm III}$. Then by (\ref{eqn:J1AJ2}), we have
\begin{align}
J(\bm\lambda^\ast)=\begin{pmatrix}J^{\rm I} & O \\ U & \Theta_2\end{pmatrix},\label{eqn:Jlambdataikakuka}
\end{align}
where $U\in{\mathbb R}^{(m_2+m_3)\times m_1}$ is an appropriate matrix. 

Now, we will prove that there exists a matrix $V\in{\mathbb R}^{(m_2+m_3)\times m_1}$ and is unique which satisfies 
\begin{align}
V\Theta_1-\Theta_2V=U\Pi.\label{eqn:equationofV}
\end{align}
Define the components of $V$ by $V\equiv\left(v_{i'i}\right),\,i'\in{\cal I}_{\rm II}\cup{\cal I}_{\rm III},\,i\in{\cal I}_{\rm I}$, then,
\begin{align}
\left(V\Theta_1\right)_{i'i}&=\ds\sum_{k\in{\cal I}_{\rm I}}v_{i'k}\theta_i\delta_{ki}=v_{i'i}\theta_i,\,i'\in{\cal I}_{\rm II}\cup{\cal I}_{\rm III},\,i\in{\cal I}_{\rm I},\label{eqn:VTheta1components}\\
\left(\Theta_2V\right)_{i'i}&=\ds\sum_{k\in{\cal I}_{\rm II}\cup{\cal I}_{\rm III}}\theta_{i'}\delta_{ki'}v_{ki}=v_{i'i}\theta_{i'},\,i'\in{\cal I}_{\rm II}\cup{\cal I}_{\rm III},\,i\in{\cal I}_{\rm I}.\label{eqn:Theta2Vcomponents}
\end{align}
Further, defining the components of $U\Pi$ by $U\Pi=\left(u^\Pi_{i'i}\right),\,i'\in{\cal I}_{\rm II}\cup{\cal I}_{\rm III},\,i\in{\cal I}_{\rm I}$, the both sides of (\ref{eqn:equationofV}) are represented by components as
\begin{align}
\left(\theta_i-\theta_{i'}\right)v_{i'i}=u^\Pi_{i'i},\,i'\in{\cal I}_{\rm II}\cup{\cal I}_{\rm III},\,i\in{\cal I}_{\rm I}.
\end{align}
By Theorem \ref{the:eigenvaluesofJ1}, the eigenvalues of $J^{\rm I}$ are less than 1 hence different from the eigenvalues 1 of $J^{\rm II}$, further, $\theta_i\neq\theta_{i'},\,i\in{\cal I}_{\rm I},\,i'\in{\cal I}_{\rm III}$ by the assumption (\ref{eqn:eigenvaluesaredistinct}), so we have
\begin{align}
v_{i'i}=\dfrac{u^\Pi_{i'i}}{\theta_i-\theta_{i'}},\,i'\in{\cal I}_{\rm II}\cup{\cal I}_{\rm III},\,i\in{\cal I}_{\rm I},
\end{align}
which shows the existence and uniqueness of $V\in{\mathbb R}^{(m_2+m_3)\times m_1}$ that satisfies (\ref{eqn:equationofV}).

Now, define
\begin{align}
\tilde{\Pi}\equiv\begin{pmatrix}\Pi & O\\ V & I\end{pmatrix}\in{\mathbb R}^{m\times m},
\end{align}
then by noting (\ref{eqn:J1taikakuka}), (\ref{eqn:Jlambdataikakuka}), (\ref{eqn:equationofV}), we have%
\begin{align}
\tilde{\Pi}^{-1}J(\bm\lambda^\ast)\tilde{\Pi}=\begin{pmatrix}\Theta_1 & O \\ O & \Theta_2\end{pmatrix},
\end{align}
which proves the lemma.\hfill$\blacksquare$
\section{Proof of Theorem \ref{the:exponentialconvergence}}
\label{sec:exponentialconvergence}
\noindent Consider the line segment with the start point $\bm\lambda^\ast$ and the end point $\bm\lambda^N$, i.e.,
\begin{align}
\bm\lambda(t)\equiv(1-t)\bm\lambda^\ast+t\bm\lambda^N,\,0\leq t\leq 1.\label{eqn:senbunlstarlN}
\end{align}
The components of (\ref{eqn:senbunlstarlN}) are $\lambda_i(t)=(1-t)\lambda^\ast_i+t\lambda^N_i,\,i=1,\ldots,m$. Let us define
\begin{align}
f(t)\equiv F(\bm\lambda(t))\in\Delta({\cal X})
\end{align}
and write its components as $f(t)=(f_1(t),\ldots,f_m(t))$. We have
\begin{align}
\ds\frac{df_i(t)}{dt}
&=\ds\sum_{i'=1}^m\ds\frac{d\lambda_{i'}(t)}{dt}\left.\ds\frac{\partial F_i}{\partial\lambda_{i'}}\right|_{\bm\lambda=\bm\lambda(t)}\\
&=\ds\sum_{i'=1}^m(\lambda^N_{i'}-\lambda^\ast_{i'})\left.\ds\frac{\partial F_i}{\partial\lambda_{i'}}\right|_{\bm\lambda=\bm\lambda(t)}\\
&=\left((\bm\lambda^N-\bm\lambda^\ast)J(\bm\lambda(t))\right)_i,\,i=1,\ldots,m,
\end{align}
thus
\begin{align}
\ds\frac{df(t)}{dt}=(\bm\lambda^N-\bm\lambda^\ast)J(\bm\lambda(t)).\label{eqn:dft}
\end{align}

Now, by the relation between the matrix norm and the maximum eigenvalue \cite{hor},\,p.347, for $\epsilon\equiv\theta-\theta_{\rm max}>0$ there exists a vector norm $\|\cdot\|'$ in $\mathbb R^m$ whose associated matrix norm $\|\cdot\|'$ satisfies
\begin{align}
\theta_{\rm max}\leq\|J(\bm\lambda^\ast)\|'<\theta_{\rm max}+\epsilon.
\end{align}
(Note that $'$ does not denote the derivative.) By the continuity of norm, for any $\epsilon_1$ with $0<\epsilon_1<\theta_{\rm max}+\epsilon-\|J(\bm\lambda^\ast)\|'$ there exists $\delta'>0$ such that if $\|\bm\lambda-\bm\lambda^\ast\|'<\delta'$ then $\left|\|J(\bm\lambda)\|'-\|J(\bm\lambda^\ast)\|'\ \right|<\epsilon_1$, especially, $\|J(\bm\lambda)\|'<\|J(\bm\lambda^\ast)\|'+\epsilon_1$. Thus,
\begin{align}
\|J(\bm\lambda)\|'&<\|J(\bm\lambda^\ast)\|'+\theta_{\rm max}+\epsilon-\|J(\bm\lambda^\ast)\|'\\
&=\theta<1.\label{eqn:Jlambdalesstheta}
\end{align}
By the mean value theorem, there exists $t^N\in[0,1]$ which satisfies 
\begin{align}
\|\bm\lambda^{N+1}-\bm\lambda^\ast\|'&=\|F(\bm\lambda^N)-F(\bm\lambda^\ast)\|'\\
&=\|f(1)-f(0)\|'\\
&\leq\left\|\left.\dfrac{df(t)}{dt}\right|_{t=t^N}\right\|'\,(1-0)\\
&=\|(\bm\lambda^N-\bm\lambda^\ast)J(\bm\lambda(t^N))\|'\ \ ({\rm by}\ (\ref{eqn:dft}))\\
&\leq\|\bm\lambda^N-\bm\lambda^\ast\|'\,\|J(\bm\lambda(t^N))\|'.\label{eqn:lessnormrecursion}
\end{align}

Here, if $\|\bm\lambda^N-\bm\lambda^\ast\|'<\delta'$ we have $\|J(\bm\lambda^N)\|'<\theta<1$ by (\ref{eqn:Jlambdalesstheta}), so $\|\bm\lambda^{N+1}-\bm\lambda^\ast\|'<\delta'$ by (\ref{eqn:lessnormrecursion}). Thus, by induction, if the initial distribution $\bm\lambda^0$ satisfies $\|\bm\lambda^0-\bm\lambda^\ast\|'<\delta'$, then $\|\bm\lambda^N-\bm\lambda^\ast\|'<\delta'$ for all $N$, and so $\|J(\bm\lambda^N)\|'<\theta<1$ by (\ref{eqn:Jlambdalesstheta}).

Therefore by (\ref{eqn:Jlambdalesstheta}),\,(\ref{eqn:lessnormrecursion}), $\|\bm\lambda^{N+1}-\bm\lambda^\ast\|'<\theta\|\bm\lambda^N-\bm\lambda^\ast\|'<\ldots<\theta^{N+1}\|\bm\lambda^0-\bm\lambda^\ast\|'$, so we have
\begin{align}
\|\bm\lambda^N-\bm\lambda^\ast\|'<(\theta)^N\|\bm\lambda^0-\bm\lambda^\ast\|',\ N=0,1,\ldots.\label{eqn:exponentialdelay}
\end{align}

By the equivalence of norms in the finite dimensional vector space \cite{rob}, we can replace the norm from $\|\cdot\|'$ to the Euclidean norm $\|\cdot\|$ to have
\begin{align}
\|\bm\lambda^N-\bm\lambda^\ast\|\leq K(\theta)^N,\,K>0,\,N=0,1,\ldots.
\end{align}
\hfill$\blacksquare$

\section{Proof of Theorem \ref{the:Hessecomponents} (Calculation of Hessian matrix $H_i(\bm\lambda^\ast)$)}
\label{sec:Hessiancomponents}
We will calculate the Hessian matrix $H_i(\bm\lambda^\ast)$ of $F_i(\bm\lambda)$ at $\bm\lambda=\bm\lambda^\ast$, i.e., $H_i(\bm\lambda^\ast)=(\partial^2F_i/\partial\lambda_{i'}\partial\lambda_{i''}|_{\bm\lambda=\bm\lambda^\ast})$.

Differentiating the both sides of (\ref{eqn:dFi}) with respect to $\lambda_{i''}$, we have

\begin{align}
\ds\frac{\partial^2F_i}{\partial\lambda_{i'}\partial\lambda_{i''}}&\underline{\ds\sum_{k=1}^m\lambda_ke^{D_k}}_{\,\star1}+\underline{\ds\frac{\partial F_i}{\partial\lambda_{i'}}}_{\,\star2}\,\underline{\ds\frac{\partial}{\partial\lambda_{i''}}\ds\sum_{k=1}^m\lambda_ke^{D_k}}_{\,\star3}+\ds\frac{\partial F_i}{\partial\lambda_{i''}}\ds\frac{\partial}{\partial\lambda_{i'}}\ds\sum_{k=1}^m\lambda_ke^{D_k}\nonumber\\
&\ \ \ +F_i\underline{\ds\frac{\partial^2}{\partial\lambda_{i'}\partial\lambda_{i''}}\ds\sum_{k=1}^m\lambda_ke^{D_k}}_{\,\star4}\nonumber\\
&=\delta_{ii'}e^{D_i}\ds\frac{\partial D_i}{\partial\lambda_{i''}}+\delta_{ii''}e^{D_i}\ds\frac{\partial D_i}{\partial\lambda_{i'}}+\lambda_ie^{D_i}\ds\frac{\partial D_i}{\partial\lambda_{i''}}\ds\frac{\partial D_i}{\partial\lambda_{i'}}
+\lambda_ie^{D_i}\ds\frac{\partial^2D_i}{\partial\lambda_{i'}\partial\lambda_{i''}}.\label{eqn:ddfi}\end{align}
In the left hand side of (\ref{eqn:ddfi}), the part $\star1$ is evaluated at $\bm\lambda=\bm\lambda^\ast$ by (\ref{eqn:lem3-1}). The part $\star2$ is the component of the Jacobian matrix which is evaluated by (\ref{eqn:theorem1-0}). The part $\star3$ is evaluated by (\ref{eqn:lem3-3}). The part $\star4$ is evaluated as follows. We have
\begin{align}
\ds\frac{\partial^2}{\partial\lambda_{i'}\partial\lambda_{i''}}\ds\sum_{k=1}^m\lambda_ke^{D_k}
&=\ds\frac{\partial}{\partial\lambda_{i''}}\left(e^{D_{i'}}+\ds\sum_{k=1}^m\lambda_ke^{D_k}\ds\frac{\partial D_k}{\partial\lambda_{i'}}\right)\\
&=e^{D_{i'}}\ds\frac{\partial D_{i'}}{\partial\lambda_{i''}}+\ds\sum_{k=1}^m\left(\delta_{ki''}e^{D_k}\ds\frac{\partial D_k}{\partial\lambda_{i'}}+\lambda_ke^{D_k}\ds\frac{\partial D_k}{\partial\lambda_{i''}}\ds\frac{\partial D_k}{\partial\lambda_{i'}}\right.\left.+\lambda_ke^{D_k}\ds\frac{\partial^2D_k}{\partial\lambda_{i'}\partial\lambda_{i''}}\right)\\
&=e^{D_{i'}}\ds\frac{\partial D_i'}{\partial\lambda_{i''}}+e^{D_{i''}}\ds\frac{\partial D_{i''}}{\partial\lambda_{i'}}+\ds\sum_{k=1}^m\lambda_ke^{D_k}\ds\frac{\partial D_k}{\partial\lambda_{i'}}\ds\frac{\partial D_k}{\partial\lambda_{i''}}+\underline{\ds\sum_{k=1}^m\lambda_ke^{D_k}\ds\frac{\partial^2D_k}{\partial\lambda_{i'}\partial\lambda_{i''}}}_{\,\star5},
\end{align}
and the part $\star5$ becomes, at $\bm\lambda=\bm\lambda^\ast$,
\begin{align}
\left.\ds\sum_{k=1}^m\lambda_ke^{D_k}\ds\frac{\partial^2D_k}{\partial\lambda_{i'}\partial\lambda_{i''}}\right|_{\bm\lambda=\bm\lambda^\ast}
&=e^C\ds\sum_{k=1}^{m_1}\lambda_k^\ast\ds\sum_{j=1}^n\ds\frac{P_j^kP_j^{i'}P_j^{i''}}{\left(Q_j^\ast\right)^2}\\
&=e^C\ds\sum_{j=1}^n\ds\frac{P_j^{i'}P_j^{i''}}{Q_j^\ast}\ds\sum_{k=1}^{m_1}\ds\frac{\lambda_k^\ast P_j^k}{Q_j^\ast}\\
&=-e^CD_{i',i''}^\ast.
\end{align}
Therefore, the part $\star4$ becomes, at $\bm\lambda=\bm\lambda^\ast$,
\begin{align}
\left.\ds\frac{\partial^2}{\partial\lambda_{i'}\partial\lambda_{i''}}\ds\sum_{k=1}^m\lambda_ke^{D_k}\right|_{\bm\lambda=\bm\lambda^\ast}
&=e^{D_{i'}^\ast}D_{i',i''}^\ast+e^{D_{i''}^\ast}D_{i',i''}^\ast+e^C\ds\sum_{k=1}^{m_1}\lambda_k^\ast D_{k,i'}^\ast D_{k,i''}^\ast-e^CD_{i',i''}^\ast.
\end{align}
Define $D_{i,i',i''}^\ast\equiv\partial^2D_i/\partial\lambda_{i'}\partial\lambda_{i''}|_{\bm\lambda=\bm\lambda^\ast}$ and $E_{i',i''}\equiv\sum_{k=1}^{m_1}\lambda_k^\ast D_{k,i'}^\ast D_{k,i''}^\ast$, then based on the above calculation, we have
\begin{align}
\left.\dfrac{\partial^2F_i}{\partial\lambda_{i'}\partial\lambda_{i''}}\right|_{\bm\lambda=\bm\lambda^\ast}e^C&+\left\{e^{D^\ast_i-C}\left(\delta_{ii'}+\lambda^\ast_iD^\ast_{i,i'}\right)+\lambda^\ast_i\left(1-e^{D^\ast_{i'}-C}\right)\right\}\left(e^{D^\ast_{i''}}-e^C\right)\nonumber\\
&+\left\{e^{D^\ast_i-C}\left(\delta_{ii''}+\lambda^\ast_iD^\ast_{i,i''}\right)+\lambda^\ast_i\left(1-e^{D^\ast_{i''}-C}\right)\right\}\left(e^{D^\ast_{i'}}-e^C\right)\nonumber\\
&+F^\ast_i\left(e^{D^\ast_{i'}}D^\ast_{i',i''}+e^{D^\ast_{i''}}D^\ast_{i',i''}+e^CE_{i',i''}-e^CD^\ast_{i',i''}\right)\nonumber\\
&\hspace{-10mm}=\delta_{ii'}e^{D^\ast_i}D^\ast_{i,i''}+\delta_{ii''}e^{D^\ast_i}D^\ast_{i,i'}+\lambda^\ast_ie^{D^\ast_i}D^\ast_{i,i'}D^\ast_{i,i''}+\lambda^\ast_ie^{D^\ast_i}D^\ast_{i,i',i''}
\end{align}
By arranging this, we obtain, using (\ref{eqn:lem3-4}) of Lemma \ref{lem:shoryou},
\begin{align}
\left.\dfrac{\partial^2F_i}{\partial\lambda_{i'}\partial\lambda_{i''}}\right|_{\bm\lambda=\bm\lambda^\ast}&=e^{D^\ast_i-C}\Big\{\delta_{ii'}D^\ast_{i,i''}+\delta_{ii''}D^\ast_{i,i'}+\lambda^\ast_i\left(D^\ast_{i,i'}D^\ast_{i,i''}+D^\ast_{i,i',i''}\right)\nonumber\\
&\ \ \ +\left(\delta_{ii'}+\lambda^\ast_iD^\ast_{i,i'}\right)\left(1-e^{D^\ast_{i''}-C}\right)+\left(\delta_{ii''}+\lambda^\ast_iD^\ast_{i,i''}\right)\left(1-e^{D^\ast_{i'}-C}\right)\Big\}\nonumber\\
&\ \ \ +2\lambda^\ast_i\left(1-e^{D^\ast_{i'}-C}\right)\left(1-e^{D^\ast_{i''}-C}\right)-\lambda^\ast_i\Big(e^{D^\ast_{i'}-C}D^\ast_{i',i''}+e^{D^\ast_{i''}-C}D^\ast_{i',i''}\nonumber\\
&\ \ \ +E_{i',i''}-D^\ast_{i',i''}\Big)
\end{align}
\hfill$\blacksquare$
\section{Proof of Step2}
\label{sec:proofofstep2}
For $i=m'+1,\ldots,m$, let $H_{i,i'i''}$ be the $(i',i'')$ component of the Hessian matrix $H_i(\bm\lambda^\ast)$, then by Theorem \ref{the:Hessecomponents},
\begin{align}
H_{i,i'i''}&=\left.\dfrac{\partial^2F_i}{\partial\lambda_{i'}\partial\lambda_{i''}}\right|_{\bm\lambda=\bm\lambda^\ast}\\
&=\delta_{ii''}\left(1-e^{D^\ast_{i'}-C}+D^\ast_{i,i'}\right)+\delta_{ii'}\left(1-e^{D^\ast_{i''}-C}+D^\ast_{i,i''}\right),\label{eqn:Hessiancomponentagain}\\
&\hspace{10mm}i=m'+1,\ldots,m,\,i',i''=1,\ldots,m.\nonumber
\end{align}
Here, for the simplicity of symbols, define
\begin{align}
S_{ii'}\equiv1-e^{D^\ast_{i'}-C}+D^\ast_{i,i'},\,i'=1,\ldots,m,
\end{align}
then (\ref{eqn:Hessiancomponentagain}) becomes
\begin{align}
H_{i,i'i''}=\delta_{ii''}S_{ii'}+\delta_{ii'}S_{ii''}.\label{eqn:Hessiancomponent2again}
\end{align}
We have $S_{ii'}=D^\ast_{i,i'}$ for $i'=m'+1,\ldots,m$, then by (\ref{eqn:Hessiancomponent2again}), the Hessian matrix $H_i(\bm\lambda^\ast)$ is
\begin{align}
H_i(\bm\lambda^\ast)
%
%
%
%
%
%
%
%
&=\begin{pmatrix}
\begin{array}{c:l}
O & 
\begin{matrix}
\hspace{5.5mm}0\hspace{5mm}&\ldots&\hspace{4.5mm}0&\hspace{6mm}S_{i1}&\hspace{7mm}0&\hspace{3.5mm}\ldots&\hspace{3.5mm}0\\
\vdots&&\hspace{3mm}\vdots&\hspace{6mm}\vdots&\hspace{6.5mm}\vdots&&\hspace{3mm}\vdots\\
\hspace{5.5mm}0\hspace{5mm}&\ldots&\hspace{4.5mm}0&\hspace{6mm}S_{im'}&\hspace{7.5mm}0&\hspace{3.5mm}\ldots&\hspace{3.5mm}0\\[1mm]
\end{matrix}\\
\hdashline\\[-3mm]
\begin{matrix}0&\ldots&0\\\vdots&&\vdots\\0&\ldots&0\\S_{i1}&\ldots&S_{im'}\\0&\ldots&0\\\vdots&&\vdots\\0&\ldots&0\end{matrix} & 
\begin{matrix}0&\ldots&0&D^\ast_{i,m'+1}&0&\ldots&0\\
\vdots&&\vdots&\vdots&\vdots&&\vdots\\
0&\ldots&0&D^\ast_{i,i-1}&0&\ldots&0\\
D^\ast_{i,m'+1}&\ldots&D^\ast_{i,i-1}&2D^\ast_{i,i}&D^\ast_{i,i+1}&\ldots&D^\ast_{i,m}\\
0&\ldots&0&D^\ast_{i,i+1}&0&\ldots&0\\
\vdots&&\vdots&\vdots&\vdots&&\vdots\\
0&\ldots&0&D^\ast_{i,m}&0&\ldots&0
\end{matrix}
\end{array}
\end{pmatrix}\\
&\equiv\begin{pmatrix}O&H^1_i\\{^t}H^1_i&H^2_i\end{pmatrix},
\end{align}
where $O\in\mathbb R^{m'\times m'}$, $H^1_i\in\mathbb R^{m'\times m_2}$, $H^2_i\in\mathbb R^{m_2\times m_2}$ and
\begin{align}
H^1_i&\equiv\left(\delta_{ii''}S_{ii'}\right),\,i'=1,\ldots,m',i''=m'+1,\ldots,m,\\
{^t}H^1_i&\equiv\left(\delta_{ii'}S_{ii''}\right),\,i'=m'+1,\ldots,m,i''=1,\ldots,m',\\
H^2_i&\equiv\left(\delta_{ii''}D^\ast_{ii'}+\delta_{ii'}D^\ast_{ii''}\right),\,i',i''=m'+1,\ldots,m.\label{eqn:H2icomponent}
\end{align}
Therefore, by (\ref{eqn:bmmuN{I,III}A1+bmmuN{II}A2=bm0}),
\begin{align}
\dfrac{1}{2}\bar{\bm\mu}^NH_i(\bm\lambda^\ast)\,{^t}\bar{\bm\mu}^N&=\dfrac{1}{2}\left(\bar{\bm\mu}^N_{\rm I,III},\bar{\bm\mu}^N_{\rm II}\right)\begin{pmatrix}O&H_i^1\\{^t}\!H_i^1&H^2_i\end{pmatrix}\begin{pmatrix}\,{^t}\bar{\bm\mu}^N_{\rm I,III}\\{^t}\bar{\bm\mu}^N_{\rm II}\end{pmatrix}\\
&=\dfrac{1}{2}\left(-\bar{\bm\mu}^N_{\rm II}A_2A_1^{-1},\bar{\bm\mu}^N_{\rm II}\right)\begin{pmatrix}O&H_i^1\\{^t}\!H_i^1&H^2_i\end{pmatrix}\begin{pmatrix}-{^t}\!A_1^{-1}\,{^t}A_2\,{^t}\bar{\bm\mu}^N_{\rm II}\\{^t}\bar{\bm\mu}^N_{\rm II}\end{pmatrix}\\
%
%
%
%
%
%
&=\dfrac{1}{2}\bar{\bm\mu}^N_{\rm II}\left(-A_2A_1^{-1}H_i^1-{^t}\!H_i^1\,{^t}\!A_1^{-1}\,{^t}\!A_2+H_i^2\right){^t}\bar{\bm\mu}^N_{\rm II}.\label{eqn:muNHitmuN}
\end{align}
Now, define
\begin{align}
G_i\equiv-A_2A_1^{-1}H_i^1\in\mathbb R^{m_2\times m_2},\label{eqn:Gidefinition}
\end{align}
and $A_1^{-1}\equiv\left(\zeta_{i'i''}\right)$. Further, let $G_{i,i'i''}$ be the $(i',i'')$ component of $G_i$, then by (\ref{eqn:Gidefinition}), we have
\begin{align}
G_{i,i'i''}&=-\ds\sum_{k,k'=1}^{m'}a_{i'k}\zeta_{kk'}H^1_{i,k'i''}\\
&=-\ds\sum_{k,k'=1}^{m'}a_{i'k}\zeta_{kk'}\delta_{ii''}S_{ik'}\\
&=-\delta_{ii''}\ds\sum_{k,k'=1}^{m'}a_{i'k}\zeta_{kk'}S_{ik'},\,i',i''=m'+1,\ldots,m.\label{eqn:Gicomponent}
\end{align}
Define
\begin{align}
T_{ii'}\equiv-\ds\sum_{k,k'=1}^{m'}a_{i'k}\zeta_{kk'}S_{ik'},\,i'=m'+1,\ldots,m,
\end{align}
then (\ref{eqn:Gicomponent}) becomes
\begin{align}
G_{i,i'i''}=\delta_{ii''}T_{ii'}.
\end{align}
Thus, we have
\begin{align}
G_i=\left(\delta_{ii''}T_{ii'}\right),\ {^t}G_i=\left(\delta_{ii'}T_{ii''}\right),\,i',i''=m'+1,\ldots,m.\label{eqn:GitGi}
\end{align}
hence by (\ref{eqn:muNHitmuN}),
\begin{align}
\dfrac{1}{2}\bar{\bm\mu}^NH_i(\bm\lambda^\ast)\,{^t}\bar{\bm\mu}^N=\dfrac{1}{2}\bar{\bm\mu}^N_{\rm II}\left(G_i+{^t}G_i+H^2_i\right)\,{^t}\bar{\bm\mu}^N_{\rm II}.\label{eqn:muNHitmuN2}
\end{align}
Define
\begin{align}
\hat{H}_i\equiv G_i+{^t}G_i+H^2_i,\label{eqn:hatH}
\end{align}
and
\begin{align}
r_{ii'}\equiv T_{ii'}+D^\ast_{i,i'},\label{eqn:hiidefinition}
\end{align}
and let $\hat{H}_{i,i'i''}$ be the $(i',i'')$ component of $\hat{H}_i$. Then, by (\ref{eqn:hatH}), (\ref{eqn:GitGi}), (\ref{eqn:H2icomponent}), (\ref{eqn:hiidefinition}), we have
\begin{align}
\hat{H}_{i,i'i''}
&=\delta_{ii''}T_{ii'}+\delta_{ii'}T_{ii''}+\left(\delta_{ii''}D^\ast_{i,i'}+\delta_{ii'}D^\ast_{i,i''}\right)\\
&=\delta_{ii''}\left(T_{ii'}+D^\ast_{i,i'}\right)+\delta_{ii'}\left(T_{ii''}+D^\ast_{i,i''}\right)\\&=\delta_{ii''}r_{ii'}+\delta_{ii'}r_{ii''}.
\end{align}
Therefore, (\ref{eqn:muNHitmuN2}) becomes
\begin{align}
\dfrac{1}{2}\bar{\bm\mu}^NH_i(\bm\lambda^\ast)\,{^t}\bar{\bm\mu}^N
&=\dfrac{1}{2}\bar{\bm\mu}^N_{\rm II}\hat{H}_i\,{^t}\bar{\bm\mu}^N_{\rm II}\\
&=\dfrac{1}{2}\ds\sum_{i',i''=m'+1}^m\bar{\mu}^N_{i'}\hat{H}_{i,i'i''}\bar{\mu}^N_{i''}\\
&=\dfrac{1}{2}\ds\sum_{i',i''=m'+1}^m\bar{\mu}^N_{i'}\left(\delta_{ii''}r_{ii'}+\delta_{ii'}r_{ii''}\right)\bar{\mu}^N_{i''}\\
&=\dfrac{1}{2}\left(\ds\sum_{i'=m'+1}^m\bar{\mu}^N_{i'}r_{ii'}\bar{\mu}^N_i+\ds\sum_{i''=m'+1}^m\bar{\mu}^N_ir_{ii''}\bar{\mu}^N_{i''}\right)\\
&=\bar{\mu}^N_i\ds\sum_{i'=m'+1}^mr_{ii'}\bar{\mu}^N_{i'}.\label{eqn:zenkashiki2ndterm}
\end{align}
Summarizing above, the recurrence formula satisfied by $\bar{\mu}^N_i,\,i=m'+1,\ldots,m$, is
\begin{align}
\bar{\mu}^{N+1}_i=\bar{\mu}^N_i+\bar{\mu}^N_i\ds\sum_{i'=m'+1}^mr_{ii'}\bar{\mu}^N_{i'},\,i=m'+1,\ldots,m.\label{eqn:zenkashiki}
\end{align}
The step 2 is achieved by (\ref{eqn:zenkashiki}).\hfill$\blacksquare$
\section{Proof of Theorem \ref{the:1/Norderconvergenceinitialfree}}
\label{sec:proofof1/Norderconvergenceinitialfree}
We can prove Theorem \ref{the:1/Norderconvergenceinitialfree} for any $m_2\geq3$, but because the symbols become complicated, we will give a proof in the case of $m_2=3$. The proof for $m_2=3$ does not lose the generality, hence the extension to $m_2\geq3$ is easy. The theorem we should prove is the following.

Consider the sequence $\{\xi_i^N\},\,i=1,2,3,\,N=0,1,\ldots$ defined by
\begin{align}
\label{eqn:3hensucanonicalform}
\xi_i^{N+1}&=\xi_i^N-\xi_i^N\sum_{i'=1}^3q_{ii'}\xi_{i'}^N,\,i=1,2,3,\\
0&<\xi_i^0\leq1/2,\,i=1,2,3,\label{eqn:0<xii0leq1/2}\\
{\rm where}\ {\bm q}_i&\equiv(q_{i1},q_{i2},q_{i3})\ {\rm\ is\ a\ probability\ vector}.
\end{align}
Further, we assume the diagonally dominant condition (\ref{eqn:taikakuyuui2}), i.e.,
\begin{align}
q_{ii}>\sum_{i'=1,i'\neq i}^3q_{ii'},\,i=1,2,3.\label{eqn:taikakuyuuiagain}
\end{align}
and the conjecture (\ref{eqn:daishoukankei}), i.e.,
\begin{align}
\xi_1^N\geq\xi_2^N\geq\xi_3^N,\,N\geq N_0,\label{eqn:daishoukankeiagain}
\end{align}
Then, under the assumptions (\ref{eqn:taikakuyuuiagain}) and (\ref{eqn:daishoukankeiagain}), we will prove
\begin{align}
\lim_{N\to\infty}N\xi_i^N=1,\,i=1,2,3.
\end{align}
\begin{lemma}
\label{lem:0<xiiN<=1/2}
$0<\xi_i^N\leq1/2,\,i=1,2,3$, holds for $N=0,1,\ldots$.
\end{lemma}
{\bf Proof:} We prove by mathematical induction. For $N=0$, the assertion holds by (\ref{eqn:0<xii0leq1/2}). Assuming that the assertion holds for $N$, by noting $1/2\leq1-\sum_{i'=1}q_{ii'}\xi_{i'}^N<1$, we have $0<\xi_i^{N+1}\leq1/2$.\hfill$\blacksquare$
\begin{lemma}
\label{lem:nu_i^Nmonotonedecreasing}
The sequence $\{\xi_i^N\},\,N=0,1,\ldots$ is strictly decreasing.
\end{lemma}
\noindent{\bf Proof:} Because $\xi_i^N-\xi_i^{N+1}=\xi_i^N\sum_{i'=1}^3q_{ii'}\xi_{i'}^N>0$ holds by Lemma \ref{lem:0<xiiN<=1/2}.\hfill$\blacksquare$
\begin{lemma}
\label{lem:nu_i^Nconvergesto0}
$\lim_{N\to\infty}\xi_i^N=0,\,i=1,2,3$.
\end{lemma}
\noindent{\bf Proof:} $\xi_i^\infty\equiv\lim_{N\to\infty}\xi_i^N\geq0$ exists by Lemmas \ref{lem:0<xiiN<=1/2} and \ref{lem:nu_i^Nmonotonedecreasing}. Then, $\xi_i^\infty=\xi_i^\infty-\xi_i^\infty\sum_{i'=1}^3q_{ii'}\xi_{i'}^\infty$ holds by (\ref{eqn:3hensucanonicalform}), hence we have $\xi_i^\infty=0$.\hfill$\blacksquare$
\begin{lemma}
$\liminf_{N\to\infty}N\xi_1^N\geq1$.\label{lem:liminf_Nnu_1^Ngeq_1}
\end{lemma}
\noindent{\bf Proof:} By (\ref{eqn:daishoukankeiagain}), for $N\geq N_0$,
\begin{align}
\xi_1^{N+1}&=\xi_1^N-\xi_1^N\sum_{i'=1}^3q_{1i'}\xi_{i'}^N\\
&\geq\xi_1^N-\xi_1^N\sum_{i'=1}^3q_{1i'}\xi_1^N\\
&=\xi_1^N-\left(\xi_1^N\right)^2.\label{eqn:nu_1^N+1geqnu_1^N-(nu_1^N)^2}
\end{align}
Now, we define a sequence $\{\hat\xi_1^N\},\,N=0,1,\ldots$ by the recurrence formula
\begin{align}
\hat\xi_1^0&=\xi_1^{N_0},\label{eqn:hatnu_1shokichi}\\
\hat\xi_1^{N+1}&=\hat\xi_1^N-\left(\hat\xi_1^N\right)^2,\,N=0,1,\ldots.
\end{align}
Then, we will prove
\begin{align}
\xi_1^{N+N_0}\geq\hat\xi_1^N,\,N=0,1,\ldots\label{eqn:nu1Ngeqhatnu1N}
\end{align}
by mathematical induction. For $N=0$, (\ref{eqn:nu1Ngeqhatnu1N}) holds by the assumption (\ref{eqn:hatnu_1shokichi}). Assume that (\ref{eqn:nu1Ngeqhatnu1N}) holds for $N$. Because the function $f(\xi)=\xi-\xi^2$ is monotonically increasing in $0<\xi\leq1/2$, by (\ref{eqn:nu_1^N+1geqnu_1^N-(nu_1^N)^2}), we have $\xi_1^{N+N_0+1}\geq\xi_1^{N+N_0}-(\xi_1^{N+N_0})^2\geq\hat\xi_1^N-(\hat\xi_1^N)^2=\hat\xi_1^{N+1}$, thus, (\ref{eqn:nu1Ngeqhatnu1N}) holds also for $N+1$. Therefore, $\liminf_{N\to\infty}N\xi_1^N\geq\liminf_{N\to\infty}N\hat\xi_1^N=\lim_{N\to\infty}N\hat\xi_1^N=1$, where the last equality is due to Lemma \ref{lem:onevariablereccurence}.\hfill$\blacksquare$
\begin{lemma}
\label{lem:limsupNnu3Nleq1}
$\limsup_{N\to\infty}N\xi_3^N\leq1$
\end{lemma}
\noindent{\bf Proof:} By (\ref{eqn:daishoukankeiagain}), for $N\geq N_0$, we have $\xi_3^{N+1}=\xi_3^N-\xi_3^N\sum_{i'=1}^3q_{3i'}\xi_{i'}^N\leq\xi_3^N-\xi_3^N\sum_{i'=1}^3q_{3i'}\xi_3^n=\xi_3^N-\left(\xi_3^N\right)^2$. Now, we define a sequence $\{\hat\xi_3^N\}$ by
\begin{align}
\hat\xi_3^0&=\xi_3^{N_0},\\
\hat\xi_3^{N+1}&=\hat\xi_3^N-\left(\hat\xi_3^N\right)^2,\,N=0,1,\ldots.
\end{align}
Then, we can prove $\xi_3^{N+N_0}\leq\hat\xi_3^N,\,N=0,1,\ldots$ in a similar way as the proof of Lemma \ref{lem:liminf_Nnu_1^Ngeq_1}. Therefore, we have $\limsup_{N\to\infty}N\xi_3^N\leq\limsup_{N\to\infty}N\hat\xi_3^N=\lim_{N\to\infty}N\hat\xi_3^N=1$, where the last equality is due to Lemma \ref{lem:onevariablereccurence}.\hfill$\blacksquare$

\begin{lemma}
\label{lem:xi_1^N-xi_3^NgeqK(nu_1^N-nu_3^N)}
Let $\tau_1^N\equiv\sum_{i'=1}^3q_{1i'}\xi_{i'}^N,\,\tau_3^N\equiv\sum_{i'=1}^3q_{3i'}\xi_{i'}^N$, then there exists a constant $K>0$ with
\begin{align}
\tau_1^N-\tau_3^N\geq K(\xi_1^N-\xi_3^N),\,N\geq N_0.
\end{align}
\end{lemma}
\noindent{\bf Proof:} We have
\begin{align}
\tau_1^N-\tau_3^N&=\sum_{i'=1}^3q_{1i'}\xi_{i'}^N-\sum_{i'=1}^3q_{3i'}\xi_{i'}^N\\
&\geq q_{11}\xi_1^N+q_{12}\xi_3^N+q_{13}\xi_3^N-q_{31}\xi_1^n-q_{32}\xi_1^N-q_{33}\xi_3^N\\
&=(q_{11}-q_{31}-q_{32})\xi_1^N-(q_{33}-q_{12}-q_{13})\xi_3^N,\,N\geq N_0,
\end{align}
and $q_{11}-q_{31}-q_{32}=q_{33}-q_{12}-q_{13}$ holds by $q_{11}+q_{12}+q_{13}=q_{31}+q_{32}+q_{33}=1$. Defining $K\equiv q_{11}-q_{31}-q_{32}=q_{33}-q_{12}-q_{13}$, we have $\tau_1^N-\tau_3^N\geq K(\xi_1^N-\xi_3^N)$. By the assumption (\ref{eqn:taikakuyuuiagain}), we have $q_{11}>1/2,q_{12}+q_{13}<1/2,\,q_{33}>1/2,q_{31}+q_{32}<1/2$, thus $K>0$. \hfill$\blacksquare$
\begin{lemma}
\label{lem:sum_of_nu_1^Nminus_nu_3N_is_finite}
$\sum_{N=0}^\infty(\xi_1^N-\xi_3^N)<\infty$
\end{lemma}
\noindent{\bf Proof:} We have
\begin{align}
\dfrac{\xi_1^N}{\xi_3^N}-\dfrac{\xi_1^{N+1}}{\xi_3^{N+1}}&=\dfrac{\xi_1^N}{\xi_3^N}-\dfrac{\xi_1^N(1-\tau_1^N)}{\xi_3^N(1-\tau_3^N)}\\[1mm]
&=\dfrac{\xi_1^N}{\xi_3^N}\cdot\dfrac{\tau_1^N-\tau_3^N}{1-\tau_3^N}\label{eqn:nu_1^N/nu_3^N-nu_1^{N+1}/nu_3^{N+1}}.
\end{align}
By (\ref{eqn:daishoukankeiagain}), we have $\xi_1^N/\xi_3^N\geq1,\,N\geq N_0$, and by Lemma \ref{lem:xi_1^N-xi_3^NgeqK(nu_1^N-nu_3^N)}, $\tau_1^N-\tau_3^N\geq K(\xi_1^N-\xi_3^N),\,K>0,\,N\geq N_0$. Further, by Lemma \ref{lem:0<xiiN<=1/2}, we have $0<\tau_3^N\leq1/2$, thus $1/(1-\tau_3^N)>1$. Therefore, by (\ref{eqn:nu_1^N/nu_3^N-nu_1^{N+1}/nu_3^{N+1}}),
\begin{align}
\dfrac{\xi_1^{N_0}}{\xi_3^{N_0}}-\dfrac{\xi_1^{N+1}}{\xi_3^{N+1}}>K\sum_{l=N_0}^N(\xi_1^l-\xi_3^l),\,N\geq N_0,\label{eqn:dfrac-dfrac}
\end{align}
hence $\sum_{l=N_0}^N(\xi_1^l-\xi_3^l)$ has an upper bound $K^{-1}(\xi_1^{N_0}/\xi_3^{N_0})$ which is unrelated with $N$, then the sum is convergent.\hfill$\blacksquare$

\bigskip

Here, we cite the following theorem.

\noindent{\bf Theorem B} (\cite{bro},\,p.31) Let $\{a_N\}_{N=0,1,\ldots,}$ be a decreasing positive sequence. If $\sum_{N=0}^\infty a_N$ is convergent, then $Na_N\to0,\ N\to\infty$.

\medskip

\begin{lemma}
\label{lem:nu_1^N-nu_3^Nismonotonedecreasing}
The sequence $\{\xi_1^N-\xi_3^N\}$ is decreasing for $N\geq N_0$.
\end{lemma}
\noindent{\bf Proof:} We have
\begin{align}
\xi_1^{N+1}-\xi_3^{N+1}&=\xi_1^N-\xi_1^N\sum_{i'=1}^3q_{1i'}\xi_{i'}^N-\xi_3^N+\xi_3^N\sum_{i'=1}^3q_{3i'}\xi_{i'}^N\\
&\leq\xi_1^N-\xi_1^N\sum_{i'=1}^3q_{1i'}\xi_3^N-\xi_3^N+\xi_3^N\sum_{i'=1}^3q_{3i'}\xi_1^N\\
&=\xi_1^N-\xi_1^N\xi_3^N-\xi_3^N+\xi_3^N\xi_1^N\\
&=\xi_1^N-\xi_3^N,\ N\geq N_0,
\end{align}
hence the assertion holds.\hfill$\blacksquare$

\begin{lemma}
\label{lem:limN(nu1N-nu3N)}
$\lim_{N\to\infty}N(\xi_1^N-\xi_3^N)=0$.
\end{lemma}
\noindent{\bf Proof:} The assertion holds by Lemmas \ref{lem:sum_of_nu_1^Nminus_nu_3N_is_finite}, \ref{lem:nu_1^N-nu_3^Nismonotonedecreasing} and Theorem B.\hfill$\blacksquare$

\bigskip

Summarizing above, we have
\begin{theorem}
\label{the:maintheoremagain}
$\lim_{N\to\infty}N\xi_i^N=1,\,i=1,2,3$.
\end{theorem}
\noindent{\bf Proof:} By Lemmas \ref{lem:limsupNnu3Nleq1} and \ref{lem:limN(nu1N-nu3N)}, we have
\begin{align}
\limsup_{N\to\infty}N\xi_1^N&=\limsup_{N\to\infty}\left(N\xi_1^N-N\xi_3^N+N\xi_3^N\right)\\
&\leq\limsup_{N\to\infty}N\left(\xi_1^N-\xi_3^N\right)+\limsup_{N\to\infty}N\xi_3^N\\
&\leq1,
\end{align}
thus, together with Lemma \ref{lem:liminf_Nnu_1^Ngeq_1}, we have
\begin{align}
\lim_{N\to\infty}N\xi_1^N=1.\label{eqn:limNtoinftyNxi1N=1}
\end{align}
Further, we have
\begin{align}
\lim_{N\to\infty}N\xi_3^N&=\lim_{N\to\infty}\left(N\xi_3^N-N\xi_1^N+N\xi_1^N\right)\\
&=-\lim_{N\to\infty}N(\xi_1^N-\xi_3^N)+\lim_{N\to\infty}N\xi_1^N\\
&=1.\label{eqn:limNtoinftyNxi3N=1}
\end{align}
Finally, by (\ref{eqn:limNtoinftyNxi1N=1}), (\ref{eqn:limNtoinftyNxi3N=1}) and the squeeze theorem, we have
\begin{align}
\ds\lim_{N\to\infty}N\xi_2^N=1.
\end{align}
\hfill$\blacksquare$
\section{Proof of Lemma \ref{lem:thetasec}}
\label{sec:proofofthetasec}
Let $0=\theta_1\leq\ldots\leq\theta_{m-1}\leq\theta_m<1$ be the eigenvalues of $J(\bm\lambda^\ast)$. We have $\theta_{\rm max}=\theta_m$, $\theta_{\rm sec}=\theta_{m-1}$. Let $\bm b_i,\,i=1,\ldots,m$, be the left eigenvector of $J(\bm\lambda^\ast)$ for $\theta_i,\,i=1,\ldots,m$. We have $\bm b_{\rm max}=\bm b_m$. Due to $(\ref{eqn:eigenvaluesaredistinct})$, we can assume that $\{\bm b_i\}_{i=1,\ldots,m}$ forms a basis of $\mathbb{R}^m$. Suppose $\bm\mu^N\in\bm b_{\rm max}^\perp$ for $N=0,1,\ldots$, then $\bm\mu^N$ is uniquely represented as
\begin{align}
\bm\mu^N=\ds\sum_{i=1}^{m-1}c^N_i\bm b_i,\,c^N_i\in{\mathbb R}\label{eqn:linearcombination}
\end{align}
in the $m-1$ dimensional subspace $\bm b_{\rm max}^\perp$. By (\ref{eqn:linearcombination}), we have
\begin{align}
\bm\mu^{N+1}&=\bm\mu^NJ(\bm\lambda^\ast)\\
&=\ds\sum_{i=1}^{m-1}c^N_i\bm b_iJ(\bm\lambda^\ast)\\
&=\ds\sum_{i=1}^{m-1}c^N_i\theta_i\bm b_i.\label{eqn:N+1coefficients}
\end{align}
Comparing the coefficients of $\bm\mu^{N+1}=\sum_{i=1}^{m-1}c^{N+1}_i\bm b_i$ and (\ref{eqn:N+1coefficients}), we have $c^{N+1}_i=\theta_ic^N_i=\ldots=\left(\theta_i\right)^{N+1}c^0_i,\,i=1,\ldots,m-1$, thus $\bm\mu^N=\sum_{i=1}^{m-1}\left(\theta_i\right)^Nc^0_i\bm b_i$. Therefore,
\begin{align}
\|\bm\mu^N\|&\leq\ds\sum_{i=1}^{m-1}\left(\theta_i\right)^N|c^0_i|\|\bm b_i\|\label{eqn:11}\\
&\leq K\left(\theta_{m-1}\right)^N\label{eqn:12}\\
&=K\left(\theta_{\rm sec}\right)^N,\ K>0.\label{eqn:13}
\end{align}
\hfill$\blacksquare$
\section{Proof of Lemma \ref{lem:migikoyuuvector}}
\label{sec:proofofmigikoyuuvector}
Suppose that ${^t}\bm b_{\rm max}$ is a right eigenvector for $\theta_{\rm max}$. For any $\bm\mu\in\bm b_{\rm max}^\perp$, $\bm\mu J(\bm\lambda^\ast){^t}\bm b_{\rm max}=\theta_{\rm max}\bm\mu{^t}\bm b_{\rm max}=0$ holds, thus we obtain $\bm\mu J(\bm\lambda^\ast)\in\bm b_{\rm max}^\perp$.

Conversely, suppose $\bm\mu J(\bm\lambda^\ast)\in\bm b_{\rm max}^\perp$ for any $\bm\mu\in\bm b_{\rm max}^\perp$. Our goal is to show $J(\bm\lambda^\ast){^t}\bm b_{\rm max}=\theta_{\rm max}{^t}\bm b_{\rm max}$, which is equivalent to 
\begin{align}
\bm\mu J(\bm\lambda^\ast){^t}\bm b_{\rm max}=\theta_{\rm max}\bm\mu{^t}\bm b_{\rm max}\ {\rm holds\ for\ any\ }\bm\mu.\label{eqn:migikoyuuchiequivalent}
\end{align}
We will prove (\ref{eqn:migikoyuuchiequivalent}). Since we can write $\bm\mu$ uniquely as $\bm\mu=K\bm b_{\rm max}+\tilde{\bm\mu}$ with $K\in{\mathbb R}$ and $\tilde{\bm\mu}\in\bm b_{\rm max}^\perp$, we have
\begin{align}
\bm\mu J(\bm\lambda^\ast){^t}\bm b_{\rm max}&=K\bm b_{\rm max} J(\bm\lambda^\ast){^t}\bm b_{\rm max}+\tilde{\bm\mu} J(\bm\lambda^\ast){^t}\bm b_{\rm max}\\
&=K\theta_{\rm max}\bm b_{\rm max}{^t}\bm b_{\rm max}+0\ \ ({\rm by\ the\ assumption})\\
&=\theta_{\rm max}K\bm b_{\rm max}{^t}\bm b_{\rm max}+\theta_{\rm max}\tilde{\bm\mu}{^t}\bm b_{\rm max}\ \ ({\text{\rm by}}\ \tilde{\bm\mu}\in\bm b_{\rm max}^\perp)\\
&=\theta_{\rm max}\bm\mu{^t}\bm b_{\rm max},
\end{align}
which proves (\ref{eqn:migikoyuuchiequivalent}).\hfill$\blacksquare$
\section{Proof of Theorem \ref{the:I(lambdaN,Phi)toC}}
\label{sec:proofofI(lambdaN,Phi)toC}
Define $Q^N\equiv\bm\lambda^N\Phi$. Noting that $\sum_{i=1}^m\lambda^\ast_iD(P^i\|Q^\ast)=C$ holds by the Kuhn-Tucker condition (\ref{eqn:Kuhn-Tucker}), we have
\begin{align}
0<C-I(\bm\lambda^N,\Phi)&=C-\sum_{i=1}^m\lambda^N_i\sum_{j=1}^nP^i_j\log\left(\dfrac{P^i_j}{Q^\ast_j}\cdot\dfrac{Q^\ast_j}{Q^N_j}\right)\\
&=C-\sum_{i=1}^m\left(\lambda^\ast_i+\mu^N_i\right)D(P^i\|Q^\ast)+D(Q^N\|Q^\ast)\\
&=-\sum_{i=1}^m\mu^N_iD(P^i\|Q^\ast)+D(Q^N\|Q^\ast).\label{eqn:sumi=1mmuNiD(Pi|Qast)+D(QN|Qast)}
\end{align}
We will evaluate $D(Q^N\|Q^\ast)$ in (\ref{eqn:sumi=1mmuNiD(Pi|Qast)+D(QN|Qast)}). Defining $R^N\equiv Q^N-Q^\ast$ with $R^N_j=Q^N_j-Q^\ast_j,\,j=1,\ldots,n$, we have
\begin{align}
D(Q^N\|Q^\ast)&=\sum_{j=1}^n\left(Q^\ast_j+R^N_j\right)\log\left(1+\dfrac{R^N_j}{Q^\ast_j}\right)\\
&=\sum_{j=1}^n\left(Q^\ast_j+R^N_j\right)\left\{\dfrac{R^N_j}{Q^\ast_j}-\dfrac{1}{2}\left(\dfrac{R^N_j}{Q^\ast_j}\right)^2+o\left(\|R^N\|^2\right)\right\}\\
&=\sum_{j=1}^nR^N_j+\sum_{j=1}^n\dfrac{\left(R^N_j\right)^2}{Q^\ast_j}-\dfrac{1}{2}\sum_{j=1}^n\dfrac{\left(R^N_j\right)^2}{Q^\ast_j}+o\left(\|R^N\|^2\right)\\
&=0+\dfrac{1}{2}\sum_{j=1}^n\dfrac{1}{Q^\ast_j}\left(\sum_{i=1}^m\mu^N_iP^i_j\right)^2+o\left(\|\bm\mu^N\|^2\right).\label{eqn:0+1/2sumj=1n1/Qastj(sumi=1mmuNiPij)2}
\end{align}
Let ${\cal I}_{\rm III}=\emptyset$, then we have $D(P^i\|Q^\ast)=C,\,i=1,\ldots,m$. By $\sum_{i=1}^m\mu^N_i=0$, the first term of (\ref{eqn:sumi=1mmuNiD(Pi|Qast)+D(QN|Qast)}) is 0. Therefore, if $\|\bm\mu^N\|<K_1\left(\theta\right)^N$, then we obtain (\ref{eqn:IlambdaPhitoC_1}) by (\ref{eqn:0+1/2sumj=1n1/Qastj(sumi=1mmuNiPij)2}). While if $\lim_{N\to\infty}N\mu^N_i=\sigma_i$, then we obtain (\ref{eqn:IlambdaPhitoC_2}) by (\ref{eqn:0+1/2sumj=1n1/Qastj(sumi=1mmuNiPij)2}).

Next, let ${\cal I}_{\rm III}\neq\emptyset$. By (\ref{eqn:sumi=1mmuNiD(Pi|Qast)+D(QN|Qast)}), we have
\begin{align}
C-I(\bm\lambda^N,\Phi)\leq\sum_{i=1}^m|\mu^N_i|D(P^i\|Q^\ast)+O\left(\|\bm\mu^N\|^2\right).\label{eqn:C-I(lambdaN,Phi)leqsumi=1m|muNi|D(Pi|Qast)}
\end{align}
Therefore, if $\|\bm\mu^N\|<K_1\left(\theta\right)^N$, then we obtain (\ref{eqn:IlambdaPhitoC_3}) by (\ref{eqn:C-I(lambdaN,Phi)leqsumi=1m|muNi|D(Pi|Qast)}). While if $\lim_{N\to\infty}N\mu^N_i=\sigma_i$, then we obtain (\ref{eqn:IlambdaPhitoC_4}) by (\ref{eqn:C-I(lambdaN,Phi)leqsumi=1m|muNi|D(Pi|Qast)}).\hfill$\blacksquare$

\end{document}